\documentclass[a4paper,oneside,final,notitlepage,onecolumn,12pt]{article}
\usepackage{amsfonts}
\usepackage{epsf}
\usepackage{graphics}
\usepackage{graphicx}
\usepackage{amssymb,eso-pic}
\usepackage{latexsym}
\usepackage{tabularx}
\usepackage{amsxtra}
\usepackage{hyperref}
\usepackage{t1enc}
\usepackage{amsmath,accents}
\allowdisplaybreaks 
\usepackage{bbm}
\usepackage{enumerate}
\usepackage{cancel}

\usepackage{ifpdf,epsfig,array,amsmath,amssymb}

\usepackage{psfrag} 

\usepackage{cite} 
\usepackage{float}

\newcounter{mnotecount}

\newcommand{\mnotex}[1]
{\protect{\stepcounter{mnotecount}}$^{\mbox{\footnotesize $\bullet$\themnotecount}}$ 
	\marginpar{
		\raggedright\tiny\em
		$\!\!\!\!\!\!\,\bullet$\themnotecount: #1} }

\usepackage{psfrag}
\psfrag{na}{\footnotesize $n^a$}
\psfrag{ta}{\footnotesize $\sigma^a$}
\psfrag{vna}{\footnotesize $\check{n}^a$}
\psfrag{vNa}{\footnotesize $\check{N}^a$}
\psfrag{scrW}{\footnotesize $\mycal{W}_{\rho_0}$}
\psfrag{t=t1}{\footnotesize $\Sigma_{\sigma_1}$}
\psfrag{t=t2}{\footnotesize $\Sigma_{\sigma_2}$}
\psfrag{r=all}{\footnotesize $\rho=const$}
\psfrag{hna}{\footnotesize $\,\widehat{n}^a$}
\psfrag{tna}{\footnotesize $\tilde{n}^a$}
\psfrag{hab,Kab}{\footnotesize $h_{ab}, K_{ab}$}
\psfrag{thab,tKab}{\footnotesize $\tilde h_{ab}, \tilde K_{ab}$}
\psfrag{hhab,hKab}{\footnotesize $\widehat h_{ab}, \widehat K_{ab}$}
\psfrag{chab,cKab}{\footnotesize $\check{h}_{ab}, \check{K}_{ab}$}
\usepackage{psfrag} 

\psfrag{x}{\footnotesize $x$}
\psfrag{y}{\footnotesize $y$}
\psfrag{z}{\footnotesize $z$} 

\psfrag{x=A}{\tiny $(A,0,0)$}
\psfrag{x=-A}{\tiny $(-A,0,0)$}
\psfrag{y=A}{\tiny $(0,A,0)$}
\psfrag{y=-A}{\tiny $(0,-A,0)$}
\psfrag{z=A}{\tiny $(0,0,A)$}
\psfrag{z=-A}{\tiny $(0,0,-A)$}

\psfrag{s}{\footnotesize $\boldsymbol{\vec{s}}$}
\psfrag{v}{\footnotesize $\boldsymbol{\vec{v}}$}
\psfrag{d}{\footnotesize $\boldsymbol{\vec{d}}$}

\usepackage[normalem]{ulem}
\usepackage{color}
\definecolor{blue}{rgb}{0,0,1}
\definecolor{red}{rgb}{1,0,0}

\definecolor{DGREEN}{rgb}{0,0.7,0.3}
\definecolor{grey1}{rgb}{0.52, 0.52, 0.51}

\newcommand{\interior}[1]{\accentset{\smash{\raisebox{-0.1ex}{$\scriptstyle\circ$}}}{#1}\rule{0pt}{2.3ex}}
\newcommand{\instar}[1]{\accentset{\smash{\raisebox{-0.12ex}{$\scriptstyle\star$}}}{#1}\rule{0pt}{2.3ex}}
\makeatletter
\def\@xfootnote[#1]{%
  \protected@xdef\@thefnmark{#1}%
  \@footnotemark\@footnotetext}
\makeatother

\DeclareFontFamily{OT1}{rsfs}{} \DeclareFontShape{OT1}{rsfs}{m}{n}{
<-7> rsfs5 <7-10> rsfs7 <10-> rsfs10}{}
\DeclareMathAlphabet{\mycal}{OT1}{rsfs}{m}{n}

\def\sc{{\hskip 3.5pt {{}^{{}^{{}_{{}_{\bowtie}}}}} \kern -8.pt{}}}  
\def\SC{{\hskip 3.5pt {{}^{{}^{{}^{{}_{{}_{\bowtie}}}}}} \kern -10.5pt{}}}

\def\d{{\rm d}}
\def\preA{\hspace{-0.1cm}^{^{(A)}}\hskip-0.5mm}

\newcommand{\be}{\begin{equation}}
\newcommand{\ee}{\end{equation}}
\newcommand{\ben}{\begin{eqnarray}}
\newcommand{\een}{\end{eqnarray}}

\newcommand{\px}{\partial_x}
\newcommand{\py}{\partial_y}

\newcommand{\pr}{\partial_\rho}
\newcommand{\pa}{\boldsymbol\partial}
\newcommand{\opa}{\overline\pa}

\newcommand{\Nh}{\widehat{\mathbb N}}
\newcommand{\smk}{\mathbbm k}
\newcommand{\osmk}{\overline\smk}
\newcommand{\KK}{\mathbb K}

\newcommand{\Nt}{\widetilde{\mathbb N}}
\newcommand{\sma}{\mathbbm a}
\newcommand{\smb}{\mathbbm b}
\newcommand{\smd}{\mathbbm d}
\newcommand{\biA}{\mathbbm A}
\newcommand{\biB}{\mathbbm B}
\newcommand{\biC}{\mathbbm C}

\newcommand{\oNt}{\overline\Nt}
\newcommand{\osmb}{\overline\smb}
\newcommand{\obiA}{\overline\biA}
\newcommand{\obiB}{\overline\biB}
\newcommand{\obiC}{\overline\biC}

\newcommand{\boka}{\boldsymbol\kappa}

\newcommand{\DNh}{\Delta\Nh}
\newcommand{\Dsmk}{\Delta\smk}
\newcommand{\Dosmk}{\Delta\osmk}
\newcommand{\DKK}{\Delta\KK}

\newcommand{\ANh}{\preA\Nh}
\newcommand{\Asmk}{\preA\smk}
\newcommand{\Aosmk}{\preA\osmk}
\newcommand{\AKK}{\preA\KK}

\DeclareMathAlphabet{\mathpzc}{OT1}{pzc}{m}{it}

\newcommand{\hoch}[1]{$\, ^{#1}$}

\newcommand{\auth}{Anna Nakonieczna \hoch{1,}\,\footnote[$\S$]{~email: Anna.Nakonieczna@fuw.edu.pl}, {\L}ukasz Nakonieczny \hoch{1,}\,\footnote[$\flat$]{~email: Lukasz.Nakonieczny@fuw.edu.pl},  Istv\'an R\'{a}cz\hoch{1,2,}\,\footnote[$\sharp$]{~email: racz.istvan@wigner.mta.hu}}

\begin{document}

\newtheorem{theorem}{Theorem}[section]
\newtheorem{lemma}{Lemma}[section]
\newtheorem{proposition}{Proposition}[section]
\newtheorem{corollary}{Corollary}[section]
\newtheorem{conjecture}{Conjecture}[section]
\newtheorem{example}{Example}[section]
\newtheorem{definition}{Definition}[section]
\newtheorem{remark}{Remark}[section]
\newtheorem{exercise}{Exercise}[section]
\newtheorem{axiom}{Axiom}[section]
\renewcommand{\theequation}{\thesection.\arabic{equation}}

\begin{center}

{\LARGE{\bf Black hole initial data by numerical integration of the parabolic-hyperbolic form of the constraints }}

\vspace{20pt}
\auth

\vspace{7pt}{\hoch{1}
	\it Institute of Theoretical Physics, Faculty of Physics, University of Warsaw} \\
	{ul. Pasteura 5, 02-093 Warszawa, Poland}

\vspace{7pt}{\hoch{2}
	\it Wigner Research Center for Physics} \\  {H-1121 Budapest, Konkoly Thege Mikl\'os \'ut 29-33. Hungary}


\vspace{10pt}

\today

\begin{abstract}
The parabolic-hyperbolic form of the constraints is integrated numerically. The applied numerical stencil is $4^{th}$ order accurate (in the spatial directions) while 'time'-integration is made by using the method of lines with a $4^{th}$ order accurate Runge-Kutta scheme. The proper implementation of the applied numerical method is verified by convergence tests and monitoring the relative and absolute errors determined by comparing numerical and analytically known solutions of the constraints involving boosted and spinning vacuum single black hole configurations. The main part of our investigations is, however, centered on construction of initial data for distorted black holes which, in certain cases, have non-negligible gravitational wave content. Remarkably the applied new method is unprecedented in that it allows to construct initial data for highly boosted and spinning black holes, essentially for the full physical allowed ranges of these parameters. In addition, the use of the evolutionary form of the constraints is free from applying any sort of boundary conditions in the strong field regime.

\end{abstract} 

\end{center}

\newpage
\tableofcontents

\newpage

\section{Introduction}\label{introduction}
\setcounter{equation}{0}

This paper is to report on a systematic numerical integration of the parabolic--hyperbolic form of the constraints \cite{racz_constraints} in studying single black hole configurations in the four dimensional vacuum case.
The proper implementation of the evolutionary form of the constraints and the applied numerical method was verified by restricting attention first to stationary black hole configurations. Considerably large part of our investigations was devoted to the construction of initial data for distorted black hole configurations. The latter type of initial data is expected to come with some amount of gravitational wave content. 
The ultimate aim of the undertaken research is constructing initial data sets for binary black hole systems, which will have a potential to compete with the outcomes obtained with the widely used elliptic methods. In the current paper a first step towards this goal is presented, that is single black hole solutions are investigated in detail.

\medskip

The constraint equations were solved using the method of lines, with a finite differences discretization in the directions cognate to the spatial ones and a Runge-Kutta `time'-integrator in one of the spatial directions playing the role of `time'. The numerical accuracy applied in both spatial and `temporal' directions was of 4$^{th}$ order.  
The employed numerical scheme enabled us to solve the constraints in two alternative ways: using a full and a  deviations based formalisms. In the applied numerical algorithm the evolutionary equations were solved on a cubical subset of the $t=0$ slice of the auxiliary Minkowski background of the Kerr-Schild black hole. This $t=0$ slice---that is horizon penetrating and going out to spacelike infinity---was covered by Cartesian coordinates $\left(x,y,z\right)$ arranged such that the (pointlike or ringlike) singularity of the pertinent Kerr-Schild black hole was always confined to the $z=0$ plane. In particular, the spin was assumed to be aligned with the $z$-axis, whereas (as indicated in Fig.\ref{kocka}) the speed and the displacement were aligned with the $x$- and $y$-axis, respectively. 
The spatial initial slice was foliated by the $z=const$ planes and hence the $z$ coordinate played the role of 'time'.

\medskip

Our numerical results were tested against analytically known data using L$^2$ norms, as well as the absolute and relative errors as accuracy indicators. It turned out that even the strictest, or in some sense the  most sensitive, criteria based on the use of relative errors confirmed the appropriateness of our numerical setup. 
Regarding the time efficiency of our numerical computations, the deviations formulation was found to be superior to the full setup  thus it will be favorable to use the deviation based {scheme} in constructing initial data for black hole binaries. 

\medskip	

In applying both the full and deviation based formulations the functional dependence of the constrained variables was monitored for a wide range of black hole configurations. These were supplemented by convergence tests and a large variety of error indicators of the constrained variables which provided an adequate measure of the appropriateness of the applied numerical schema. To have a better understanding of the  investigated vacuum black hole initial data specifications some gauge invariant quantities were also depicted.

\medskip	
	
The most important property of the applied method--ensured by the use of the parabolic-hyperbolic form of the constraints--is that, in contrast to other proposals (see, e.g.~\cite{pfeiffercookteukolsky,yocookshapirobaumgarte,jaramillogourgoulhonmarugan,cookpfeiffer,hannamcook,cook,gourgoulhon,alcubierre} and references therein), our method does not require {using} any sort of assumption, e.g.~some boundary data on the constrained fields, in the strong field regime. Indeed, in the applied new formalism the constrained variables emerge everywhere, in particular in the vicinity of the singularity, by solving the pertinent evolutionary system.	

\medskip

It is worth stressing that the combination of the parabolic-hyperbolic formulation of the constraints and the applied numerical method was found to be capable {of investigating} black holes even when their spin and speed were close to the maximal values of their physically adequate ranges. Despite the great efforts invested in constructing initial data for highly spinning black holes within the conformal approach~\cite{lovelaceowenpfeifferchu,loustonakanozlochowermundimcampanelli}, up to now, merely some  boost-restricted black hole initial data could be deduced~\cite{ruchlinhealyloustozlochower}. As opposed to this, to our best knowledge, our proposal is the first one which essentially comes without any limitation on the physical range of the allowed boosts. 
It has to be stressed that the referenced articles concentrate on black hole binaries, while the current paper addresses the issue of single black holes. The above statement on the lack of limitations on the black hole parameters definitely requires a future confirmation in the binary case.

\medskip

The parabolic-hyperbolic formulation of the constraint equations has been up to now employed in investigating the vacuum binary black hole initial data without spin, i.e. binary Schwarzschild solutions, with the boundary conditions set in the strong field region~\cite{beyer2019}. The spatial domain was foliated with 2-spheres, the equations were integrated outwards towards the spacelike infinity and apart from the construction of the data the asymptotic flatness of the solutions was discussed. The issue of asymptotic flatness of the near Schwarzschild vacuum solutions within the parabolic-hyperbolic formulation was further discussed in the case of boundary conditions imposed on an arbitrary 2-sphere located between the strong field region and spatial infinity~\cite{csukasracz2020}. Unrestricted single and binary black hole initial data sets were elaborated on using an explicit PDE solver of the parabolic-hyperbolic set of constraints in~\cite{doulis2019}. As was explained in detail in the previous paragraphs, our approach discusses single black hole solutions admitting all possible black hole parameters (mass, spin, boost and displacement) within their full physically admissible ranges. It concentrates on the construction of initial data with the planar foliation of the spatial domain and using two formalisms -- full and deviations based. The boundary conditions were posed in the weak field regime and the adequate equations were integrated inwards, towards the strong field region. Therefore, new aspects of employing the parabolic-hyperbolic formulation of the initial data were introduced in the current research, namely the lack of restrictions on the black hole parameters, the use of an implicit PDE solver and most notably investigating the deviations based formalism of the formulation under study.

\medskip

The paper is organized as follows. Section~\ref{sec:theorback} provides the theoretical background of the undertaken problem. In particular, the parabolic-hyperbolic form of the constraints, along with the  applied basic variables, is recalled briefly. 
Some useful set of variables, with a definite conformal weight, is introduced in Section~\ref{sec:constr}. Applying these variables the parabolic-hyperbolic equations are recast and their full and deviations based forms are given explicitly. Section~\ref{sec:numcomp} is to introduce our numerical setup. Sections~\ref{sec:fullform-num} and~\ref{sec:devs-num} contain a detailed discussion on the code accuracy tests and solutions of the constraint equations within the full and deviation {based} formulations, for single non-distorted and distorted black holes, respectively. Section~\ref{sec:concl} {is to} summarize our results{, while} in the Appendix some of the more technical numerical expressions, including the explicit form of the applied derivative operators are presented.

\section{The theoretical background
	}\label{sec:theorback}
\setcounter{equation}{0}
\subsection{The parabolic--hyperbolic form of the constraints}\label{Parabolic-hyperbolic-system}

Recall first that{, in the vacuum case,} the initial data consists of a Riemannian metric $h_{ij}$ and a symmetric tensor field $K_{ij}$, both given on a three-dimensional manifold $\Sigma$ \cite{choquet,wald}. These fields are not free, they are subject to the Hamiltonian and momentum constraints, which---for the conventional choice of variables---are known to form a quasi-linear elliptic system \cite{Lichnerowicz,York0}. Nevertheless, whenever $\Sigma$ can be foliated by the level surfaces of a smooth function $\rho: \Sigma \rightarrow \mathbb{R}$, i.e.~by a one-parameter family of homologous two-surfaces $\mycal{S}_\rho$, for a suitably chosen set of dependent variables, the constraints happen to form a parabolic--hyperbolic system \cite{racz_constraints}. (The specific choice of the foliation used in the current studies along with graphical explanations of the intrinsic black hole characteristics is shown in Fig.~\ref{kocka} and will be discussed in detail in Section~\ref{sec:boundinival} below.) Indeed, using the scalar, vector and tensor projections
\begin{equation}\label{hij}
h_{ij}=\widehat \gamma_{ij}+\widehat  n_i \widehat n_j\,,
\end{equation}
\begin{equation}
K_{ij}= \boka \,\widehat n_i \widehat n_j  + \left[\widehat n_i \,{\rm\bf k}{}_j  
+ \widehat n_j\,{\rm\bf k}{}_i\right]  + {\rm\bf K}_{ij}\,,
\end{equation}
of the fields $h_{ij}$ and  $K_{ij}$, where $\widehat \gamma_{ij}$ is the metric induced on the level surfaces and $\boka=\widehat  n^k \widehat n^l K_{kl}$, ${\rm\bf k}{}_i=\widehat \gamma^k_i \widehat n^l K_{kl}$, ${\rm\bf K}_{ij}=\widehat \gamma^k_i \widehat \gamma^l_j K_{kl}$, along with the decomposition of the unit normal to the $\mycal{S}_\rho$ level surfaces
\begin{equation}\label{nhat}
\widehat n^i={\,\widehat{N}}^{-1}\,[\, \rho^i-{\widehat N}{}^i\,]
\end{equation}
---given in terms of the `lapse' $ \widehat N$ and `shift' $\widehat N^i$ of {an arbitrary but fixed} `time evolution' vector field $\rho^i$, satisfying the relation $\rho^i \partial_i \rho=1$--- the Hamiltonian and momentum constraints can be written as \cite{racz_constraints} (see also  \cite{racz_geom_det,racz_geom_cauchy,racz_tdfd})
\begin{align}
{}& \instar{K}\,[\,(\partial_{\rho} \widehat N) - \widehat N{}^l(\hat D_l\widehat N) \,] - \widehat N^{2} (\hat D^l \hat D_l \widehat N) - \mathcal{A}\,\widehat N - \mathcal{B}\,\widehat N{}^{3} = 0 \,, \label{bern_pde} \\ 
{}& \mycal{L}_{\hat n} {\rm\bf k}{}_{i} - \tfrac12\,\hat D_i ({\rm\bf K}^l{}_{l}) - \hat D_i\boka + \hat D^l \interior{\rm\bf K}{}_{li} + \widehat N{}^{-1}\instar{K}\,{\rm\bf k}{}_{i}  + [\,\boka-\tfrac12\, ({\rm\bf K}^l{}_{l})\,]\,\dot{\hat n}{}_i - \dot{\hat n}{}^l\,\interior{\rm\bf K}_{li} = 0 \label{par_const_n}, \\
{}& \mycal{L}_{\hat n}({\rm\bf K}^l{}_{l}) - \hat D^l {\rm\bf k}_{l} - \widehat N{}^{-1}\instar{K}\,[\,\boka-\tfrac12\, ({\rm\bf K}^l{}_{l})\,]  + \widehat N{}^{-1}\interior{\rm\bf K}{}_{kl}\instar{K}{}^{kl}  + 2\,\dot{\hat n}{}^l\, {\rm\bf k}_{l}  = 0\,, \label{ort_const_n}
\end{align}
where ${\rm\bf K}^l{}_{l}$ {and $\interior{\rm\bf K}_{ij}$} denote the trace $\widehat\gamma^{kl}{\rm\bf K}_{kl}$ {and trace free part ${\rm\bf K}_{ij}-\tfrac12\,\widehat \gamma_{ij}\,{\rm\bf K}^l{}_{l}$} of ${\rm\bf K}_{kl}$, respectively, $\hat D_i$ stands for the covariant derivative operator associated with $\hat \gamma_{ij}$ and $\dot{\hat n}{}_k={\hat n}{}^lD_l{\hat n}{}_k=-{\hat D}_k(\ln{\widehat N})$, and the notation 
\begin{align}
{}& \instar{K}_{ij}=\tfrac12\mycal{L}_{\rho} {\hat \gamma}_{ij} -\hat  D_{(i}\widehat N_{j)}, \\ 
{}&  \instar{K}  =\tfrac12\,{\hat \gamma}^{ij}\mycal{L}_{\rho} {\hat \gamma}_{ij} -  \hat D_j\widehat N^j\,,\label{eq:trhatext} \\
{}& \mathcal{A} =(\partial_{\rho} \instar{K}) - \widehat N{}^l (\hat D_l \instar{K}) + \tfrac{1}{2}[\,\instar{K}^2 + \instar{K}{}_{kl} \instar{K}{}^{kl}\,], \label{A}  \\
{}& \mathcal{B} =  -\tfrac12\,\bigl[\widehat{R} + 2\,\boka\,({\rm\bf K}^l{}_{l})+\tfrac12\,({\rm\bf K}^l{}_{l})^2 
-2\,{\rm\bf k}{}^{l}{\rm\bf k}{}_{l}  - \interior{\rm\bf K}{}_{kl}\,\interior{\rm\bf K}{}^{kl}\,\bigr] \label{B}
\end{align}
was applied. 

\medskip

Recall that (\ref{bern_pde}) stands for the Hamiltonian constraint and it is a Bernoulli-type parabolic partial differential equation in those subsets of $\Sigma$ where $\instar{K}$ can be guaranteed to be strictly positive or negative. Whenever this happens equation (\ref{bern_pde}) is uniformly parabolic such that $\rho$ plays the role of `time' and  $\rho^i$ gets indeed to be a `time evolution' vector field (for more details see \cite{racz_constraints}). It is also important that the subsystem (\ref{par_const_n})--(\ref{ort_const_n}) comprises a first order symmetrizable hyperbolic system that is also linear in the dependent variables ${\rm\bf k}{}_{i}$ and ${\rm\bf K}^l{}_{l}$ \cite{racz_constraints}. 

\medskip

More importantly, the coupled parabolic--hyperbolic system (\ref{bern_pde})--(\ref{ort_const_n}) possesses a well-posed initial value problem for the dependent variables $\widehat N, {\rm\bf k}{}_{i}, {\rm\bf K}^l{}_{l}$ that can always be solved locally \cite{racz_constraints}.
Note also that in virtue of (\ref{bern_pde})--(\ref{ort_const_n}) the variables $\widehat N, {\rm\bf k}{}_{i}, {\rm\bf K}^l{}_{l}$ are subject to the constraints whereas the remaining {four} fields 
$\widehat N^i,\widehat \gamma_{ij}, \boka, \interior{\rm\bf K}_{ij}$ are
freely specifiable throughout $\Sigma$.

\subsection{The constraints in new variables}\label{sec:constr}

In solving the parabolic--hyperbolic system (\ref{bern_pde})--(\ref{ort_const_n}) we shall use foliation of $\Sigma$ by planes. For this reason it turns out to be rewarding to apply a reference flat metric $q _{ab}$ on $\mathbb{R}^2$ determined as \cite{i_jeff_3}
\begin{equation}\label{eq:flat_metric}
q _{ab} = q_{(a} \,\overline q_{b)}\,,  
\quad q^{ab} = q^{(a} \,\overline q{}^{\,b)}\,, \quad q^{ae} q_{eb} =\delta^a{}_b\,
\end{equation} 
in terms of a complex dyad 
\begin{equation}\label{eq:dual-dyad}
	q_a = (\d x)_a + \mathbbm{i}\,(\d y)_a\,, \quad q^a = (\partial_{x})^a + \mathbbm{i}\,(\partial_{y})^a\,,
\end{equation}
which is normalized as
\begin{equation}\label{eq:flat_metric-norm}
q^a \,\overline q_a =2\,, \quad q^a q_a =0\,. 
\end{equation}

\medskip

Consider now the contraction 
\begin{equation}\label{parc-def3}
\mathbb{L}=q^{a_1}\dots q^{a_w}\, \mathbf{L}_{{a_1}\dots{a_w}}\,,
\end{equation} 
where $\mathbf{L}_{{a_1}\dots{a_w}}$ is a totally symmetric traceless tensor field on $\mathbb{R}^2$. $\mathbb{L}$ can be seen to be a conformal weight $w$ function  on $\mathbb{R}^2$. Using the standard complex structure defined there, analogs of the $\eth$ and $\overline{\eth}$ operators can be defined by making use of the torsion free covariant derivative operator associated with $q_{ab}$. This is indeed the partial derivative operator $\partial_a$ with respect to the chosen Descartes type coordinates  $(x,y)$---corresponding also to the holomorphic coordinate $x+\mathbbm{i}\,y$---on $\mathbb{R}^2$ defined as
\begin{align}
\pa\,\mathbb{L} = {}& q^b q^{a_1}\dots q^{a_s}\, \partial_b\mathbf{L}_{({a_1}\dots{a_s})}\,,\\
\,\opa\,\mathbb{L} = {}& {\,\overline q}^b q^{a_1}\dots q^{a_s}\, \partial_b\mathbf{L}_{({a_1}\dots{a_s})}\,.
\end{align}

\medskip

This, along with the relation $q^i=(\partial_{x})^i+ \mathbbm{i}\,(\partial_{y})^i$, implies, in particular, that for any symmetric tensor field $\boldsymbol{t}_{ij}$ the contractions $q^i\overline{q}{}^j{\bf t}_{ij}$ and $q^iq^j{\bf t}_{ij}$ can be evaluated as 
\begin{align}
q^i \overline{q}{}^j{\bf t}_{ij}= {}& {\bf t}_{xx}+{\bf t}_{yy}, \\
q^i q^j{\bf t}_{ij}= {}& {\bf t}_{xx}-{\bf t}_{yy} + 2\,\mathbbm{i}\,{\bf t}_{xy}\,.
\end{align}

\medskip

In proceeding note first that the complex dyad $q_a$ defined merely on a $\rho=const$ level surface can be extended onto the other $\rho=const$ level surfaces by Lie dragging  $q_a$ along the $\rho$-streamlines. If this is done, for instance, the metric $\widehat\gamma_{ab}$, induced on the $\rho=const$ level surfaces, can be decomposed throughout $\Sigma$ as
\begin{equation}\label{ind_metr}
\widehat\gamma_{ab}=\sma\, q_{ab}+\tfrac12\left[ \smb \,\overline q_a \,\overline q_b + \,\osmb\, q_a q_b \right]\,, 
\end{equation}
where
\begin{equation}
\sma=\tfrac12\,\widehat\gamma_{ab}\,q^a \,\overline q^b
\end{equation}
is a positive (conformal-weight zero) function on the $\rho=const$ level surfaces, whereas the contraction
\begin{equation}
\smb=\tfrac12\,\widehat\gamma_{ab}\,q^a q^b\,
\end{equation}
is a (conformal-weight $2$) function. 

\medskip

The inverse $\widehat\gamma^{ab}$ metric can then be given as 
\begin{equation}\label{inv_ind_metr}
\widehat\gamma^{ab}=\smd^{-1}\left\{\sma\, q^{ab}-\tfrac12\left[ \smb \,\overline q^a \,\overline q^b
+ \,\osmb\, q^a q^b \right]\right\}\,, 
\end{equation}
where 
\begin{equation}
\smd=\sma^2-\smb\,\osmb
\end{equation}
is equal to the determinant $\det(\widehat\gamma_{ab})$ of $\widehat\gamma_{ab}$ as $\det(q_{ab})=1$.

\subsubsection{The full form of the constraint equations}
\label{sec:fullform}

Following the generic procedure applied in \cite{i_jeff_2,i_jeff_3} conformal-weighted variables, as collected in Table~\ref{table:data}, can be introduced. This section is to give the explicit form of the parabolic--hyperbolic form of the constraints by making use of these variables.
\begin{table}[H]
	\centering  \hskip-.15cm
	\begin{tabular}{|c|c|c|} 
		\hline notation &  definition  & \hskip-0.7cm$\phantom{\frac{\frac12}{A}_{B_D}}$ 
		spin-weight \\ \hline \hline
		
		$\mathbbm{a}$ &  $\tfrac12\,q^i\,\overline q^j\,\widehat\gamma_{ij}$  & \hskip-0.7cm$  
		\phantom{\frac{\frac12}{A}_{B_D}}$ $0$ \\  \hline 
		
		$\mathbbm{b}$ &  $\tfrac12\,q^i q^j\,\widehat\gamma_{ij}$ 
		& \hskip-0.7cm$\phantom{\frac{\frac12}{A}_{B_D}}$ $2$ \\  \hline 
		
		$\mathbbm{d}$ &  $\mathbbm{a}^2-\mathbbm{b}\,\overline{\mathbbm{b}}$  
		& \hskip-0.7cm$\phantom{\frac{\frac12}{A}_{B_D}}$ $0$ \\  \hline 
		$\mathbbm{k}$ &  $q^i {\rm\bf k}{}_{i}$  & 
		\hskip-0.7cm$\phantom{\frac{\frac12}{A}_{B_D}}$ $1$
		\\  \hline 
		
		$\mathbb{A}$ &  $q^a q^b {C^e}{}_{ab}\,\overline q_e
		= \mathbbm{d}^{-1}\left\{ \mathbbm{a}\left[2\,\eth\,\mathbbm{a}
		-\,\overline{\eth}\,\mathbbm{b}\right] -  \,\overline{\mathbbm{b}}\,\eth\,\mathbbm{b} \right\} $ 
		& \hskip-0.7cm$\phantom{\frac{\frac12}{A}_{B_D}}$ $1$ \\  \hline 
		
		$\mathbb{B}$ &  $\,\overline q^a q^b {C^e}{}_{ab}\,q_e
		= \mathbbm{d}^{-1}\left\{ \mathbbm{a}\,\overline{\eth}\,\mathbbm{b}
		- \mathbbm{b}  \,\eth\,\overline{\mathbbm{b}}\right\}$ 
		& \hskip-0.7cm$\phantom{\frac{\frac12}{A}_{B_D}}$ $1$ \\  \hline 
		
		$\mathbb{C}$ &  $q^a q^b {C^e}{}_{ab}\,q_e 
		= \mathbbm{d}^{-1}\left\{ \mathbbm{a}\,\eth\,\mathbbm{b}
		-  \mathbbm{b}\left[2\,\eth\,\mathbbm{a} 
		-\,\overline{\eth}\,\mathbbm{b}\right] \right\}$ 
		& \hskip-0.7cm$\phantom{\frac{\frac12}{A}_{B_D}}$ $3$ \\  \hline 
		
		$\,\widehat{\mathbb{R}}$ &  $\tfrac12\, {\mathbbm{a}}^{-1}\left(\,2\, \mathbb{R}
		- \left\{ \,  \eth\,\overline{\mathbb{B}} - \overline{\eth}\,\mathbb{A}
		- \tfrac12\,\left[\, \mathbb{C}\,\overline{\mathbb{C}} 
		- \mathbb{B}\,\overline{\mathbb{B}} \,\right]\, \right\}\,\right)$ 
		& \hskip-0.7cm$\phantom{\frac{\frac12}{A}_{B_D}}$ $0$ \\  \hline 
		
		$\,\widehat{\mathbb{N}}$ &  $\widehat N$  
		& \hskip-0.7cm$\phantom{\frac{\frac12}{A}_{B_D}}$ $0$ \\  \hline
		$\mathbb{N}$ &  $q^i\widehat N_i= q^i \widehat\gamma{}_{ij} {\widehat N}{}^{j}$ 
		& \hskip-0.7cm$\phantom{\frac{\frac12}{A}_{B_D}}$ $1$ \\  \hline
		
		$\widetilde{\mathbb{N}}$ &  $q_i\widehat N^i = q_i \,\widehat\gamma{}^{ij} {\widehat N}{}_{j}
		=\mathbbm{d}^{-1} (\mathbbm{a}\,\mathbb{N} - \mathbbm{b}\,\overline{\mathbb{N}})$  
		& \hskip-0.7cm$\phantom{\frac{\frac12}{A}_{B_D}}$ $1$ \\  \hline
		
		$\mathbb{K}$ &  $ \widehat\gamma^{kl} \,{\rm\bf K}{}_{kl}
		$  & \hskip-0.7cm$\phantom{\frac{\frac12}{A}_{B_D}}$ $0$ \\  \hline 
		
		$\interior{\mathbb{K}}{}_{qq}$ &  $q^kq^l\,\interior{\rm\bf K}{}_{kl}
		$  & \hskip-0.7cm$\phantom{\frac{\frac12}{A}_{B_D}}$ $2$ \\  \hline 
		
		$\interior{\mathbb{K}}{}_{q\overline{q}}$ &  $q^k\,\overline{q}^l\,\interior{\rm\bf K}{}_{kl}
		= (2\,\mathbbm{a})^{-1} [\,\mathbbm{b}\,\overline{\interior{\mathbb{K}}{}_{qq}} 
		+  \overline{\mathbbm{b}}\,\interior{\mathbb{K}}{}_{qq} \,] 
		$  & \hskip-0.7cm$\phantom{\frac{\frac12}{A}_{B_D}}$ $0$ \\  \hline   
		
		

		
		$\,\instar{\mathbb{K}}$ &  ${\instar{K}}{}^l{}_{l} = \widehat\gamma^{ij} \instar{K}_{ij} $  
		& \hskip-0.7cm$\phantom{\frac{\frac12}{A}_{B_D}}$ $0$ \\  \hline 
		
		$\instar{\mathbb{K}}{}_{qq}$ &  $q^i q^j\instar K_{ij} 
		= \tfrac12\,\left\{2\,\partial_\rho\mathbbm{b} - 2\,\eth\,\mathbb{N}
		+ {\mathbb{C}}\,\overline {\mathbb{N}} +\mathbb{A} \,{\mathbb{N}} \,  \right\} $ 
		& \hskip-0.7cm$\phantom{\frac{\frac12}{A}_{B_D}}$ $2$ \\  \hline   
		
		$\instar{\mathbb{K}}{}_{q\overline{q}}$ &  $q^k\,\overline{q}^l\,\instar{K} {}_{kl} 
		=  {\mathbbm{a}}^{-1}\{\,\mathbbm{d}\cdot\instar{\mathbb{K}} 
		+ \tfrac12 \,[\,\mathbbm{b}\,\overline{\instar{\mathbb{K}}{}_{qq}} 
		+  \overline{\mathbbm{b}}\,\instar{\mathbb{K}}{}_{qq}  \,]\,\}
		$  & \hskip-0.7cm$\phantom{\frac{\frac12}{A}_{B_D}}$ $0$ \\  \hline                  
		
	\end{tabular}
	\caption{\small The most fundamental conformal-weighted quantities applied in (\ref{bern_pde2})--(\ref{eq:B2}) below. Note that as the metric $q_{ab}$ is flat its scalar curvature $\mathbb{R}$, in the expression for $\widehat{\mathbb{R}}$, vanishes.
	}\label{table:data} 
\end{table}

\medskip

By applying the above introduced conformal-weighted variables the parabolic-hyperbolic system (\ref{bern_pde})--(\ref{ort_const_n}) reads as
\begin{align}
\instar\KK & \,\Big[ \pr\Nh - \tfrac12 \Nt \big(\opa\Nh\big) - \tfrac12 \oNt \big(\pa\Nh\big) \Big]  
- \tfrac12 \smd^{-1} \Nh^2 \bigg\{ \sma \Big[ \pa\opa\Nh - \biB\big(\opa\Nh\big) - \smb\big(\opa^2\Nh\big) \nonumber \\ & \hskip3.4cm - \tfrac12 \obiA\big(\opa\Nh\big)
- \tfrac12 \obiC\big(\pa\Nh\big) \Big] + ``\,CC\," \bigg\}   - \mathcal{A}\Nh - \mathcal{B}\Nh^3 = 0\,,
\label{bern_pde2} \\ 
\pr\smk &- \tfrac12 \Nt\big(\opa\smk\big) - \tfrac12 \oNt\big(\pa\smk\big) - \tfrac12 \Nh\big(\pa\KK\big) + \mathbbm{f}_\smk = 0\,,
\label{eq:eth_constr_mom1} \\
\pr\KK &- \tfrac12 \Nt\big(\opa\KK\big) - \tfrac12 \oNt\big(\pa\KK\big) - \tfrac12 \Nh\smd^{-1} \Big[ \sma\left( \pa\osmk + \opa\smk \right)
- \smb\big(\opa\osmk\big) - \osmb\big(\pa\smk\big) \Big] + \mathbb{F}_\KK = 0\,.
\label{eq:eth_constr_mom2}
\end{align}
In equations (\ref{bern_pde2})--(\ref{eq:eth_constr_mom2}) the source terms, $\mathbbm{f}_\smk$,  $\mathbb{F}_\KK$, and the coefficients, $\mathcal{A}$, $\mathcal{B}$, are smooth functions of the dependent and freely specifiable variables---these are $\Nh$, $\KK$, $\smk$ and 
$\sma$, $\smb$, $\Nh$, $\boka$, $\interior\KK$, respectively,---and the $\pa$, $\opa$ and $\rho$-derivatives of the freely specifiable variables. The{ir} explicit forms  can be given as
\begin{align}
\hskip-0.5cm 
\mathbbm{f}_\smk = & -\tfrac12 \left[ \smk \big(\pa\oNt\big) + \osmk\big(\pa\Nt\big) \right]
- \left( \boka - \tfrac12 \KK \right) \pa\Nh\label{eq:bff}
\\
& \hskip4.2cm + \Nh \left( -\pa\boka + \Nh^{-1} \instar\KK \smk - q^i \dot{\widehat n}^l \interior{\bf K}_{li}
+ q^i \widehat{D}^l \interior{\bf K}_{li} \right)\,, \nonumber  \\
\mathbb{F}_\KK = & \tfrac14\, \Nh\, \smd^{-1} \Big[ 2\sma\biB\osmk - \smb \left( \obiC\smk + \obiA\osmk \right)
+ ``\,CC\," \Big]
\\
& \hskip1.2cm  - \smd^{-1} \left[ \left( \sma\osmk - \osmb\smk \right) \pa\Nh + ``\,CC\," \right] 
+ \left[ \interior{\bf K}_{ij} \instar{K}{}^{ij} - \left( \boka - \tfrac12 \KK \right) \instar\KK \right]\nonumber 
\,, \\
\mathcal{A} = & \,\pr\instar\KK - \tfrac12 \Nt \big(\opa\instar\KK\big) - \tfrac12 \oNt \big(\pa\instar\KK\big) 
+ \tfrac12 \left( \instar\KK{}^2 + \instar\KK_{kl} \instar\KK{}^{kl} \right) 
\,,
\label{A2} \\
\mathcal{B} = & -\tfrac12 \left[ \widehat{R} + 2\boka\KK + \tfrac12 \KK^2 
- \smd^{-1} \left( 2\sma\smk\osmk - \smb\osmk{}^2 - \osmb\smk^2 \right) - \interior{\bf K}_{kl} \interior{\bf K}{}^{kl} 
\right] \,,
\label{eq:B2}
\end{align}
where the explicit forms of some of the terms abbreviated in \eqref{eq:bff}--\eqref{eq:B2} read as 
\begin{align}
\hskip -0.15cm
q^{i\,} \dot{\widehat n}{}^{k\,} \interior{\rm\bf K}{}_{ki}	  = {}&
-\tfrac12 ( {\Nh}\,{\smd})^{-1} \left[
\sma\,(\opa\, \Nh ) \, \interior{\KK}_{qq}    
+  \sma\,( \pa {\Nh} ) \, \interior{\KK}_{q\bar q}  
-\smb\,(\opa {\Nh} ) \, \interior{\KK}_{q\bar q} 
- \osmb\,(\pa {\Nh} ) \,
\interior{\KK}_{qq}   \right]\,, \\
\hskip -0.3cm q^i \widehat D^{k\,} \interior{\rm\bf K}{}_{ki}	 = {}& {} 
\frac{ 1} {2\smd}\left(
\sma\,\opa \, \interior{\KK}_{qq}    
+  \sma\,\pa \, \interior{\KK}_{q\bar q}  
-\smb \,\opa \, \interior{\KK}_{q\bar q} 
- \osmb\, \pa \, \interior{\KK}_{qq}   \right)+ \frac{\overline{ \smb}} {2\smd} \left(
\biA\,\interior{\KK}_{qq}  +\biC\, \interior{\KK}_{q\bar q}  \right) \nonumber 
\\
& \hskip -2.1cm - \frac{ \sma} {4\smd} \left(
3\,\obiB \,\interior{\KK}_{qq} +3\,{\biB}\, \interior{\KK}_{q\bar q} 
+{\biA} \,\interior{\KK}_{q\bar q} +{\biC}\,
\overline{\interior{\KK}_{qq}} \right)
 + \frac{ \smb} {4\smd} \left(
\obiC \,\interior{\KK}_{qq} +\obiA \,\interior{\KK}_{q\bar q} 
+\obiB \,\interior{\KK}_{q\bar q} +{\biB} \,\overline{\interior{\KK}_{qq}} \right)\label{eq:divKnull}  
\,, \\
\hskip -0.4cm\interior{\rm\bf K}{}_{ij} \instar{K}{}^{ij}   =  {}& \tfrac14\, {\smd}^{-2} 
               \left\{ \,2\, \interior{\KK}_{q\bar q}\left[ \left(\,{\sma}^2+{\smb}\,\overline {\smb}\right)\,{\instar{\KK}_{q\bar q} } - {\sma}\left(\,\osmb\,{\instar{\KK}_{qq}}  
               + {\smb} {\,\overline{\instar{\KK}_{qq}} } \right)\,\right] \right. \nonumber \\ 
               {}& \left.\hskip2.2cm + \left[ \,{\overline{\interior{\KK}_{qq}}} 
               \left(\, {\sma}^2\,{\instar{\KK}_{qq}}  
               + {\smb}^2 {\,\overline{\instar{\KK}_{qq}} } 
               -2\,\sma \,\smb \,\instar{\KK}_{q\bar q} \right)
               + ``\,CC\," \right] \right\}\,,  \\
\hskip -1.1cm \instar{K}{}_{ij} \instar{K}{}^{ij}  =  {}&  \tfrac14\, {\smd}^{-2} 
               \left[ \,\overline{\,\instar{\KK}_{qq}}
               \left( {\sma}^2\,\instar{\KK}_{qq} 
               + {\smb}^2 \,\overline{\,\instar{\KK}_{qq} } 
               -4\,\sma \,\smb \,\instar{\KK}_{q\bar q} \right)
               + ``\,CC\," \right] 
               +  \tfrac12\, {\smd}^{-2} 
               ({\sma}^2+{\smb}\,\overline {\smb} )\,\instar{\KK}_{q\bar q}^2 \,, \\
\hskip -0.7cm  \interior{\rm\bf K}{}_{ij}   \interior{\rm\bf K}{}^{ij}  
 = {}& {} \tfrac14  \,{\smd}^{-2} \left\{
\left[ \, {\overline { \interior{\KK}_{qq}}} \,
(\, {\sma}^2\,{ \interior{\KK}_{qq}}   
+ {\smb}^2  {\,\overline { \interior{\KK}_{qq}}}  
-4\,\sma \,\smb \,\interior{\KK}_{q\bar q}\, )
+ ``\,CC\," \right] 
+  2\,  ({\sma}^2+{\smb}\,\overline {\smb} )
\,\interior{\KK}_{q\bar q}^2\right\} \,.
\label{eq:abb-last}
\end{align}

\subsubsection{Solving the constraints in terms of deviations}
\label{sec:devs} 

Note that some of the constrained fields, as well as some of the freely specifiable variables, blow up at the ring singularity--confined to the $z=0$ plane---of the considered Kerr black holes. Thereby, it turned out to be rewarding to investigate the evolution of the deviations 
\begin{equation}\label{deviations}
\DNh=\Nh-\ANh,\quad  \DKK=\KK-\AKK,\quad \Dsmk=\smk-\Asmk
\end{equation}
of the dependent variables $\Nh$, $\KK$, $\smk$ from some analytic background ones $\ANh$, $\AKK$, $\Asmk$. In {the} case of a single black hole configuration the most obvious choice for  $\ANh$, $\AKK$, $\Asmk$ is the one that can be deduced from the  Kerr-Schild form of the Kerr black hole. As it is verified in Section~\ref{sec:devs-num}, the desired regularization occurs when these decompositions are substituted into  (\ref{bern_pde2})--(\ref{eq:eth_constr_mom2}) and into the source terms $\mathbbm{f}_\smk$ and $\mathbb{F}_\KK$. 

\medskip

In deriving the evolution equations $\Delta[{}^{(\Nh)}\hspace{-0.07cm}E]$, $\Delta[{}^{(\KK)}\hspace{-0.07cm}E]$, $\Delta[{}^{(\smk)}\hspace{-0.07cm}E]$ 
for the deviations we shall use the following simple observations. Denote by $^{(\Nh)}\hspace{-0.07cm}E$, $^{(\KK)}\hspace{-0.07cm}E$ and $^{(\smk)}\hspace{-0.07cm}E$ the left hand sides of (\ref{bern_pde2})--(\ref{eq:eth_constr_mom2}), respectively. Analogously, denote by ${}^{(\Nh,A)}\hspace{-0.07cm}E$, ${}^{(\KK,A)}\hspace{-0.07cm}E$ and ${}^{(\smk,A)}\hspace{-0.07cm}E$ the left hand sides of (\ref{bern_pde2})--(\ref{eq:eth_constr_mom2}) when they are evaluated at their analytically known background values $\ANh$, $\AKK$, $\Asmk$, respectively. These expressions can be seen to be related as 
\begin{equation}\label{separation}
^{(\Nh)}\hspace{-0.07cm}E={}^{(\Nh,A)}\hspace{-0.07cm}E + \Delta[{}^{(\Nh)}\hspace{-0.07cm}E],\quad 
^{(\KK)}\hspace{-0.07cm}E={}^{(\KK,A)}\hspace{-0.07cm}E + \Delta[{}^{(\KK)}\hspace{-0.07cm}E],\quad 
^{(\smk)}\hspace{-0.07cm}E={}^{(\smk,A)}\hspace{-0.07cm}E + \Delta[{}^{(\smk)}\hspace{-0.07cm}E] 
\end{equation}
and our task is to solve these equations for the deviations $\DNh$, $\DKK$ and $\Dsmk$.

\medskip

Below the explicit forms of $\Delta[{}^{(\Nh)}\hspace{-0.07cm}E]$, $\Delta[{}^{(\KK)}\hspace{-0.07cm}E]$, $\Delta[{}^{(\smk)}\hspace{-0.07cm}E]$ are given. Note that they all are quasilinear and homogeneous in the differences $\DNh$, $\DKK$, $\Dsmk$ that are the chosen dependent variables in the present setup.  
\begin{align}\label{Delta_bern_pde2}
\hskip-.35cm\Delta[{}^{(\Nh)}\hspace{-0.07cm}E] =\instar\KK \Big[ \pr\DNh & - \tfrac12 \Nt\big(\opa\DNh\big) 
- \tfrac12 \oNt\big(\pa\DNh\big) \Big]  \nonumber\\
&- \tfrac12 \smd^{-1} \bigg\{ \left( \DNh^2 + 2\:\ANh\DNh \right) \Big[ \sma \left( \pa\opa\DNh - \biB\big(\opa\DNh\big) \right)  \nonumber\\
&- \smb \left( \opa^2\DNh - \tfrac12 \obiA\big(\opa\DNh\big) - \tfrac12 \obiC\big(\pa\DNh\big) \right) + ``\,CC\," \Big]  \nonumber\\
&+ \left( \DNh^2 + 2\:\ANh\DNh \right) \Big[ \sma \left( \pa\opa\:\ANh - \biB\big(\opa\:\ANh\big) \right)  \nonumber\\
&- \smb \left( \opa^2\:\ANh - \tfrac12 \obiA\big(\opa\:\ANh) - \tfrac12 \obiC\big(\pa\:\ANh\big) \right) + ``\,CC\," \Big]  \nonumber\\
&+ \ANh^2 \Big[ \sma \left( \pa\opa\DNh - \biB\big(\opa\DNh\big) \right)  \nonumber\\
&- \smb \left( \opa^2\DNh - \tfrac12 \obiA\big(\opa\DNh\big) - \tfrac12 \obiC\big(\pa\DNh\big) \right) + ``\,CC\," \Big] \bigg\} - \preA\hskip-0.15cm\mathcal{A}\,\DNh \nonumber\\
& - \ANh^3\Delta\mathcal{B} - \left( \DNh^3 + 3\:\ANh\DNh^2 + 3\:\ANh^2\DNh \right) \left( \Delta\mathcal{B} + \preA\mathcal{B} \right),
\end{align}
where
\begin{align}
\mathcal{A} =\ \preA\hskip-0.15cm\mathcal{A}, \hskip1.1cm & \\
\Delta\mathcal{B} = \,\mathcal{B}-\preA\mathcal{B} = &  -\boka\DKK - \tfrac14 \left( \DKK^2 + 2\:\AKK\DKK \right)  \nonumber\\
&+ \tfrac12 \smd^{-1} \Big[ 2\sma \left( \Dsmk\Dosmk + \Asmk\Dosmk + \Dsmk\:\Aosmk \right) \nonumber\\ 
&- \smb \left( \Dosmk{}^2 + 2\:\Aosmk\Dosmk \right) - \osmb \left( \Dsmk^2 + 2\:\Asmk\Dsmk \right) \Big].
\label{Delta_B2}
\end{align}
\begin{align}\label{Delta_constr_mom1}
\Delta[{}^{(\smk)}\hspace{-0.07cm}E] =\pr\Dsmk &  - \tfrac12 \,\Nt\big(\opa\Dsmk\big) -\tfrac12 \oNt\big(\pa\Dsmk\big) \nonumber\\
&- \tfrac12 \left[ \DNh\big(\pa\DKK\big) + \ANh\big(\pa\DKK\big) + \DNh\big(\pa\:\AKK\big) \right] + \Delta\mathbbm{f}_\smk\,,
\end{align}
where
\begin{align}
\Delta\mathbbm{f}_\smk = \mathbbm{f}_\smk & -\preA\mathbbm{f}_\smk = -\tfrac12 \left[ \Dsmk \big(\pa\oNt\big) + \Dosmk \big(\pa\Nt\big) \right]  \nonumber\\
& + \tfrac12 \left[ \DKK\big(\pa\DNh\big) + \DKK\big(\pa\:\ANh\big) + \AKK\big(\pa\DNh\big) \right] 
- \boka\,(\pa\DNh) - \big(\pa\boka\big)\DNh + \instar\KK\Dsmk  \nonumber\\
& + \tfrac12 \smd^{-1} \left[ \sma\big(\opa\DNh\big)\interior{\KK}_{qq} + \sma\big(\pa\DNh\big)\interior{\KK}_{q\bar q} - \smb\big(\opa\DNh\big)\interior{\KK}_{q\bar q} 
- \osmb\big(\pa\DNh\big)\interior{\KK}_{qq} \right]  \nonumber\\
& + \left(q^i\widehat{D}^l \interior{\bf K}_{li}\right)\DNh
\end{align}
and the explicit form of $q^i\widehat{D}^l \interior{\bf K}_{li}$ is given by \eqref{eq:divKnull}.

\medskip

Finally,
\begin{align}\label{Delta_constr_mom2}
\Delta[{}^{(\KK)}\hspace{-0.07cm}E] =\pr\DKK & - \tfrac12 \Nt\big(\opa\DKK\big) -\tfrac12 \oNt\big(\pa\DKK\big) \nonumber\\
&- \tfrac12 \smd^{-1} \bigg\{ \DNh \left[ \sma \Big( \pa\Dosmk + \opa\Dsmk \Big)
- \smb \big(\opa\Dosmk\big) - \osmb\big(\pa\Dsmk\big) \right] \nonumber\\
&+ \ANh \left[ \sma \Big( \pa\Dosmk + \opa\Dsmk \Big)
- \smb\big(\opa\Dosmk\big) - \osmb\big(\pa\Dsmk\big) \right] \nonumber\\
&+ \DNh \left[ \sma\left(\pa\:\Aosmk + \opa\:\preA\smk\right) - \smb\big(\opa\:\Aosmk\big) - \osmb\big(\pa\:\Asmk\big) \right] \bigg\}
+ \Delta\mathbb{F}_\KK \,,
\end{align}
where
\begin{align}\label{Delta_constr_mom2FF}
\Delta\mathbb{F}_\KK =\mathbb{F}_\KK -\preA\mathbb{F}_\KK =  &
\tfrac14 \smd^{-1} \bigg\{ \DNh \Big[ 2\sma\biB\Dosmk - \smb \left( \obiC\Dsmk + \obiA\Dosmk \right) \nonumber\\
&+ 2\sma\biB\:\Aosmk - \smb \left( \obiC\:\Asmk + \obiA\:\Aosmk \right) + ``\,CC\," \Big] \nonumber\\
&+ \ANh \Big[ 2\sma\biB\Dosmk - \smb \left( \obiC\Dsmk + \obiA\Dosmk \right) + ``\,CC\," \Big]\bigg\} \nonumber\\
&- \smd^{-1} \Big[ \left(\sma\Dosmk - \osmb\Dsmk\right) \left( \pa\DNh + \pa\:\ANh \right) \nonumber\\
&+ \left( \sma\:\Aosmk -\osmb\:\Asmk \right) \pa\DNh + ``\,CC\," \Big] + \tfrac12 \,\instar\KK\,\DKK\,.
\end{align}

Note that in the considered single black hole case the Kerr-Schild metric provides us an ideal analytically known
background, thereby, the vanishing of $\Delta[{}^{(\Nh)}\hspace{-0.07cm}E]$, $\Delta[{}^{(\KK)}\hspace{-0.07cm}E]$ and $\Delta[{}^{(\smk)}\hspace{-0.07cm}E]$ is equivalent to the vanishing of the left hand sides ${}^{(\Nh)}\hspace{-0.07cm}E$, ${}^{(\KK)}\hspace{-0.07cm}E$ and ${}^{(\smk)}\hspace{-0.07cm}E$ of (\ref{bern_pde2})--(\ref{eq:eth_constr_mom2}), provided that the variables $\ANh$, $\AKK$, $\Asmk$ are chosen to take their Kerr-Schild form. 

\medskip

If, in addition, the freely specifiable background and initial-boundary data are synchronized, i.e.~both are deduced by making use of the very same Kerr-Schild solution, then the fields $\DNh$, $\DKK$ and $\Dsmk$ all vanish throughout $\Sigma$ as they vanish at the boundary and initial data surfaces, and, in addition, the deviation equations are at least linear and homogeneous in the basic variables $\DNh$, $\DKK$ and $\Dsmk$. Therefore, the physically interesting cases ar{i}se whenever the freely specifiable background data on $\Sigma$ and the initial-boundary data for the fields $\DNh$, $\DKK$ and $\Dsmk$ on the boundary of $\Sigma$ does not completely match. Note that the latter is also freely specifiable but only on the boundary of $\Sigma$. 

Indeed, in most of our numerical simulations the latter type of distorted Kerr black hole configurations were investigated. The discrepancies in the choices made for the freely specifiable background data and for the initial-boundary data were tuned in a wide range, from tiny to considerably large discrepancies. This way it is demonstrated that numerical solutions of the parabolic--hyperbolic form of the constraints can be determined, with considerably high accuracy, even for extreme circumstances when the yielded data is supposed to store considerably large amount of gravitational radiation.


\subsection{Black holes in their Kerr-Schild form}
\label{sec:boo:spin:bh}

Our numerical investigations are centered on single black hole configurations. These are either exact Kerr-Schild black holes or distorted ones. The approach to setting the black hole initial data using the Kerr-Schild metric was introduced in~\cite{BishopIsaacsonMaharajWinicour2020}.

\medskip

A Lorentzian metric $g_{\alpha\beta}$ is of the Kerr-Schild type if it is of the form 
\begin{equation}\label{eq:ksm}
g_{\alpha\beta}=\eta_{\alpha\beta}+2 H \ell_{\alpha} \ell_{\beta}\,, 
\end{equation}
or equivalently, in inertial coordinates $(t,x^i)$ adapted to the background Minkowski metric $\eta_{\alpha\beta}$, it can be given as
\begin{align}
g_{\alpha\beta}\,dx^{\alpha} dx^{\beta}= {}&  (-1+2H{\ell_0}^2)\,dt^2 + 4H \ell_0\ell_i \,dt dx^i  + (\delta_{ij}+ 2 H \ell_i \ell_j)\,dx^i dx^j\,, 
\end{align}
where $H$, apart from singularities, is a smooth function on $\mathbb{R}^4$ and
$\ell_{\alpha}$ is null with respect to both $g_{\alpha\beta}$ and $\eta_{\alpha\beta}$. In particular, for $\ell^{\alpha}=g^{\alpha\beta}\ell_{\beta} = \eta^{\alpha\beta}\ell_{\beta}$ the relations $g^{\alpha\beta}\ell_{\alpha}\ell_{\beta} = \eta^{\alpha\beta}\ell_{\alpha}\ell_{\beta}=-(\ell_0)^2+\ell^i\ell_i=0$ and $\ell^{\beta} \partial_{\beta} \, \ell^{\alpha}=0$ hold.  

\medskip

The induced metric and the extrinsic curvature on a $t=const$ hypersurface in $(\mathbb{R}^4,g_{\alpha\beta})$ can be written as \cite{i_jeff_2}
\begin{align}
\hskip1cm h_{ij} = {} & \delta_{ij}+2 H\ell_i \ell_j  \, ,\label{SymS2} \\
N^{-1} K_{ij}= {}& - \ell_t \left[ \partial_i  (H\ell_j)+\partial_j (H\ell_i \right) ] 
+N^{-2} \partial_t (H\ell_i \ell_j)    \nonumber \\ 
{}& +2 H \ell^t \ell^k \partial_k(H\ell_i \ell_j) -H(\ell_i\partial_j \ell_t+\ell_j \partial_i \ell_t)  \,,
\label{eq:kerrxc}
\end{align}
where $N=1/\sqrt{1+2\, H \,(\ell_t)^2}$.

\medskip

The most general stationary axisymmetric asymptotically flat vacuum black hole, i.e.~the Kerr black hole does also possess the Kerr-Schild form with
\begin{equation}\label{H-ell-kerr}
H=\frac{r^3M}{r^4+{a^2z^2}} \,\,\, {\rm and} \,\,\, \ell_{\alpha}=\left(1, \frac{r\,x+a\,y}{r^2+a^2},
\frac{r\,y-a\,x}{r^2+a^2},\frac{z}{r}  \right)\,,
\end{equation}
where the Boyer-Lindquist radial coordinate $r$ is related to 
the spatial part of the inertial coordinates $x^i =(x,y,z)$ as 
\begin{equation}\label{imp-r-def}
\frac{x^2+y^2}{r^2+a^2}+\frac{z^2}{r^2}=1\,.
\end{equation}
The ADM mass, centre of mass, linear and angular momenta of asymptotically flat solutions can be determined by applying asymptotic expansions. In particular, for the Kerr black hole, given by (\ref{H-ell-kerr}) and (\ref{imp-r-def}), the ADM mass is $M$, the centre of mass is represented by the origin of the background Euclidean space, the linear momentum vanishes (the latter means that the black hole is at rest with respect to the background reference frame), while the ADM angular momentum is $\vec{J}=a M \vec{e}_z$, where the unit vector $\vec{e}_z$ points to the positive $z$ direction. 

\subsection{Boosted and spinning black holes}

The most important advances related to the use of the above form of the Kerr black hole come with the form-invariance of the Kerr-Schild metric under Lorentz transformations. More concretely, if a Lorentz transformation
\begin{equation}
x'{}^{\alpha}=\Lambda^{\alpha}{}_{\beta}\,x^{\beta}
\end{equation}
is performed, the metric retains its distinguished form $g'_{\alpha\beta}=\eta_{\alpha\beta}+2 H' \ell'_{\alpha} \ell'_{\beta}$, where $H'=H'(x'{}^{\alpha})$ and $\ell'_{\beta}=\ell'_{\beta}(x'{}^{\varepsilon})$ are given as
\begin{align}
H'(x'{}^{\alpha})=  {}& H\left([\Lambda^{\alpha}{}_{\beta}]^{-1}x'{}^{\beta}\right),
\\ \ell'_{\beta}(x'{}^{\varepsilon})= {}& \Lambda^{\alpha}{}_{\beta}\, \ell_{\alpha}\left([\Lambda^\varepsilon{}_\varphi]^{-1}x'{}^\varphi\right)\,.
\end{align}

\smallskip

As boosts and rotations are special Lorentz transformations it is straightforward to construct moving and rotating black holes with preferably oriented speed and spin by performing a suitable sequence of boosts and rotations starting with a Kerr black hole. 

To provide a simple example start with a Kerr black hole that is at rest with respect to some reference system $x'{}^{\alpha}$. Then, $H(x^{\alpha})$ and $\ell_{\alpha}(x^{\varepsilon})$, relevant for a black hole that is displaced by distance $d$ in the positive $y$ direction and moving with velocity $0< v < 1$ in the positive $x$ direction of a reference system $x{}^{\alpha}$, are obtained by substituting $x'=\gamma\, x - \gamma v\, t$, $y'= y - d$ and $z'=z$ into
\begin{align}\label{eq:Kerr-alt0}
H= {}& \frac{r'{}^3M}{r'{}^4+{a^2 z'{}^2}}  \quad {\rm and}\\ \ell_{\beta}={}& \left(\gamma\,\ell'_0 - \gamma v \, \ell'_1, \gamma \,\ell'_1 - \gamma v \,\ell'_0, \ell'_2,\ell'_3\right) \,,
\end{align}
where $\gamma=1/\sqrt{1-v^2}$, while $\ell'_{\beta}$ and $r'{}$ are determined by the primed variant of (\ref{H-ell-kerr}) and (\ref{imp-r-def}), respectively,
\begin{equation}\label{eq:Kerr-alt1}
\ell'_{\beta}=\left(1,\frac{r'x'+a y'}{r'{}^2+a^2},\frac{r'y'-a x'}{r'{}^2+a^2},\frac{z'}{r'}\right)\,.
\end{equation}

Asymptotic expansions, in accordance with the transformations performed, verify that for the considered displaced, boosted and spinning black holes the ADM mass, centre of mass, linear and angular momenta can be given as $\gamma\,M$, $\vec{d}$, $\gamma\,M\,\vec{v}$ and $\gamma\,M\{\vec{d} \times\vec{v} + a\,\vec{e}_z \}$, respectively, where $\vec{d}=d\,\vec{e}_y$, $\vec{v}=v\,\vec{e}_x$, and the unit vectors  $\vec{e}_x$ and $\vec{e}_y$ are aligned to the positive $x$ and $y$ directions, respectively.

\subsection{The applied initial--boundary value problem}
\label{sec:boundinival}

In returning to our parabolic--hyperbolic system  (\ref{bern_pde})--(\ref{ort_const_n}) recall that the level surfaces $\mycal{S}_\rho$ have not been fixed yet. As in most of the numerical approaches the initial data surface $\Sigma$ is chosen to be a sufficiently large but compact subset of $\mathbb{R}^3$ we shall also adopt such a scheme here. In order to ensure the product structure of the initial data surface $\Sigma \approx \mathbb{R}\times \mycal{S}$---this is essential for the evolutionary setup proposed in \cite{racz_constraints}---the leaves of the foliations have to be diffeomorphic to a closed disk in $\mathbb{R}^2$. 
Accordingly, we may ensure $\Sigma$ to be a compact subset in $\mathbb{R}^3$, which, however, requires the parabolic-hyperbolic system (\ref{bern_pde})--(\ref{ort_const_n}) to be solved as an initial-boundary value problem. It is important that if (\ref{bern_pde}) is uniformly parabolic well-posedness of such an initial-boundary value problem is guaranteed (see, e.g.~\cite{kreissl}). Note, however, that this requires a suitable splitting of the boundary of $\Sigma$ into disjoint subsets on which the initial and boundary values can be specified, respectively. 

\medskip

\begin{figure}[htb]
	\begin{center}
		\includegraphics[width=9.cm]{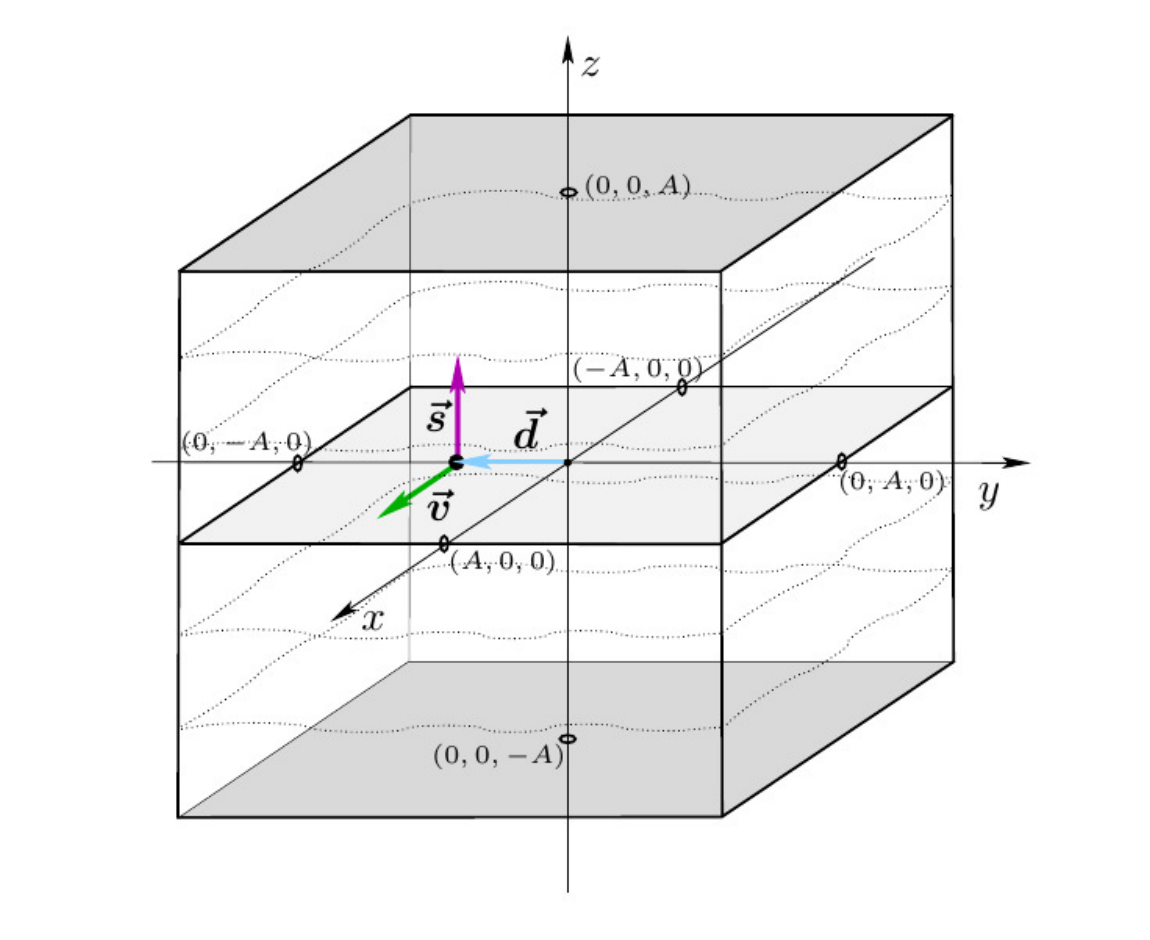}
	\end{center}
	\vskip-.5cm\caption{\footnotesize{(color online). The initial data surface $\Sigma$ is chosen to be {a} cube centered at the origin in $\mathbb{R}^3$ with edges $2A$. It will be argued below that initial data 
			are to be specified on the horizontal squares, with $z=\pm A$, bounding the cube from above and below, whereas boundary values have to be given on the complementary part of the boundary comprised by four vertical squares.}}
	\label{kocka}
\end{figure}

Here we choose $\Sigma$ to be {a} cube centered at the origin in $\mathbb{R}^3${---covered by the Cartesian coordinates $\left(x,y,z\right)$---and} with edges $2A$ {(see Fig.\,\ref{kocka})}, which for sufficiently large value of $A$ contains the black hole with a reasonable size of margin.

\medskip

Before splitting the boundary of $\Sigma$, consisting of six squares, into suitable parts where initial and boundary values are to be specified recall that (\ref{bern_pde}) is uniformly parabolic only in those subsets of $\Sigma$, where $\instar{K}$ is strictly negative or positive. Indeed, it is the sign of $\instar{K}$ that decides whether the coupled system (\ref{bern_pde})--(\ref{ort_const_n}) evolves in the positive or negative $\rho$-direction. It propagates aligned {with} the vector field $\rho^i$ for positive $\instar{K}$, while anti-aligned for negative $\instar{K}$.

\medskip

\begin{figure}[htb]
	\begin{center}
		\includegraphics[width=9.4cm]{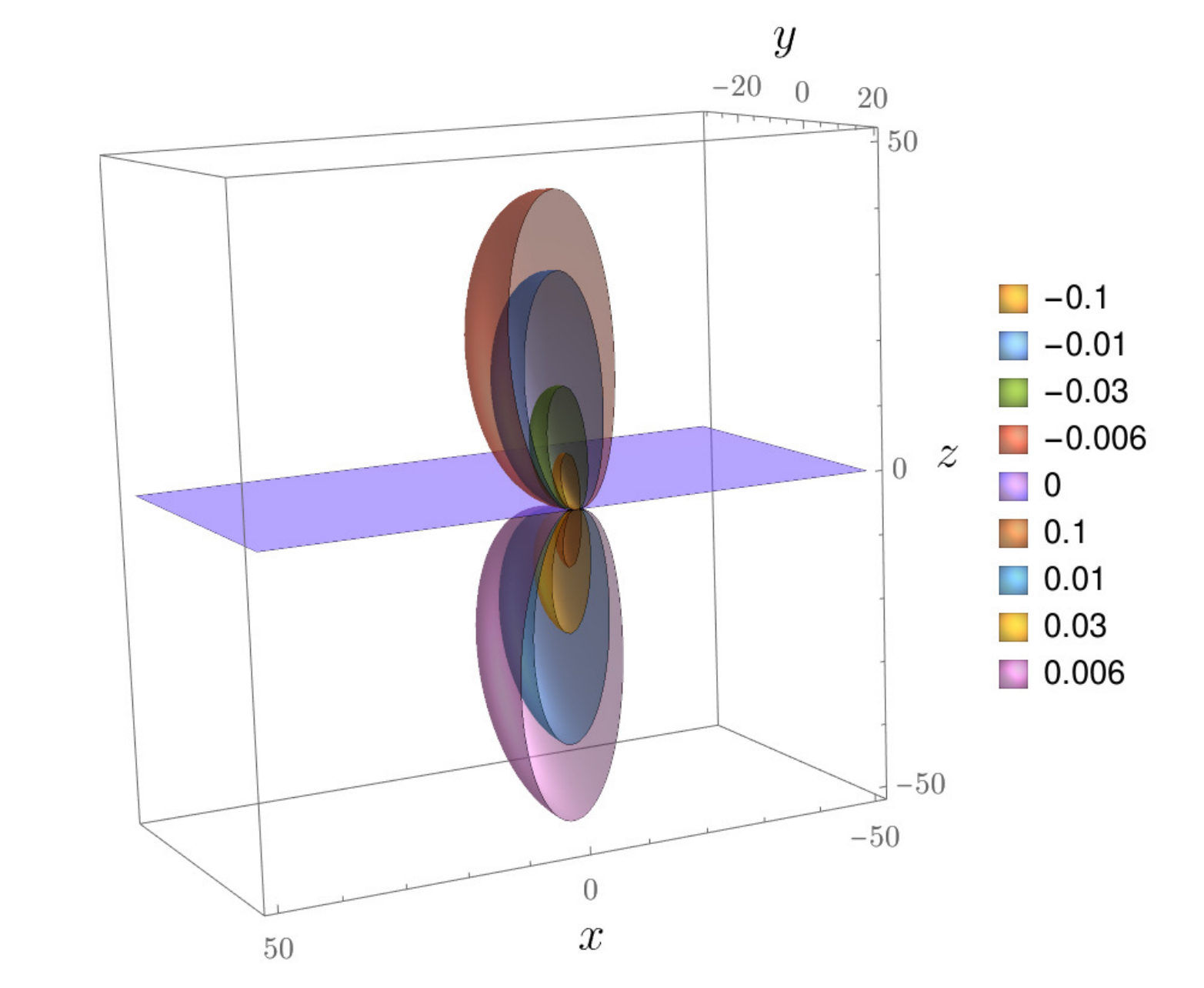} 
	\end{center}
	\vskip-.5cm\caption{\footnotesize{(color online).  $\overset{\hskip.07cm{}_{\star}}{K}=const$ level surfaces are depicted in the $y-d<0$ half of the cube with edges $2 A=100$. The parameters of the considered Kerr-Schild black hole are: $M{}=1, {d}=20, {v}=0.9$ and $a =0.99$. The positive and negative $\overset{\hskip.07cm{}_{\star}}{K}=const$ level surfaces are well separated by the $z=0$ plane that also coincides with $\overset{\hskip.07cm{}_{\star}}{K}=0$ level surface.}}
	\label{instarK}
\end{figure}

Restrict now considerations to a Kerr-Schild black hole with displacement, speed and spin---as indicated in Fig.\,\ref{kocka}---aligned parallel to the $x,y$ and $z$-ax{e}s, respectively. Consider then a foliation of $\Sigma$ by $z=const$ level surfaces, and determine the function $\instar{K}$ using (\ref{eq:trhatext}). 
{A d}irect calculation verifies then that $\instar{K}$ can be given as the product of a strictly negative function and the $z$-coordinate. This means that $\instar{K}$ is positive everywhere below the $z=0$ plane while it is negative above that plane. This behavior is verified by plotting $\instar{K}=const$ level surfaces for a specific choice of physical parameters in Fig.\,\ref{instarK}. 
It is clearly visible that the $\instar{K}=const$ level surfaces get concentrated in a neighborhood of the singularity while the absolute value of $\instar{K}$ is increasing. Note also that the ring-like singularity of the considered Kerr black hole is always confined to the $z=0$ plane.

\medskip

Then, for a black hole arranged as indicated in Fig.\,\ref{kocka}, the evolutionary equations (\ref{bern_pde})--(\ref{ort_const_n}) are well-posed on the disjoint domains, $\Sigma^+$ and $\Sigma^-$, above and below the $z=0$ plane. 
In particular, they can be solved by propagating initial values specified on the horizontal $z=\pm A$ squares, along the $z$-streamlines, meanwhile the $z=const$ `time' level surfaces approach the orbital plane 
from above and below. The boundary values are to be given on the four vertical sides of the cube (see Fig.\,\ref{kocka}). As the fields $\widehat N$,  ${\rm\bf k}{}_{i}$ and ${\rm\bf K}^l{}_{l}$ are developed on the $z>0$ and $z<0$ domains, denoted by $\Sigma^+$ and $\Sigma^-$, separately, provided that they exist up to the $z=0$ plane, their proper matching there, at their common Cauchy horizon, is of fundamental importance. 

\medskip

The first important payback of the specific choice we made for the freely specifiable part of the data comes now. Indeed, for the considered class of black hole configurations, the auxiliary metric (\ref{eq:ksm}), with constituents given by \eqref{eq:Kerr-alt0}-\eqref{eq:Kerr-alt1}, possesses a $z \rightarrow -z$ reflection symmetry which guarantees that the solutions {obtained within $\Sigma^+$ and $\Sigma^-$ separately} will match at $z=0$.  
Notably, using deviations of the fields $\widehat N$,  ${\rm\bf k}{}_{i}$ and ${\rm\bf K}^l{}_{l}$ from some suitably chosen analytic background ones, an argument---analogous to the one applied in \cite{Racz1605.01669}  to verify that (at least) $C^2$ solutions exist on the closure of the union of $\Sigma^+$ and $\Sigma^-$---can be used to show that solutions to the considered initial-boundary value problems exist on the closure of $\Sigma^+$ and $\Sigma^-$ separately. To verify the existence of at least $C^2$ solutions the argument in the Supplementary material of \cite{Racz1605.01669} can be repeated. In particular, the constrained fields can be shown to possess well-defined values (apart from the `ring singularity') in the $z\rightarrow 0$ limit, and that the fields $\widehat N$,  ${\rm\bf k}{}_{i}$ and ${\rm\bf K}^l{}_{l}$, and at least their first two $z$-derivatives match through the $z=0$ plane.

\section{The applied numerical method}
\label{sec:numcomp}
\setcounter{equation}{0}

As reported in the previous sections the constraint equations are treated in two alternative forms---in their full, (\ref{bern_pde2})--(\ref{eq:abb-last}), and in their deviation based, (\ref{Delta_bern_pde2})--(\ref{Delta_constr_mom2FF}), forms. In both cases the corresponding sets of equations were solved numerically. To this end, the complex variables $\smk$ and $\Dsmk$, as well as the quantities associated with them through $\mathbbm{f}_\smk$ and $\Delta\mathbbm{f}_\smk$, were split into their real and imaginary parts. For the variables, the split resulted in $\smk=\smk_1+\mathbbm{i}\cdot\smk_2$ and $\Dsmk=\Dsmk_1+\mathbbm{i}\cdot\Dsmk_2$. Moreover, as indicated in Section~\ref{sec:boundinival}, we applied a Cartesian coordinate system $\left(x,y,z\right)$. Readers interested in some of the technical terms of the applied numerical schema should communicate with the Appendix of the present paper. 

As explained in Section~\ref{Parabolic-hyperbolic-system}, in the full form of the constraints the set of dependent variables {is} $\mathcal{F}=\big\{\smk_1,\smk_2,\KK,\Nh\big\}$. In the {form based on} deviations the {set of basic} variables reads as $\mathcal{F}^\Delta=\big\{\Dsmk_1,\Dsmk_2,\DKK,\DNh\big\}$. These variables will {frequently} be {referred} to as constrained {variables}. {In both cases the equations relevant} for $\smk_1$, $\smk_2$ and $\KK$ (or $\Dsmk_1$, $\Dsmk_2$ and $\DKK$) {form a first order symmetric hyperbolic system}, whereas the equation for $\Nh$ (or $\DNh$) {is parabolic}. The background analytic variables applied in the deviation setup will collectively be referred to as $\preA\hspace{-0.07cm}\mathcal{F}=\big\{\,\Asmk_1,\Asmk_2,\AKK,\ANh\big\}$. Both setups, the full and deviation {based} ones share the freely specifiable variables $\mathcal{F}^f=\big\{\sma,\smb,\kappa,\mathbb{N},\interior\KK\big\}$ (for their definitions see Table \ref{table:data}). 

\medskip

The {time integration of the parabolic-hyperbolic form of the constraint} equations {was} {performed} on a cubical grid $\left(x,y,z\right)$ centered at the point $\left(0,0,0\right)$ using the method of lines {using a $4^{th}$ order accurate Runge-Kutta scheme}. Each of the systems of PDEs was discretized in the $(x,y)$-plane with $4^{th}$ (and next to the edges $6^{th}$) order accurate derivatives whose {explicit expressions} can be found in the Appendix.
 
The initial-boundary value problem was solved in two halves of the {cubical region,} separately. {As explained in more details in Subsection \ref{sec:boundinival}, the cubical region was bounded by the $x_{MIN}=y_{MIN}=z_{MIN}=-A$ and $x_{MAX}=y_{MAX}=z_{MAX}=A$ planes. The initial data was specified at corresponding parts of the $z=-A\equiv z_{MIN}$ and $z=A\equiv z_{MAX}$ planes such that the boundary data {was} specified at the vertical boundaries of the cubical region.} 

The actual value of $A$ applied in our numerical computations was $10$, i.e.~we had $x_{MIN}=y_{MIN}=z_{MIN}=-10$ and $x_{MAX}=y_{MAX}=z_{MAX}=10$. {This choice meant that the length of edges {of} the squares {which} bounded the cubical region was} either $10M$ or $20M$, {where} $M$ {denotes the} mass of {the} black hole {located in} the investigated three-dimensional {domain}. 
A uniform grid was applied in the $x$ and $y$ directions. Almost exclusively, the number of the grid points in each of these directions was $N_x=N_y=30$ corresponding to the {grid spacing} $h_x=h_y=\tfrac{20}{29}$. Computations involving big boosts in integrating the full form of the constraints required the use of a much finer grid in the $x$ and $y$ directions in order to achieve the desired accuracy. In doing so in this exceptional case we used a grid with spacing $h_x=h_y=\tfrac{5}{11}$, corresponding to $N_x=N_y=45$. 

\medskip

The {use of a much finer} discretization in {the} $z$ direction was {required by} the parabolic character of the equations for $\Nh$ and $\DNh$. The stability criterion, which interrelates the involved spatial grid spacing and the {`time'-}integration step {size}, for the parabolic equations is more severe than the {conventional} Courant-Friedrichs-Lewy (CFL) condition for the hyperbolic equations. Indeed, this dominated the relative sizes of the integration steps for the investigated hyperbolic-parabolic systems. In both of the considered cases the parabolic equations required~\cite{GustafssonKreissOliger}
\ben\label{eqn:stab-parab}
 h_z \leqslant C\cdot \Big[ \min\left(h_x,h_y\right) \Big]^2\,,
\een
where $h_z$ stands for the {`}time{'}-integration step in $z$ and $C=\tfrac12\,\instar\KK\,\Nh^{-1}$. Since $\Nh$ is of the order of unity, $C\sim\instar\KK$, whose behavior is indicated in Fig.~\ref{instarK}.
The minimal absolute value of $\instar\KK$ on the consecutive $z=const$ {level surfaces{, $\instar\KK_{min}$,} tends to} zero as $z=0$ is approached. Since it {bounds} the integration step {size} in $z$---see (\ref{eqn:stab-parab})---the use of an adaptive $z$-step {size} {turned out to be rewarding with $h_z= 0.1\cdot\instar\KK{_{min}}$. It is important to keep in mind that $\instar\KK$, and hence $h_z$,  strongly depend on the values of the black hole's boost. Typically, for medium values {of the speed, i.e.~}close to $0.5$, the {value of $h_z$ initially (close to $z_{MAX}$ or $z_{MIN}$)} was {of} the order of $10^{-4}$, {whereas it was required to be} of the order of $10^{-7}$ close to the {central} $z=0$ level surface. The value of $h_z$ decreased even more significantly {as the integration along $z$-streamlines proceeded}---increasing thereby the computational time considerably---for black holes with large boosts.

\subsection{The initial-boundary data specifications}
\label{ssec:inibound}

{As discussed in Section~\ref{sec:boo:spin:bh}} the initial-boundary data for the variables $\mathcal{F}$ and $\mathcal{F}^\Delta$ were specified on the faces of the computational domain. They were given by referring to the four {physical} parameters, the mass $M$, angular momentum $a$, boost $v$ and displacement $d$ {of  a single black hole in its Kerr-Schild form} (see Section~\ref{sec:boo:spin:bh} and~\cite{Racz1605.01669}). The considered black holes were either non-spinning or with spin parallel to the $z$ direction.  Boosts and displacements were aligned with the $x$ and $y$ directions, respectively. A sketch of the black hole location in the computational domain was shown in Fig.~\ref{kocka} and discussed in more detail in Section~\ref{sec:boundinival}. {Due to this arrangement} the central singularity {was} either  pointlike for a Schwarzschild {black hole} or ringlike for a Kerr black hole confined in both cases to the $z=0$ plane.


The physically adequate ranges of the black hole parameters are: $a\in[0,M]$, $v\in[0,1]$, while $M$ and $d$ are restricted only by the size of the grid so that its boundary does not have any common points with the event horizon. Naturally, it is desirable to place the black hole in an appropriate distance from the boundary in order to guarantee that this boundary is located in the weak field regime. The spins and displacements will be given as relative values, in units of the particular black hole's mass.

In the full setup the initial-boundary values and the freely specifiable variables were synchronized such that both referred to the same Kerr-Schild black hole. In the deviation setup, the information on a black hole was embedded within the background functions set $\preA\hspace{-0.09cm}\mathcal{F}$. The initial-boundary variables values in this case ought to be equal to zero in principle, as the idea behind this formulation is that on the boundary the solution does not deviate from the assumed Kerr-Schild form. Since the single black hole cases are analytic solutions of the formulated set of equations, the deviations vanish within the whole computational domain, what was indeed observed during numerical calculations. Nevertheless, when the initial-boundary {and the freely specifiable parts of the} data are mismatched a distorted black hole {is generated.} Of course, deviations {were} assumed to be small. It allowed to draw conclusions on the accuracy and appropriateness of the deviation code and on physical implications stemming from the proposed formulation of the problem.

\subsection{Error quantifiers}
\label{ssec:errquant}

The formulation of the undertaken problem in the full form has the advantage that the outcomes of the numerical computations can be compared with the analytically known expressions. To this end, as well as to check the correctness of the deviation setup, a set of error quantifiers was used.

The overall accuracy of the numerical method can be estimated as
\ben\label{eqn:accuracy}
 \mathcal{F}=\mathcal{F}^a+\mathcal{O}\big(h^n\big),
\een
where $\mathcal{F}^a=\big\{\smk_1^{\: a},\smk_2^{\: a},\KK^a,\Nh^a\big\}$ are the analytically {known}  variables {in} the full setup. They are equal to the background variables in the deviation setup, $\mathcal{F}^a={}\preA\hspace{-0.07cm}\mathcal{F}$. The quantity $h\equiv\max(h_x,h_y,h_z)$ is the biggest integration step and $n$ is the lowest order of accuracy of the employed numerical methods. In the case of considering the global truncation error (what is reasonable because of having exact analytic results), $n$ was equal to $4$.

As a semi-global {measure of the} accuracy the two-dimensional L$^2$~(Euclidean) matrix norm
\ben\label{eqn:L2}
 \big\Arrowvert {\mycal{F}} \big\Arrowvert_{\textrm{L}^2} = \sqrt{\sum_{i,j=1}^N \big|{\mycal{F}}_{ij}\big|^2}
\een
{is applied,} where the $N\times N$ matrix $\mycal{F}_{ij}$ stands for the grid values, on a $z=const$ slice, of either of the variables $\mathcal{F}-\mathcal{F}^{a}$ or $\mathcal{F}^\Delta$.}

Point-wise errors of specific constrained variables were {also monitored} via the absolute 
\ben\label{eqn:ae}
 {\textrm{E}_\textrm{abs}}(\mathcal{F})=\big|\mathcal{F}-\mathcal{F}^{a}\big|
\een
and relative 
\ben\label{eqn:re}
{\textrm{E}_\textrm{rel}}(\mathcal{F})=\frac{\big|\mathcal{F}-\mathcal{F}^{a}\big|}{\big|\mathcal{F}^{a}\big|}
\een
{errors.}

\begin{figure}[H]
\centering
(a)\hspace{7cm}(b)\\
\includegraphics[width=0.45\textwidth]{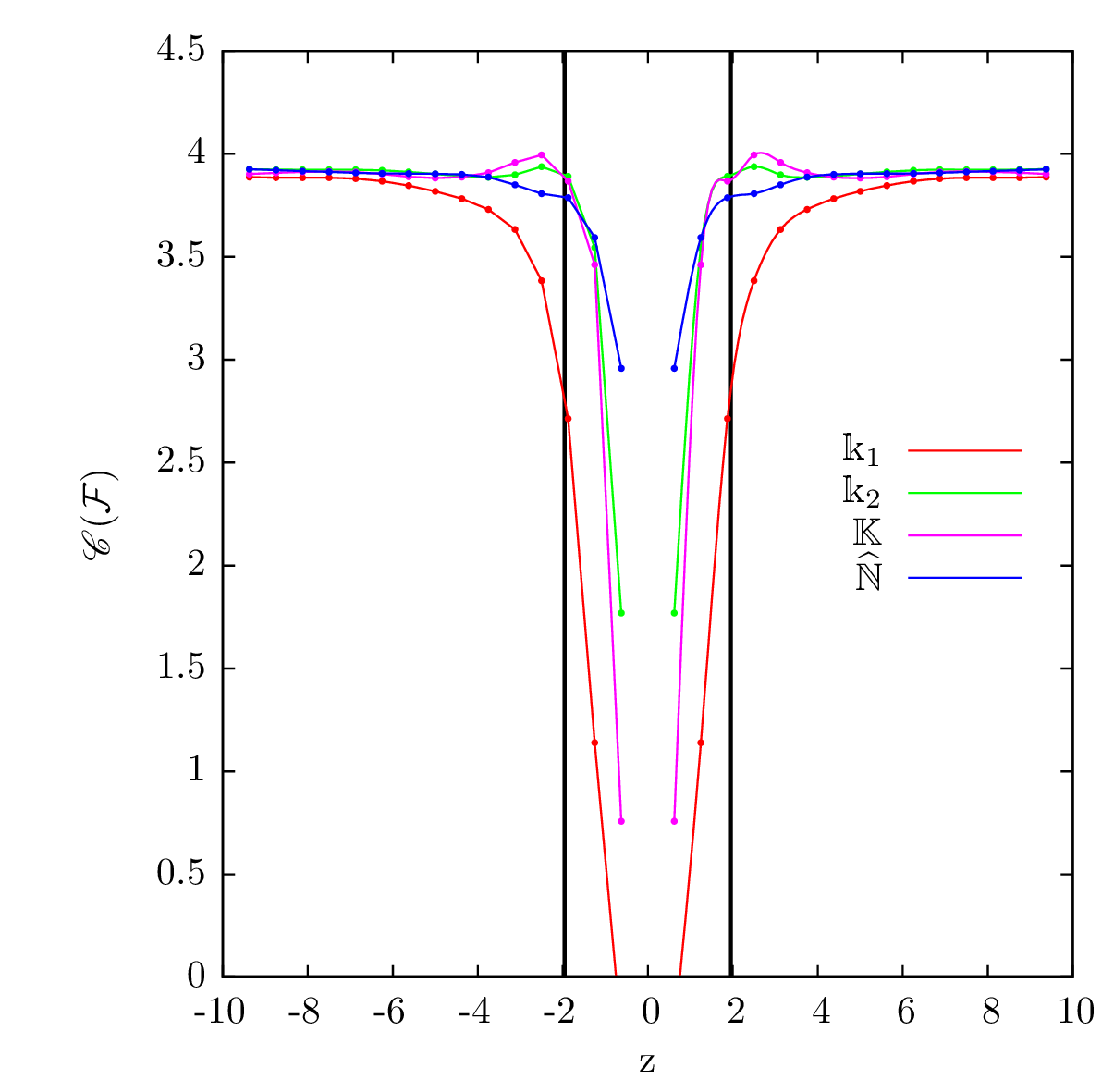}
\includegraphics[width=0.45\textwidth]{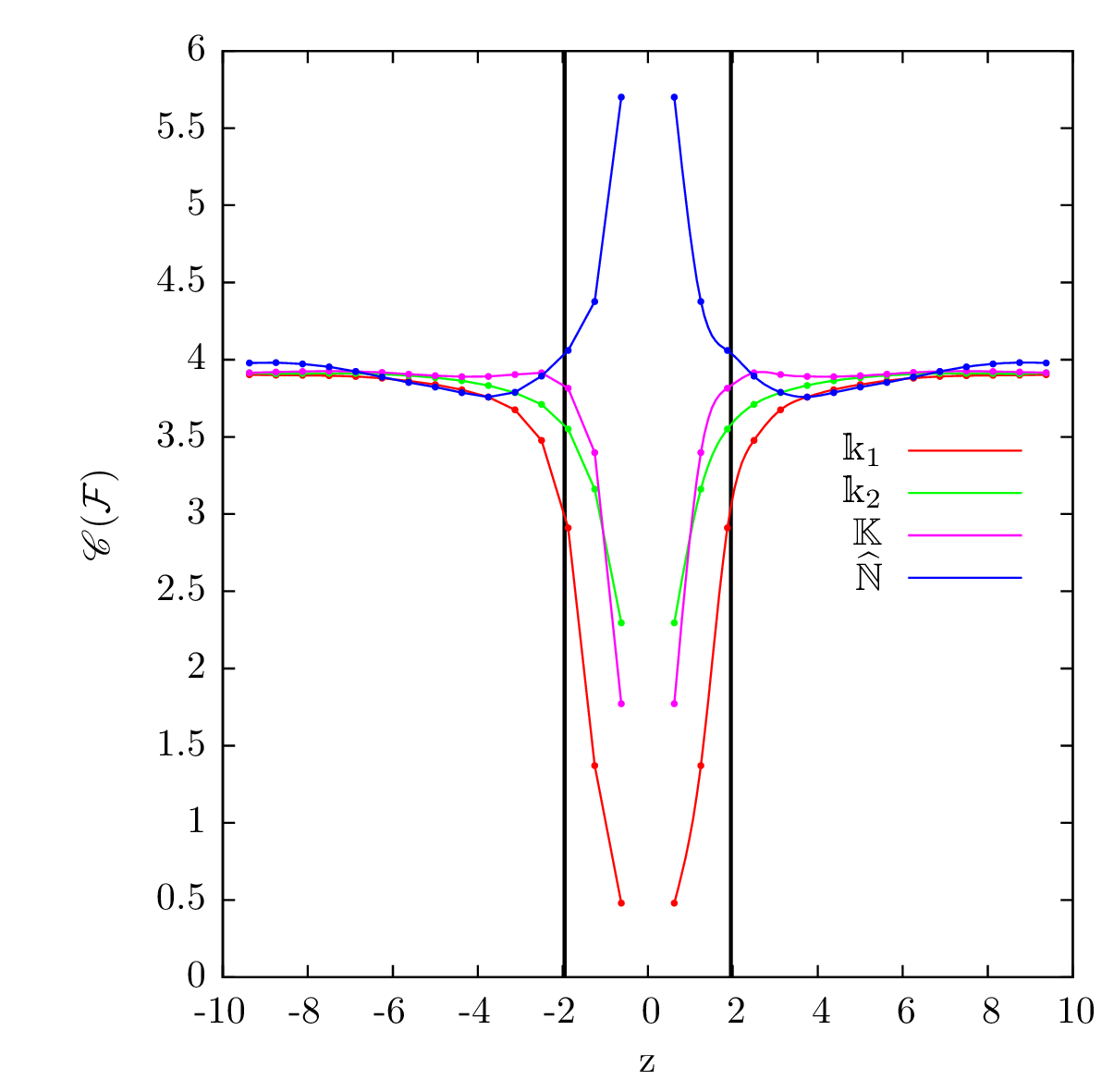}
\caption{The $z$-dependence of the convergence rates for (a)~Schwarzschild black hole with $M=1$, $v=0.7$, $d=3M$ and (b)~Kerr black hole with $M=1$, $a=0.3M$, $v=0.6$, $d=5M$. The spacings of the two employed grids were $h^<=\frac{5}{11}$ and $h^>=\frac{20}{29}$. Thick black vertical lines are to indicate the location of event horizons in the $z$-direction.}
\label{fig:conv-full}
\end{figure}

The convergence of the numerical program was tested with semi-global error estimations on two grids of different resolutions made against analytically known expressions. The convergence rate of the code was defined by
\ben
\mycal{C} \left( \mathcal{F} \right) = \log^{-1} \left(\frac{h^<}{h^>}\right) 
\log \left(\frac{\big\Arrowvert {\mycal{F}} \big\Arrowvert^<_{\textrm{L}_n^2}}
{\big\Arrowvert {\mycal{F}} \big\Arrowvert^>_{\textrm{L}_n^2}}\right)
\een
where $h$ stands for $h_x=h_y$ and the two-dimensional normalized L$^2$~(Euclidean) matrix norm is
\ben\label{eqn:L2normalized}
 \big\Arrowvert {\mycal{F}} \big\Arrowvert_{\textrm{L}_n^2} = \frac{1}{\sqrt{N_x \cdot N_y}} \big\Arrowvert {\mycal{F}} \big\Arrowvert_{\textrm{L}^2}
\een
with $\big\Arrowvert {\mycal{F}} \big\Arrowvert_{\textrm{L}^2}$ defined by (\ref{eqn:L2}). The superscripts $<$ and~$>$ indicate whether the particular quantity is related to a grid of a higher or lower resolution, respectively. The outcomes of convergence tests for two single black hole cases are shown in Fig.~\ref{fig:conv-full}. Regarding the facts that the applied spatial numerical stencil is $4^{th}$ order accurate and the 'time'-integration proceeds with the $4^{th}$ order accurate Runge-Kutta scheme, the semi-global convergence rate is close to $4$ up to about the event horizon of a black hole, as expected.

\section{Integration of the full form of the constraints}
\label{sec:fullform-num}
\setcounter{equation}{0}

\subsection{Verifications of the applied numerical schema}
\label{ssec:fullform-relevance}

As it has already been mentioned in Sections~\ref{sec:devs} and~\ref{ssec:errquant} the Kerr-Schild form of {a} single black hole is an exact solution to Einstein equations. This fact provides an excellent opportunity to test both the numerical algorithm and the code implementation, as the outcomes of computations can be compared to analytically known data. This allows to replace the conventional convergence tests by more robust {numerical verifications}. The scanning of the parameter ranges of the computational grid, which were the integration steps in $x$ and $y$ directions and the coefficient $C_z$ defining the integration step in the $z$ direction, led to an optimal set of values {as} listed in Section~\ref{sec:numcomp}. They ensured satisfactory outcomes of the numerical computations, as described below, in a reasonable computational time.

Comparisons between numerical and analytic results were performed for a {wide range} of Schwarzschild and Kerr black holes, with various displacements---chosen such {that} the event horizons were reasonably away from the boundary of the computational domain---while choosing the boosts and spins {parameters} from the{ir} {entire} admissible ranges. In the present section, a set of representative examples will be {introduced} enabling us to draw conclusions concerning the performance of the employed numerical setup.

\begin{figure}[H]
\centering
(a)\hspace{6cm}(b)\\
\includegraphics[width=0.45\textwidth]{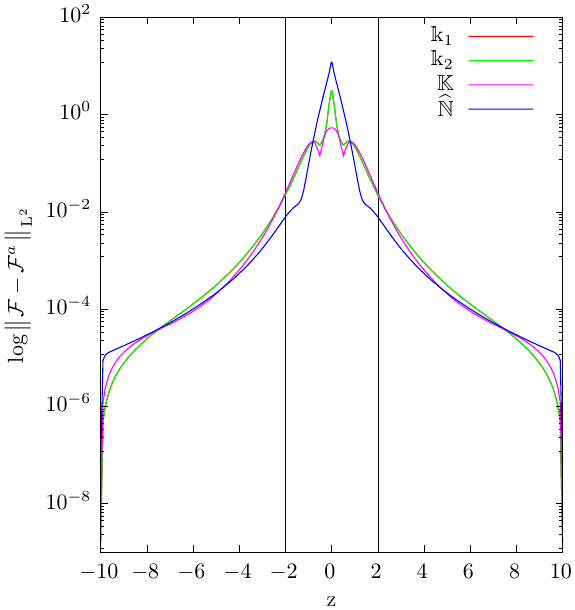}
\includegraphics[width=0.45\textwidth]{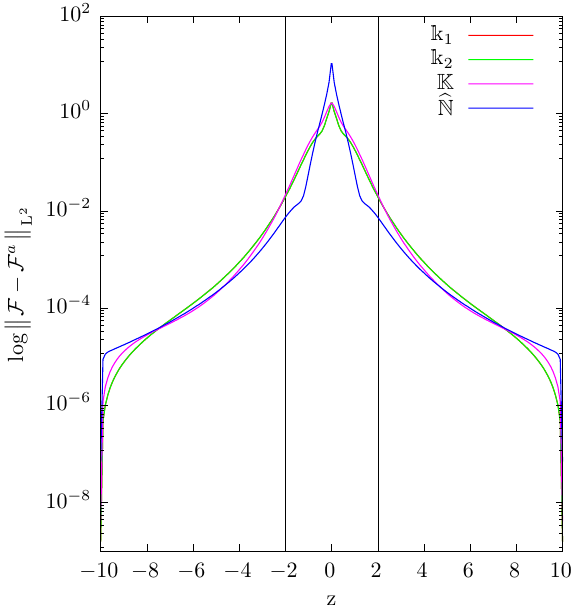} \\
(c)\hspace{6cm}(d)\\
\includegraphics[width=0.45\textwidth]{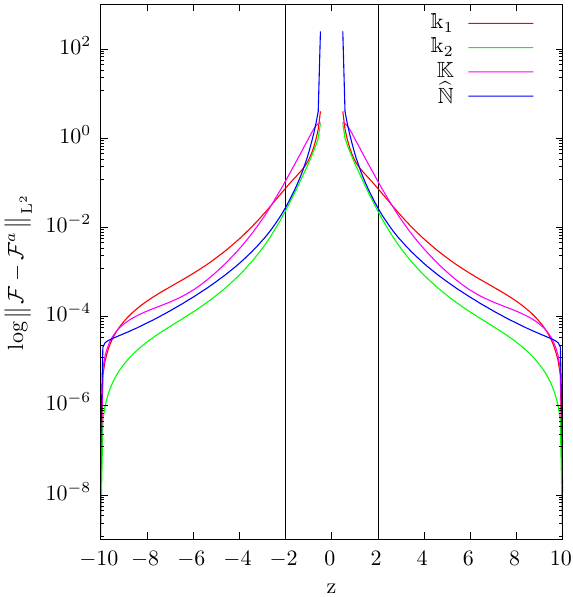}
\includegraphics[width=0.45\textwidth]{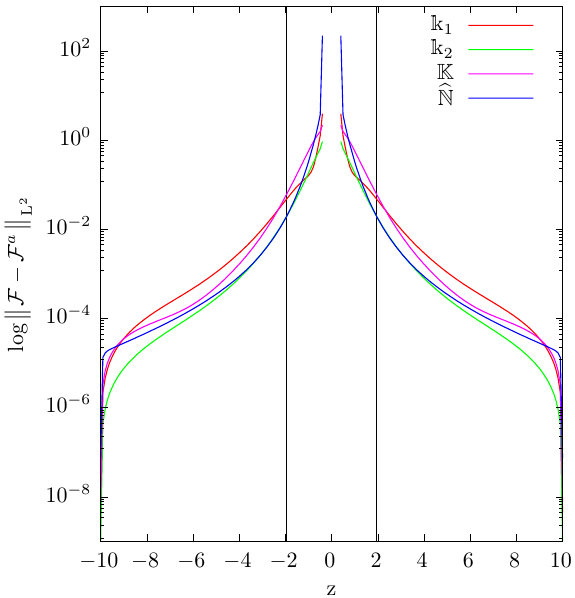}
\caption{The $z$-dependence of logarithms of L$^2$ norms of matrices representing differences between the numerically obtained and analytic values of the variables, \mbox{$\mathcal{F}-\mathcal{F}^{a}$}, for (a)~a~Schwarzschild black hole with $M=1$, (b)~a~Kerr black hole with $M=1$, $a=0.3M$, as well as boosted and displaced (c)~Schwarzschild black hole with $M=1$, $v=0.7$, $d=3M$ and (d)~Kerr black hole with $M=1$, $a=0.3M$, $v=0.6$, $d=5M$. Black vertical lines {are to} indicate {the} location of event horizons {in the $z$-direction}.}
\label{fig:norms-sing}
\end{figure}

\begin{figure}[H]
\centering
(a)\hspace{6cm}(b)\\
\includegraphics[width=0.45\textwidth]{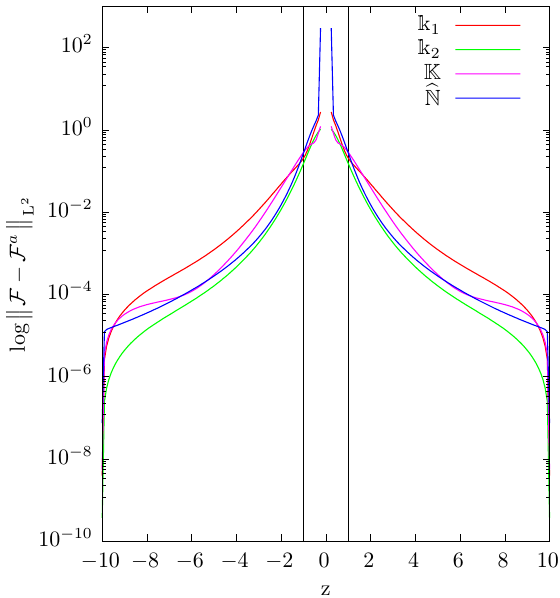}
\includegraphics[width=0.45\textwidth]{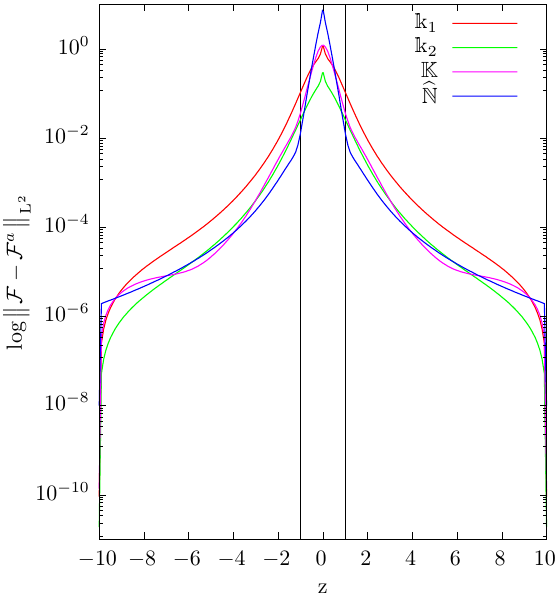}
\caption{Plots analogous to the ones in Fig.~\ref{fig:norms-sing}, for boosted and displaced (a)~Schwarzschild black hole with $M=0.5$, $v=0.7$, $d=3M$ and (b)~Kerr black hole with $M=0.5$, $a=0.3M$, $v=0.6$, $d=5M$.}
\label{fig:norms-sing-smM}
\end{figure}

Figs. \ref{fig:norms-sing} and \ref{fig:norms-sing-smM} {depict} the logarithms of L$^2$ norms---{as} defined {in}~(\ref{eqn:L2})---of differences between numerical and analytical values of the constrained {variables} versus $z$ for several single black holes. {It is visible that} the norms increase as the slice $z=0$ is approached, what is expected due to the fact that this surface contains a black hole singularity, in the vicinity of which the numerical solution fails to be reliable. Moreover, a comparison between the plots \ref{fig:norms-sing}c-d with \ref{fig:norms-sing-smM}a-b, respectively, {verifies} that for smaller black holes the norms are smaller. However, if one wishes to refer the computational accuracy to the event horizon $z$-location, the computations for bigger black holes are more accurate at the respective $z$.

\begin{figure}[H]
\centering
\includegraphics[width=0.45\textwidth]{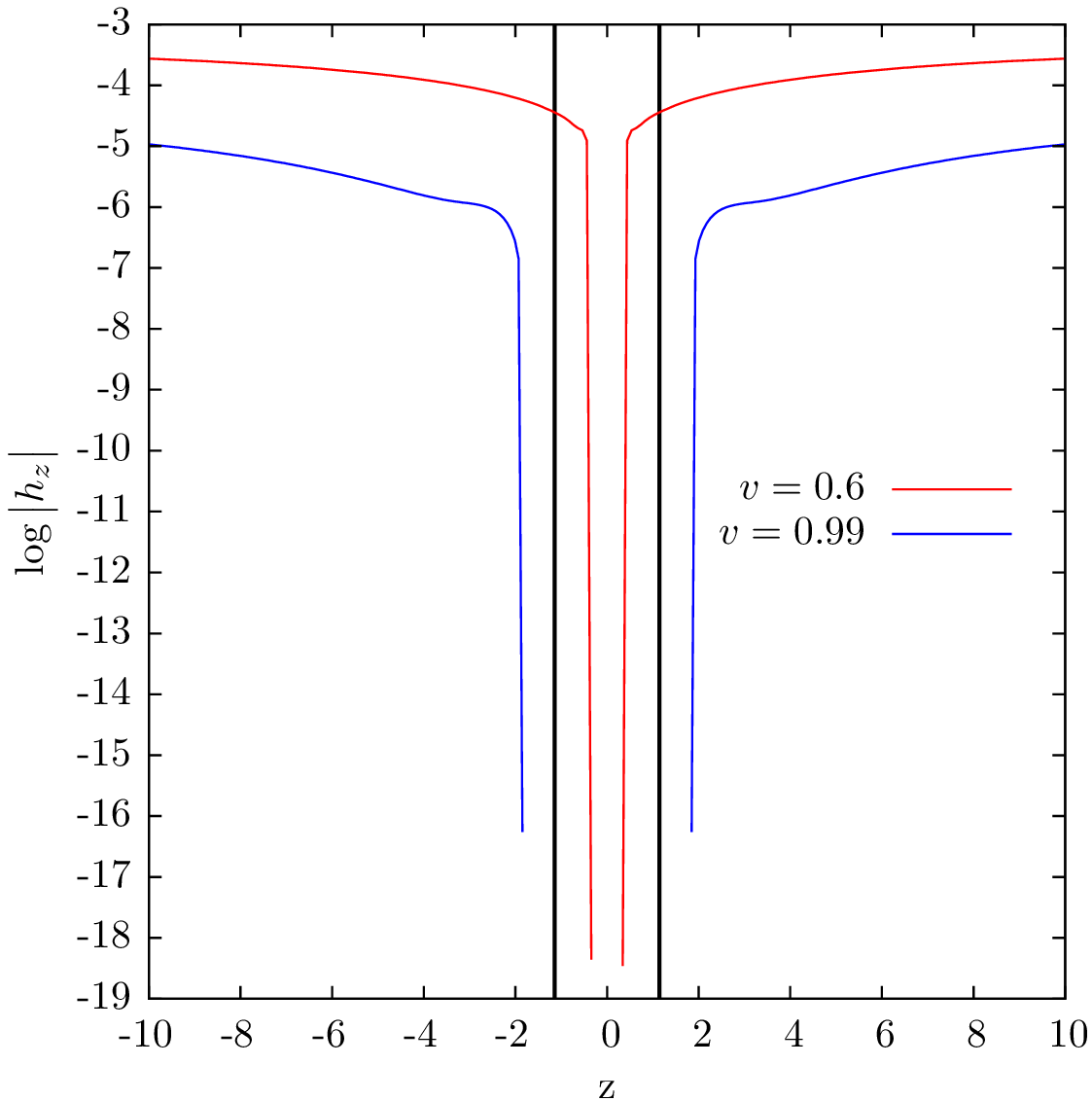}
\caption{The {logarithm of the} adaptive $z$-step {size}, $h_z$, versus $z$ is plotted for a highly boosted, highly spinning and displaced Kerr black hole with {parameters} $M=1$, $a=0.99M$, $v=0.99$, $d=3M$ and {for} a black hole characterized by the same set of parameters but {with} a smaller speed $v=0.6$. Thick black vertical lines {are to} indicate the location of event horizons of the black holes {in the $z$-direction}.}
\label{fig:norms-zsteps-sing-Bav}
\end{figure}

The choice of the black hole parameters as such does not {seem to} have significant influence on the performance of the numerical results. {Accordingly,} the proposed method seems to be very robust, without any intrinsic limitations regarding the physical parameters of the examined objects. The performance of the numerical schema was sensitive to the applied boost in the sense that a finer grid was required by higher values of the applied boost. The behavior of the $z$-step in the case of a highly boosted and highly spinning black hole in comparison with a black hole with a smaller boost is shown in Fig.~\ref{fig:norms-zsteps-sing-Bav}. The proposed parabolic-hyperbolic formulation of the initial data along with the proposed numerical treatment enabled us to investigate black holes of arbitrarily large spins and boosts. Values of $a$ and $v$ close to the upper bounds of their admissible ranges do not decrease the accuracy of numerical computations. However, as has already been mentioned in Section~\ref{sec:numcomp}, big boosts enforce{d} {a} smaller $h_z$, as well as {required} smaller $x$ and $y$ {grid {spacings}.} {This increased} the computation{al time} considerably when the full form of the equations {was} integrated. This time-inefficiency {was} overcome {by} using the deviations form of the constraints, as it will be demonstrated in Section~\ref{ssec:devs-codetests}.

\begin{figure}[H]
\centering
(a)\hspace{7cm}(b)\\
\includegraphics[width=0.495\textwidth]{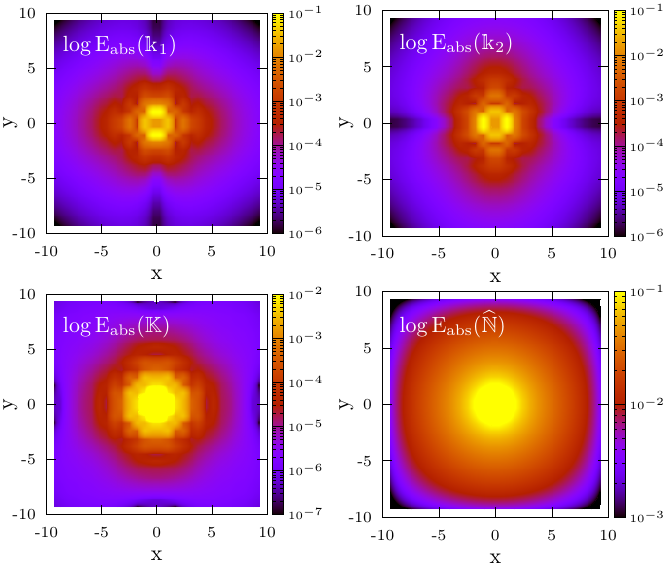}
\includegraphics[width=0.495\textwidth]{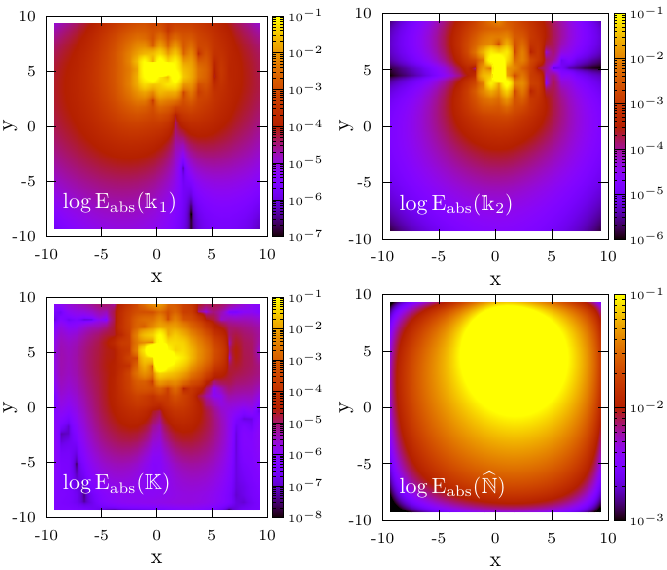}
\caption{Logarithms of absolute values of differences between numerical and analytic values of the {indicated} variables, $\textrm{E}_\textrm{abs}\left(\mathcal{F}\right)$, at $z=\tfrac14 r_H$ for (a)~a~Schwarzschild black hole with $M=1$ and (b)~a~Kerr black hole with $M=1$, $a=0.3M$, $v=0.6$, $d=5M$.}
\label{fig:AE-sing}
\end{figure}

\begin{figure}[H]
\centering
(a)\hspace{7cm}(b)\\
\includegraphics[width=0.495\textwidth]{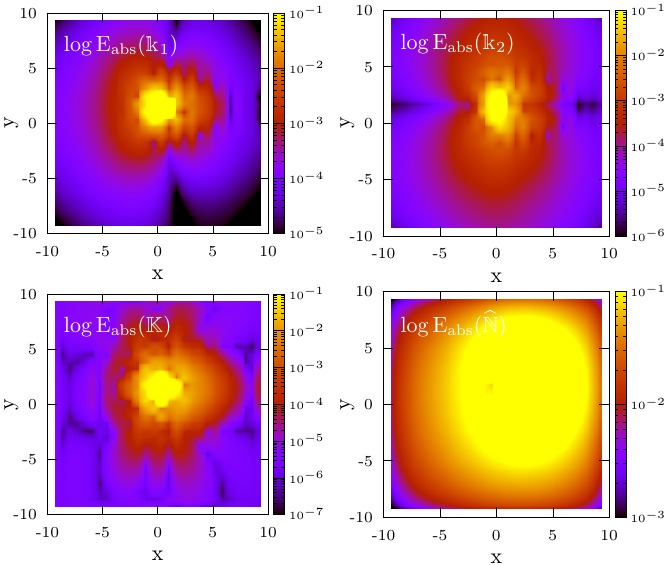}
\includegraphics[width=0.495\textwidth]{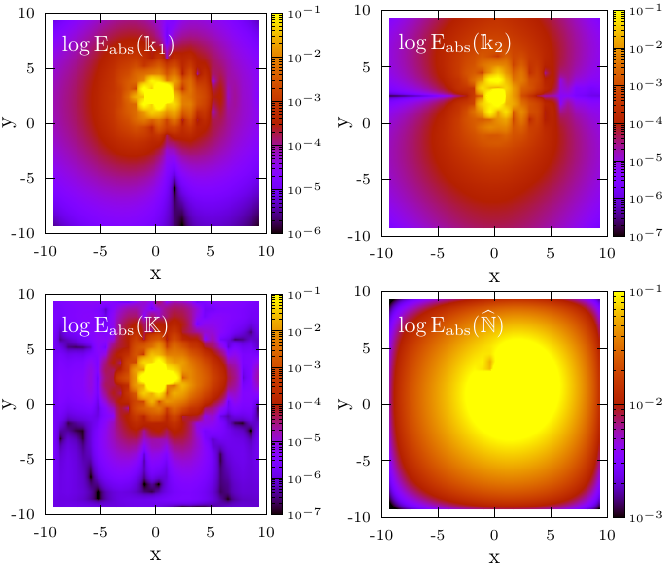}
\caption{Plots analogous to the ones in Fig.~\ref{fig:AE-sing}, for boosted and displaced (a)~Schwarzschild black hole with $M=0.5$, $v=0.7$, $d=3M$ and (b)~Kerr black hole with $M=0.5$, $a=0.3M$, $v=0.6$, $d=5M$.}
\label{fig:AE-sing-smM}
\end{figure}

{The inspection of} absolute errors---determined by comparing the corresponding numerical and analytic values (\ref{eqn:ae})---{on {some specific} $z=const$ slices} enabled us to determine the domain of reliability of the applied numerical scheme. Figs.~\ref{fig:AE-sing} and \ref{fig:AE-sing-smM} are to show logarithms of the absolute errors on a $z=const$ slice, which corresponds to a quarter of the particular black hole's event horizon, $r_H$. The errors are bigger for smaller black holes, due to the fact that their event horizon is located at smaller $z$ {values}. The critical threshold $\textrm{E}_\textrm{abs}\left(\mathcal{F}\right)=10\cdot h^n$, which stems from (\ref{eqn:accuracy}) and is equal to $2.262$ in our case, was not crossed within the region of calculations. It {means that} the accuracy of the overall numerical approach {is of the 4$^{th}$ order as expected} in the entire considered computational domain.

\begin{figure}[H]
\centering
(a)\hspace{7cm}(b)\\
\includegraphics[width=0.495\textwidth]{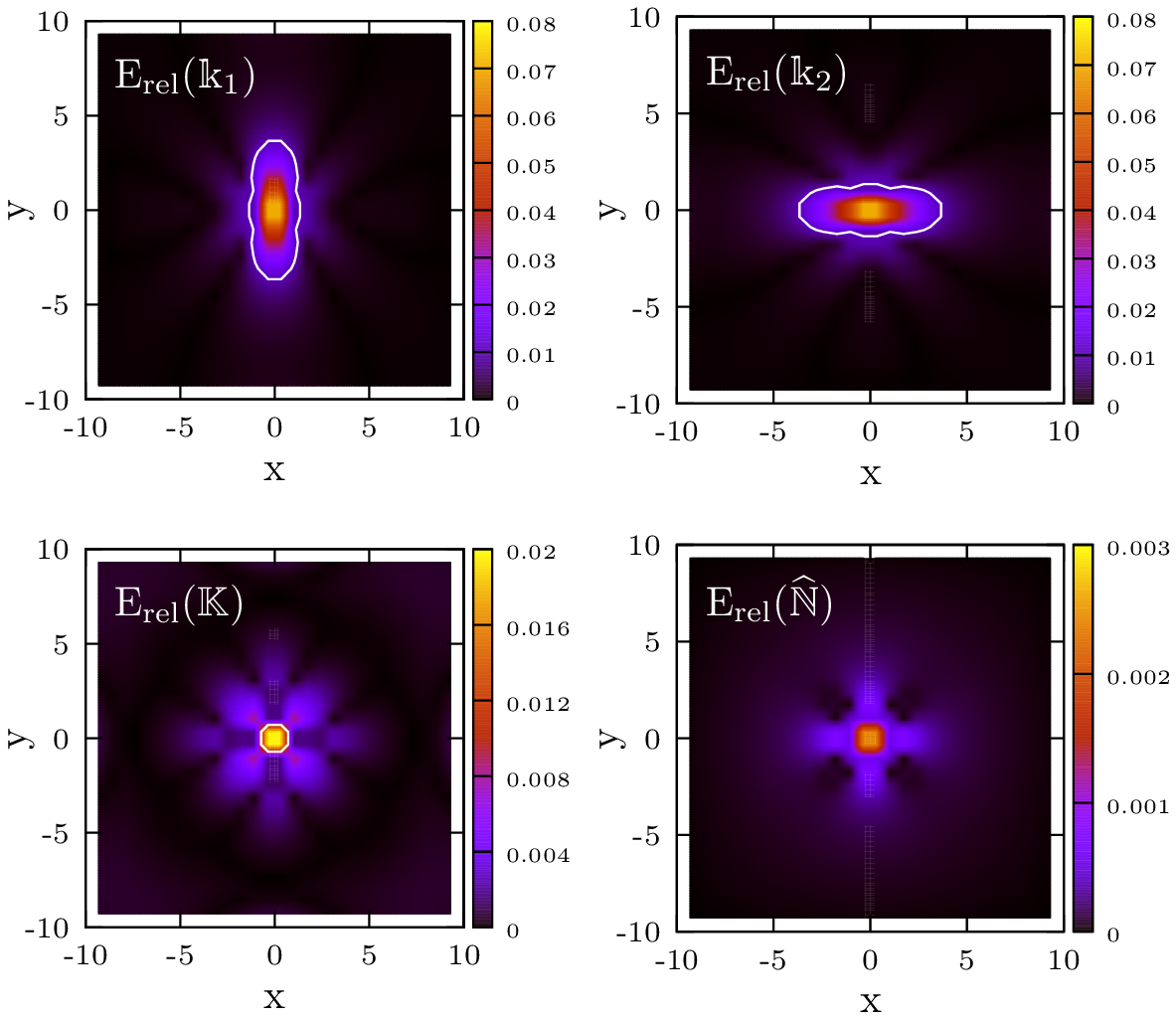}
\includegraphics[width=0.495\textwidth]{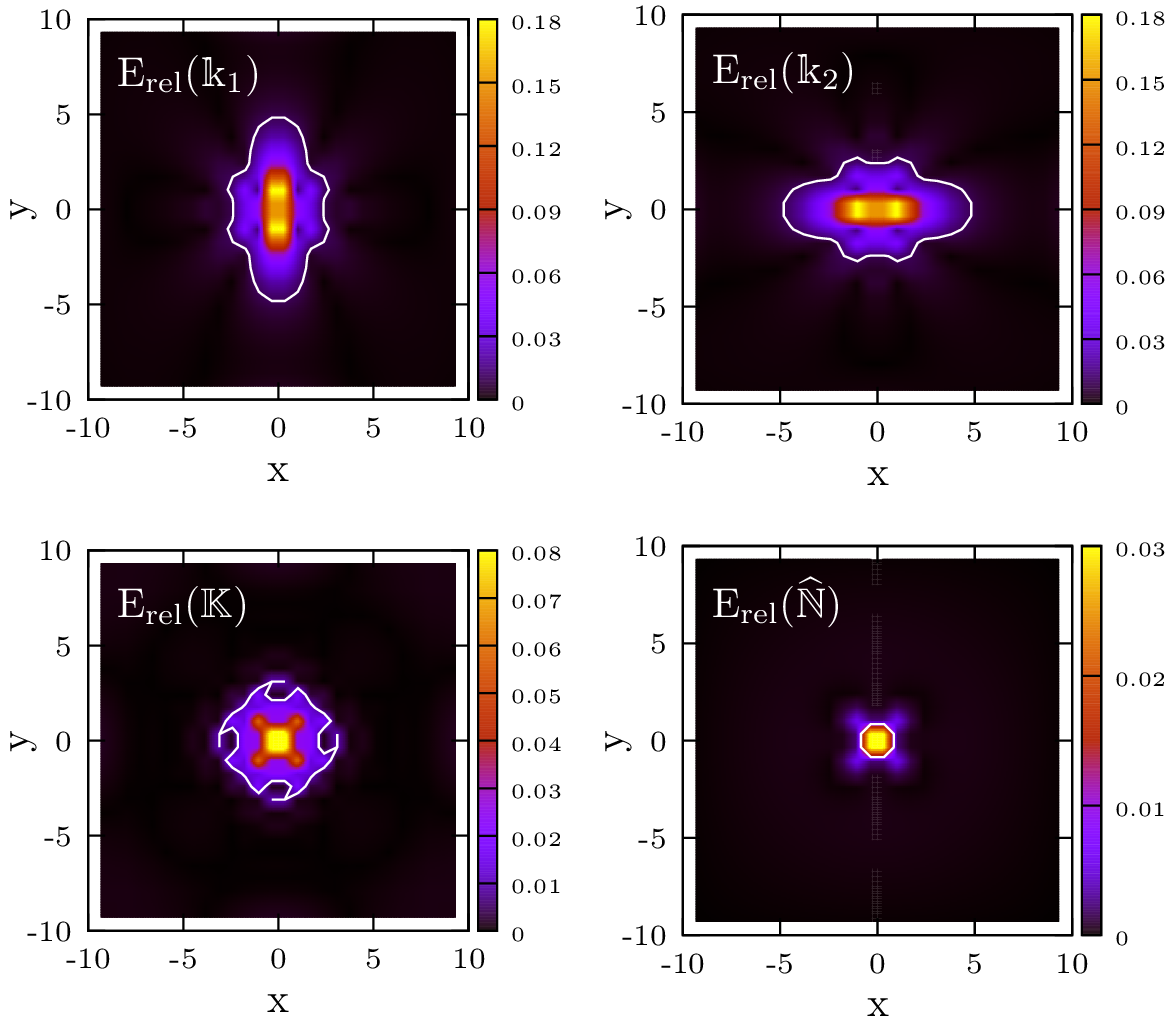}
\caption{Relative errors of the constrained variables, $\textrm{E}_\textrm{rel}\left(\mathcal{F}\right)$, are indicated at (a)~$z=r_H$ and (b)~$z=\tfrac12 r_H$, for a Schwarzschild black hole with $M=1$. The thin white lines are to indicate the error threshold at which $\textrm{E}_\textrm{rel}\left(\mathcal{F}\right)=0.01$.}
\label{fig:RE-sing-S}
\end{figure}

\begin{figure}[H]
\centering
(a)\hspace{7cm}(b)\\
\includegraphics[width=0.495\textwidth]{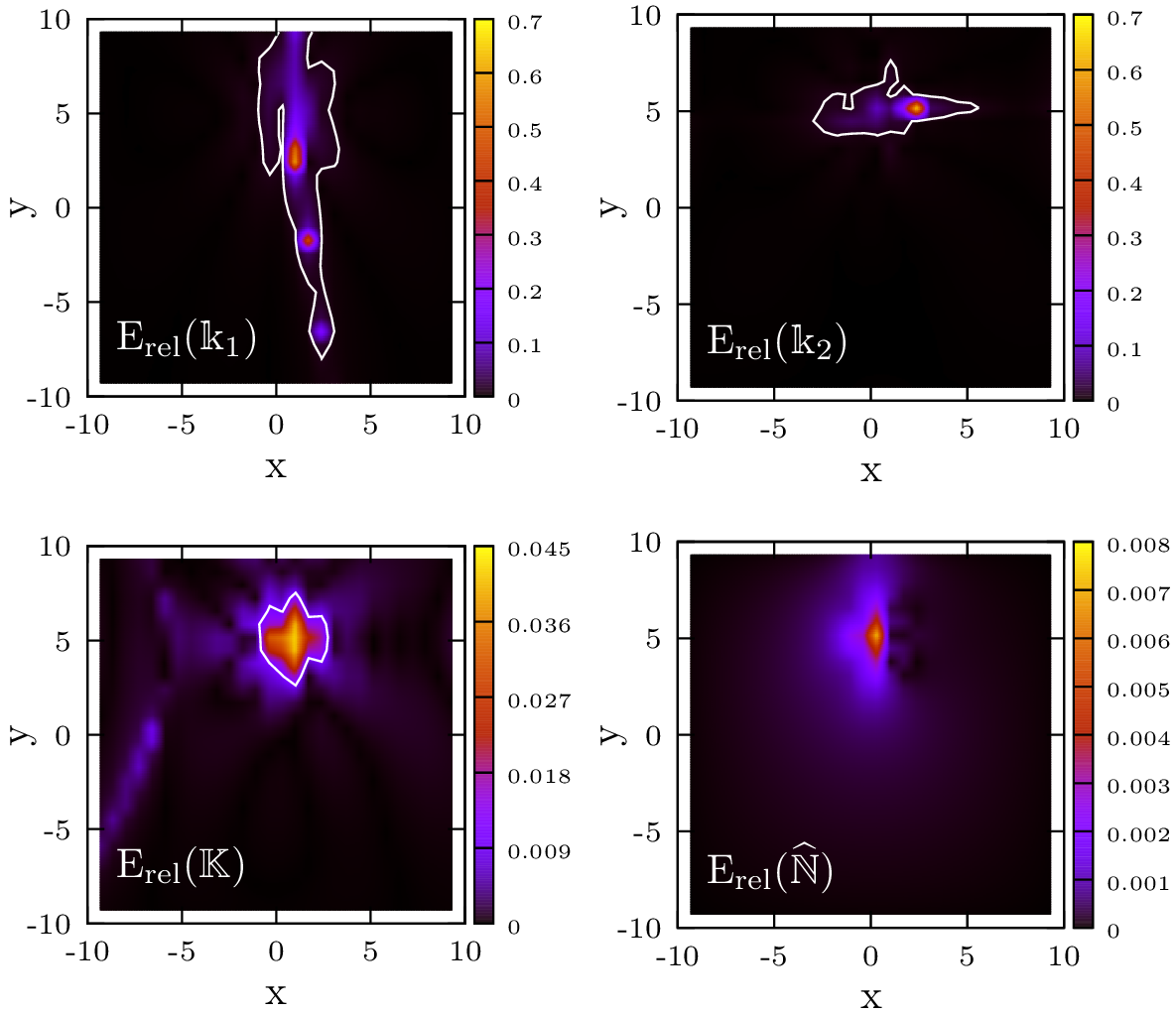}
\includegraphics[width=0.495\textwidth]{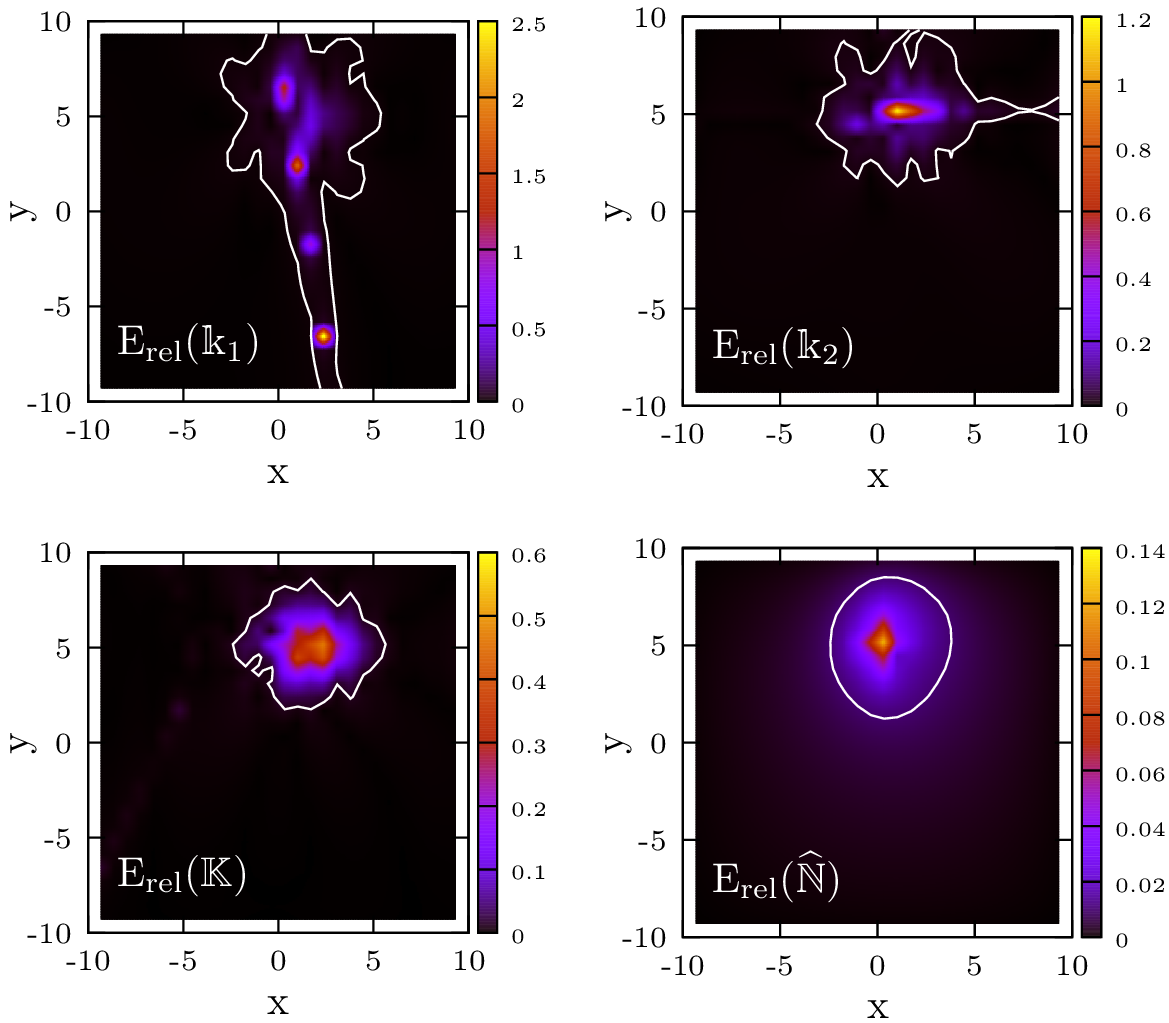}
\caption{Plots analogous to the ones in Fig.~\ref{fig:RE-sing-S}, for a displaced and boosted Kerr black hole with $M=1$, $a=0.3M$, $v=0.6$, $d=5M$.}
\label{fig:RE-sing-Kdb}
\end{figure}

\begin{figure}[H]
\centering
(a)\hspace{7cm}(b)\\
\includegraphics[width=0.495\textwidth]{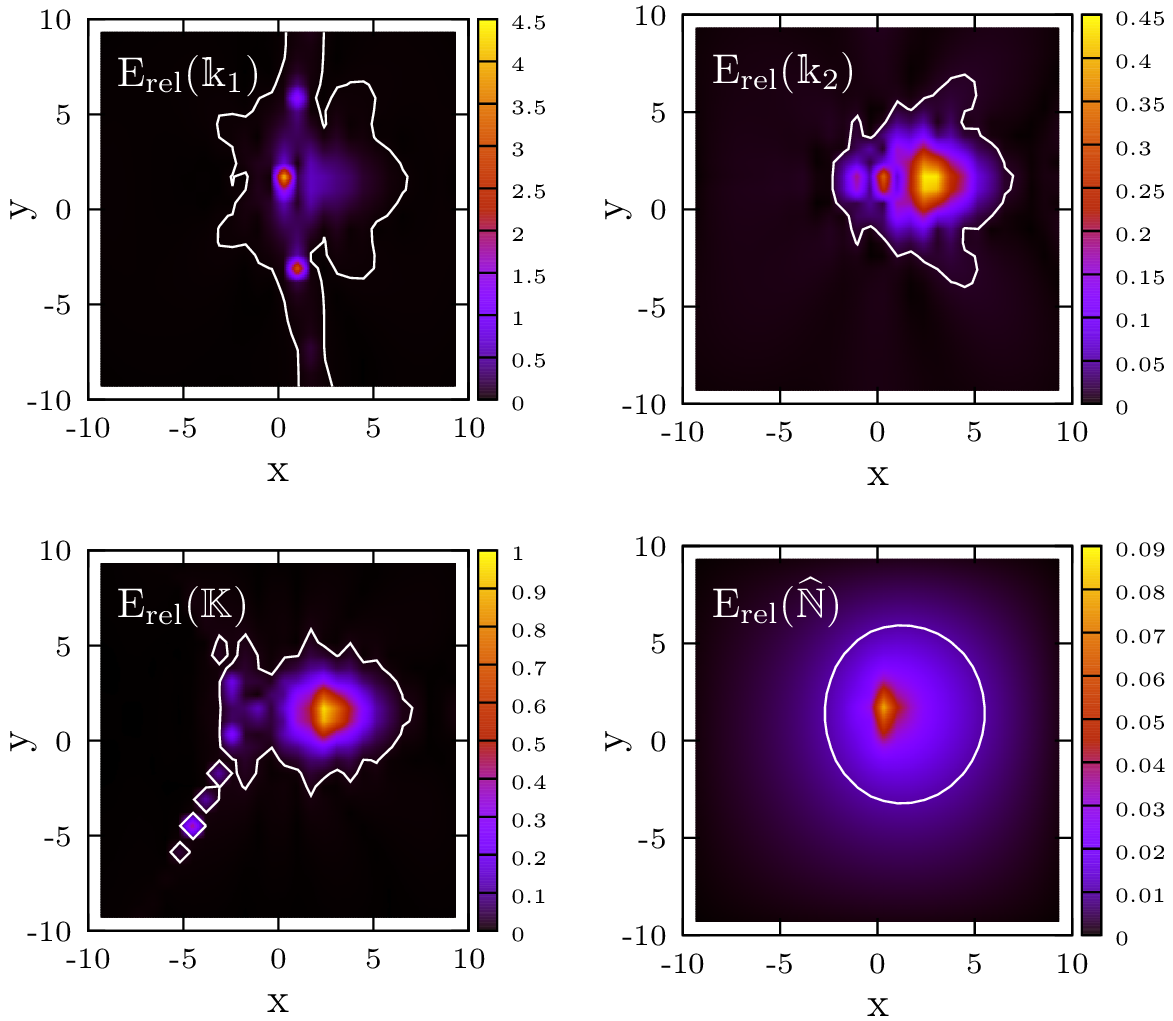}
\includegraphics[width=0.495\textwidth]{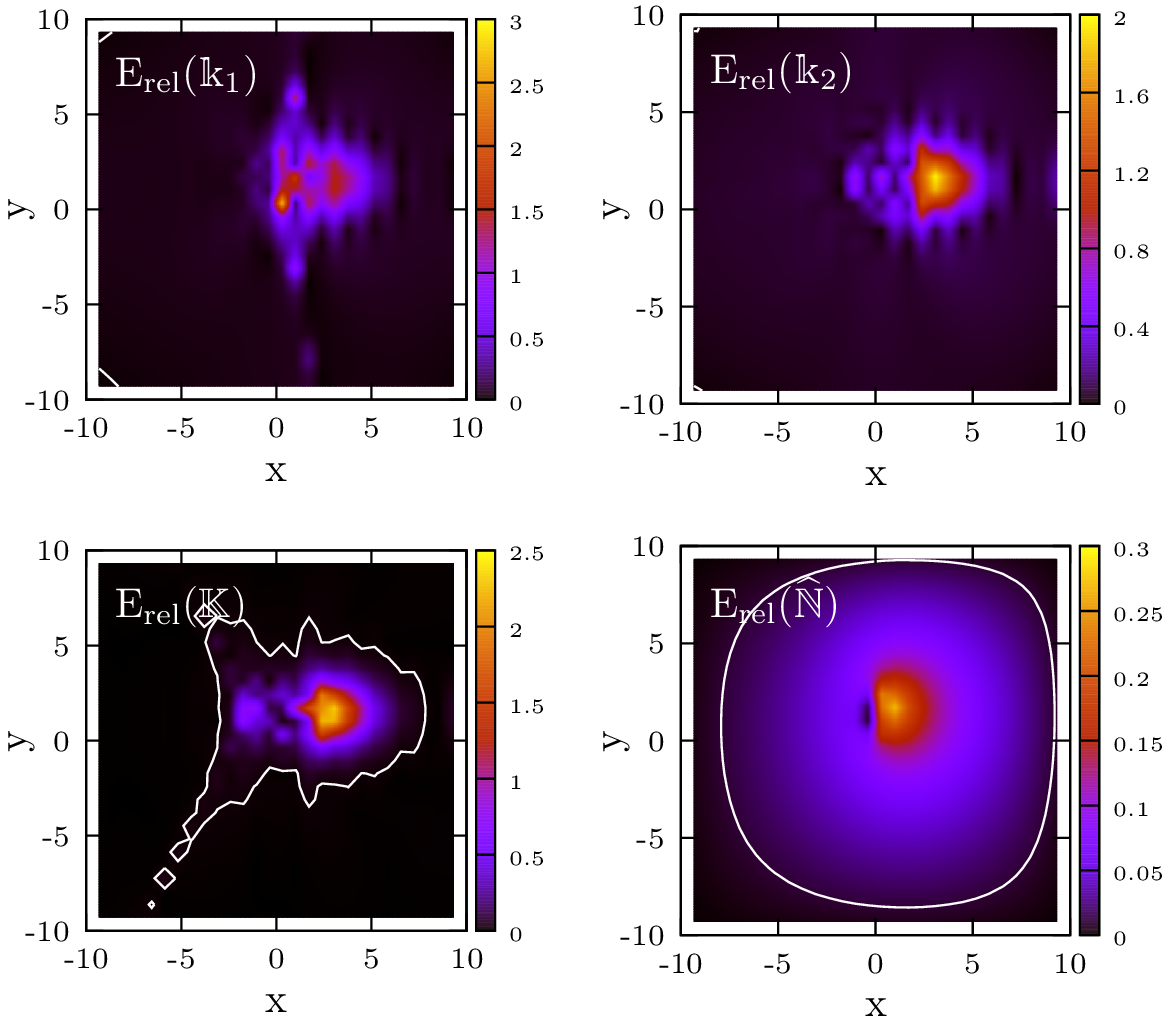}
\caption{Plots analogous to the ones in Fig.~\ref{fig:RE-sing-S}, for boosted and displaced Schwarzschild black hole with $M=0.5$, $v=0.7$, $d=3M$.}
\label{fig:RE-sing-S-smM}
\end{figure}

\begin{figure}[H]
\centering
(a)\hspace{7cm}(b)\\
\includegraphics[width=0.495\textwidth]{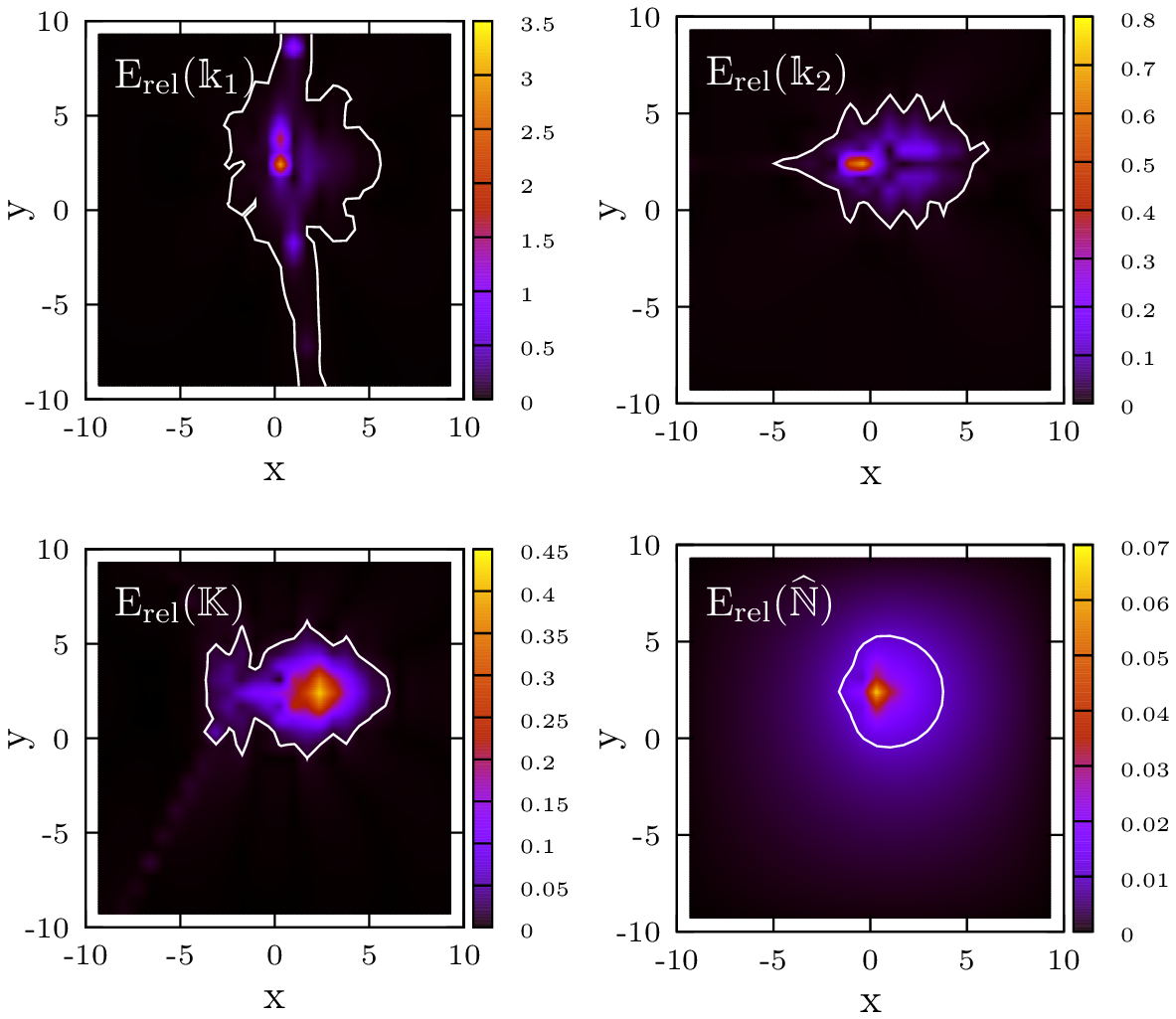}
\includegraphics[width=0.495\textwidth]{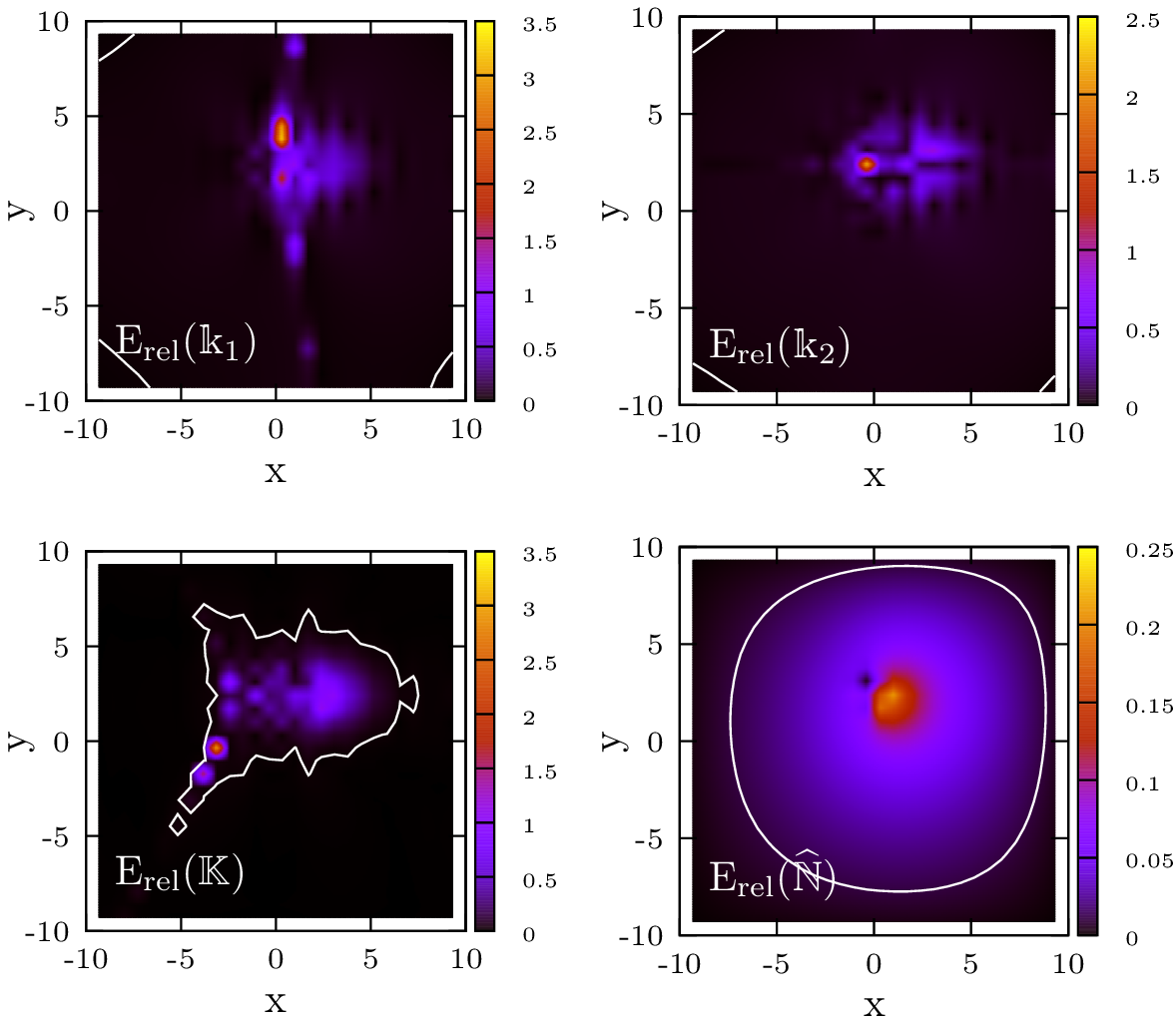}
\caption{Plots analogous to the ones in Fig.~\ref{fig:RE-sing-S}, for boosted and displaced Kerr black hole with $M=0.5$, $a=0.3M$, $v=0.6$, $d=5M$.}
\label{fig:RE-sing-K-smM}
\end{figure}

The {most severe} accuracy check was performed by {monitoring the} relative errors of the numerically obtained data with respect to the analytic ones~(\ref{eqn:re}). {The corresponding results} are shown in Figs.~\ref{fig:RE-sing-S}--\ref{fig:RE-sing-K-smM} by depicting the values of the relative errors on various $z=const$ level surfaces---in particular, at the location of the black hole event horizon and at {a} half of that value---for several black holes. {On each of these plots} the {thin white color} line indicates the accuracy equal to $1\%$. Likewise the above-discussed absolute error, the point-wise relative errors are bigger for black holes with smaller mass, but this is rooted in the fact that the grid was not made finer for these smaller black holes {and the errors were investigated at slices corresponding to smaller $z$ values}. The errors are {again} bigger for boosted black holes. 

{It {is} worth emphasiz{ing}} that the relative error as such is very sensitive to the {absolute} value {of the variable} present in the denominator {(see the definition \eqref{eqn:ae})} {which in the present case was the} analytic value of the {investigated} variable. Thus, a part of the relatively big values of the error, especially visible along the almost vertical line for $\textrm{E}_\textrm{rel}\left(\smk_1\right)$ and an arc in the region of $x< -2.5$ for $\textrm{E}_\textrm{rel}\left(\KK\right)$, is related to small values of the respective constrained functions there, {rather than yielded by} the {inappropriateness of the} numerical solution. This observation is confirmed by the fact that the absolute error discussed above does not indicate any sort of undesirable behavior in these regions. It {is} worth underlying that the relative error test provided an extremely robust verification of the accuracy of the applied numerical code.

\subsection{Single black hole solutions in the full schema}
\label{ssec:fullform-sing}

\begin{figure}[H]
\centering
(a)\hspace{7cm}(b)\\
\includegraphics[width=0.475\textwidth]{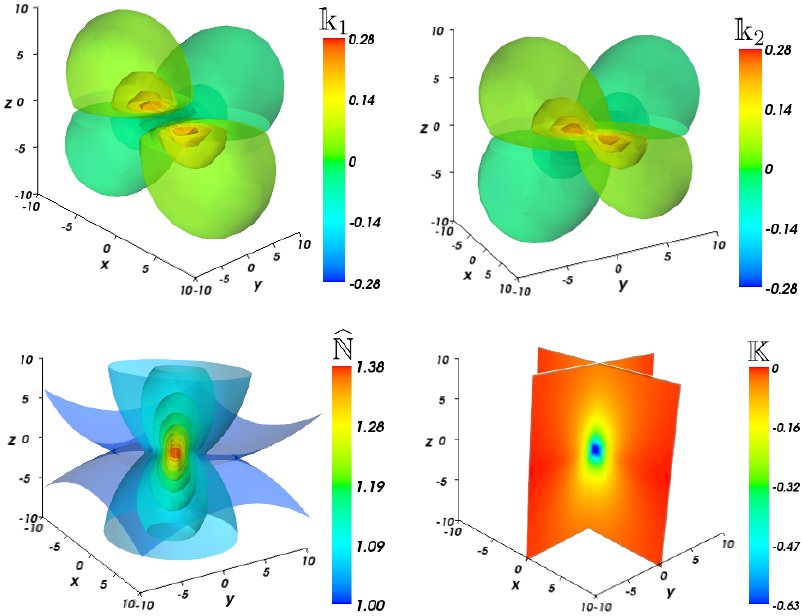}\hspace{0.5cm}
\includegraphics[width=0.475\textwidth]{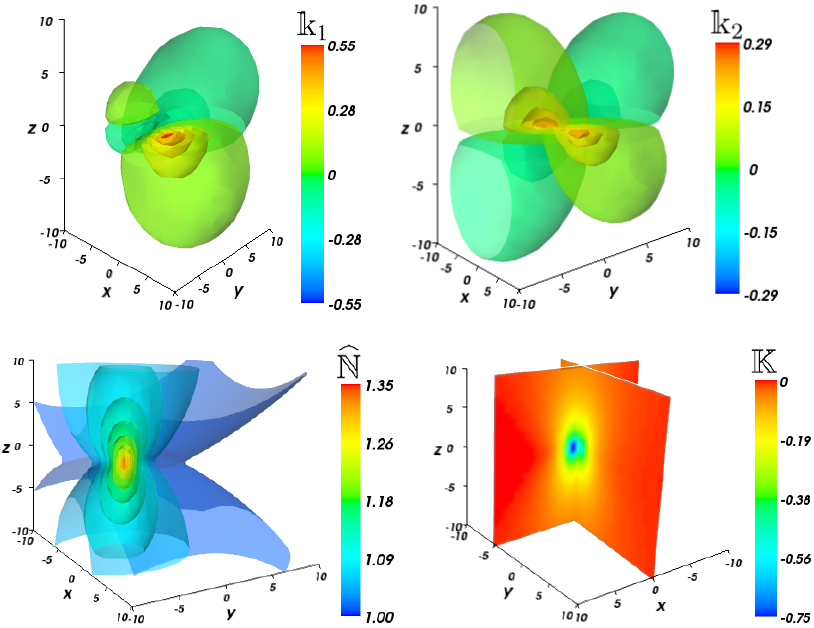}
\caption{The constrained {variables}, $\mathcal{F}$, are depicted for black holes with $M=1$ and (a)~$a=0.3M$, (b)~$a=0.3M$, $v=0.6$, $d=5M$.}
\label{fig:full-constr}
\end{figure}

After performing the above-described tests of the {applied} numerical code, the constraints were solved for specific choices of {the} physical parameters $M$, $v$, $d$ and $a$. The corresponding constrained variables are shown in Fig.~\ref{fig:full-constr}.

\begin{figure}[H]
\centering
(a)\hspace{5cm}(b)\\
\includegraphics[width=0.3\textwidth]{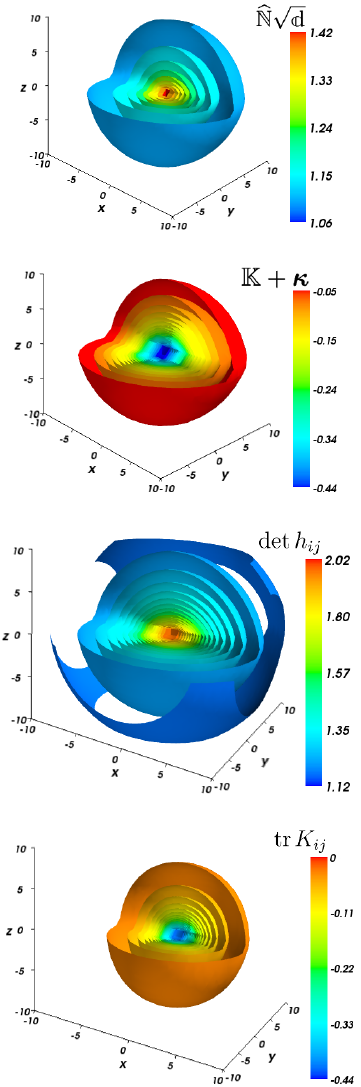}\hspace{0.5cm}
\includegraphics[width=0.3\textwidth]{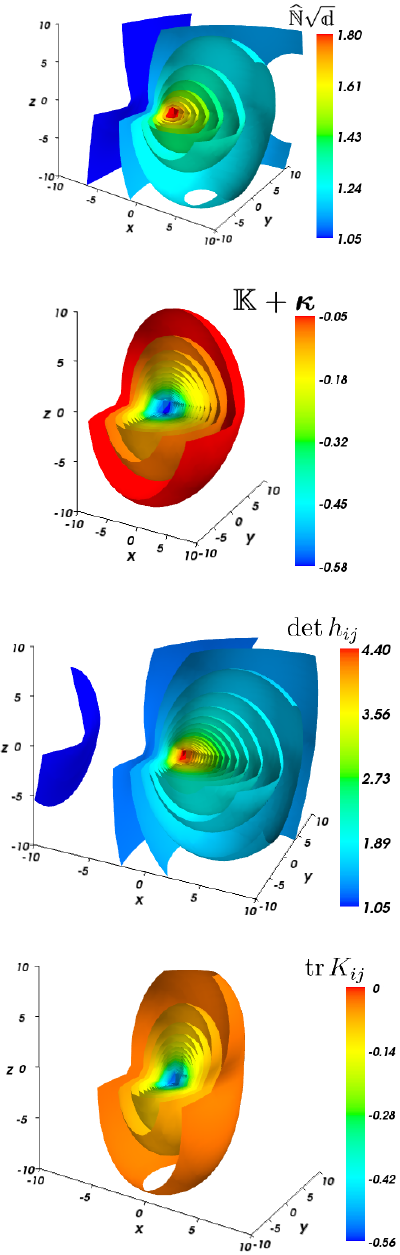}
\caption{{The gauge independent} quantities {$\Nh\sqrt{\mathbbm{d}}$}, {$\KK+\boldsymbol{\kappa}$}, $\det h_{ij}$ and $\textrm{tr}K_{ij}$ {are plotted} for black holes with {physical} parameters {as} listed in the caption of Fig.~\ref{fig:full-constr}.}
\label{fig:full-quant}
\end{figure}

{In} Fig.~\ref{fig:full-quant} some gauge independent quantities characterizing the obtained initial data are plotted. These are {$\Nh\sqrt{\mathbbm{d}}$}, {$\KK+\boldsymbol{\kappa}$}, the determinant of the three-dimensional metric tensor of the slice, $\det h_{ij}$, and the trace of its extrinsic curvature, $\textrm{tr}K_{ij}$.

\section{Integration of the deviation form of the constraints}
\label{sec:devs-num}
\setcounter{equation}{0}

\subsection{Verifications of the applied numerical schema}
\label{ssec:devs-codetests}

The accuracy of our approach based on deviations was checked by solving the parabolic-hyperbolic constraints while examining distorted black holes as it was outlined in Section~\ref{ssec:inibound}. By a distorted black hole initial data we mean a solution of the parabolic-hyperbolic form of the constraints such that the complete set of the freely specifiable variables on the entire initial data surface {and the initial-boundary values of the constrained variables} {are deduced from slightly different} Kerr-Schild black hole solution{s}. The size of the corresponding discrepancies characterize{s} the strength {of} distortion.  

\medskip

\begin{figure}[H]
\centering
\includegraphics[width=0.45\textwidth]{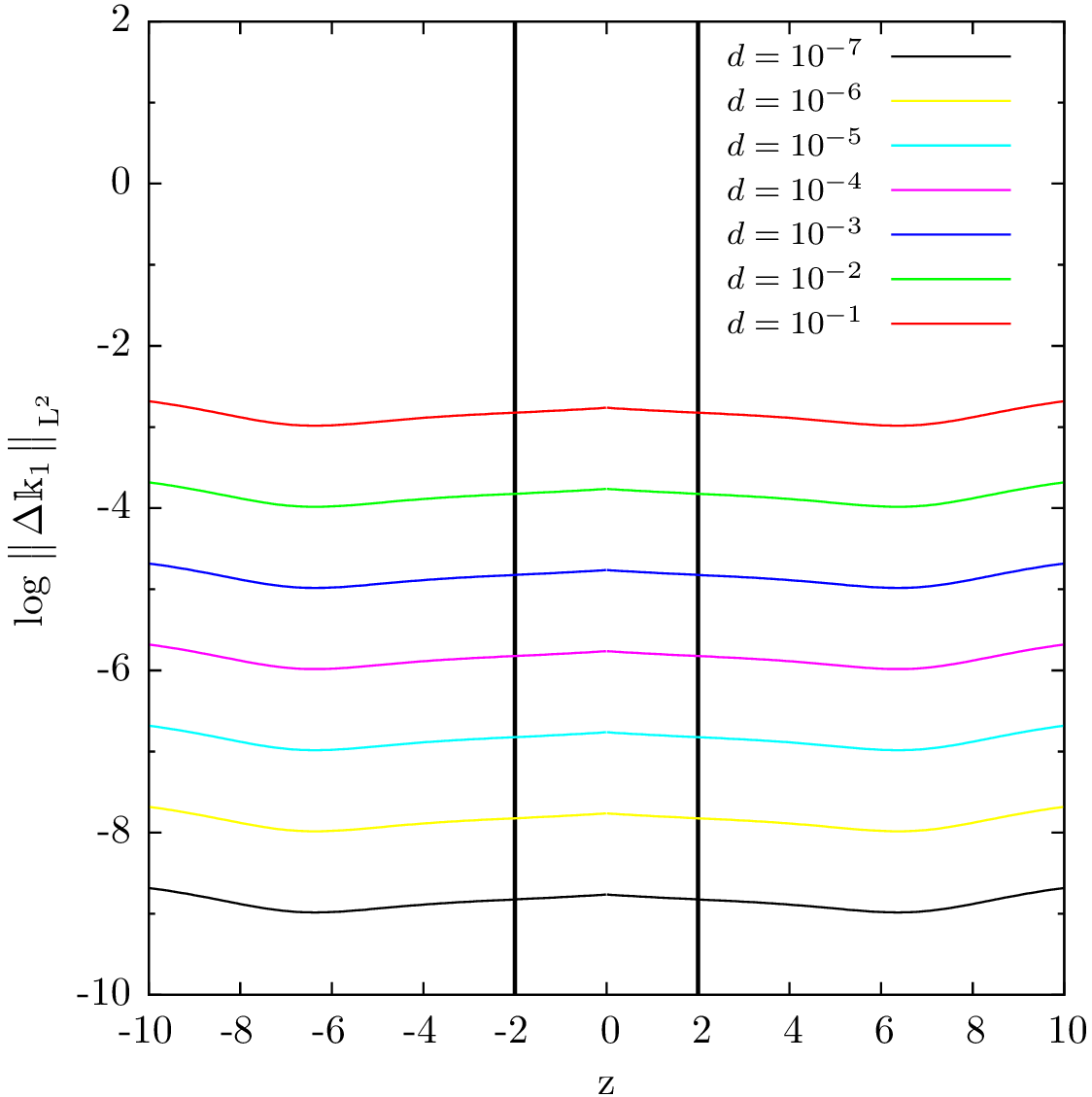}
\includegraphics[width=0.45\textwidth]{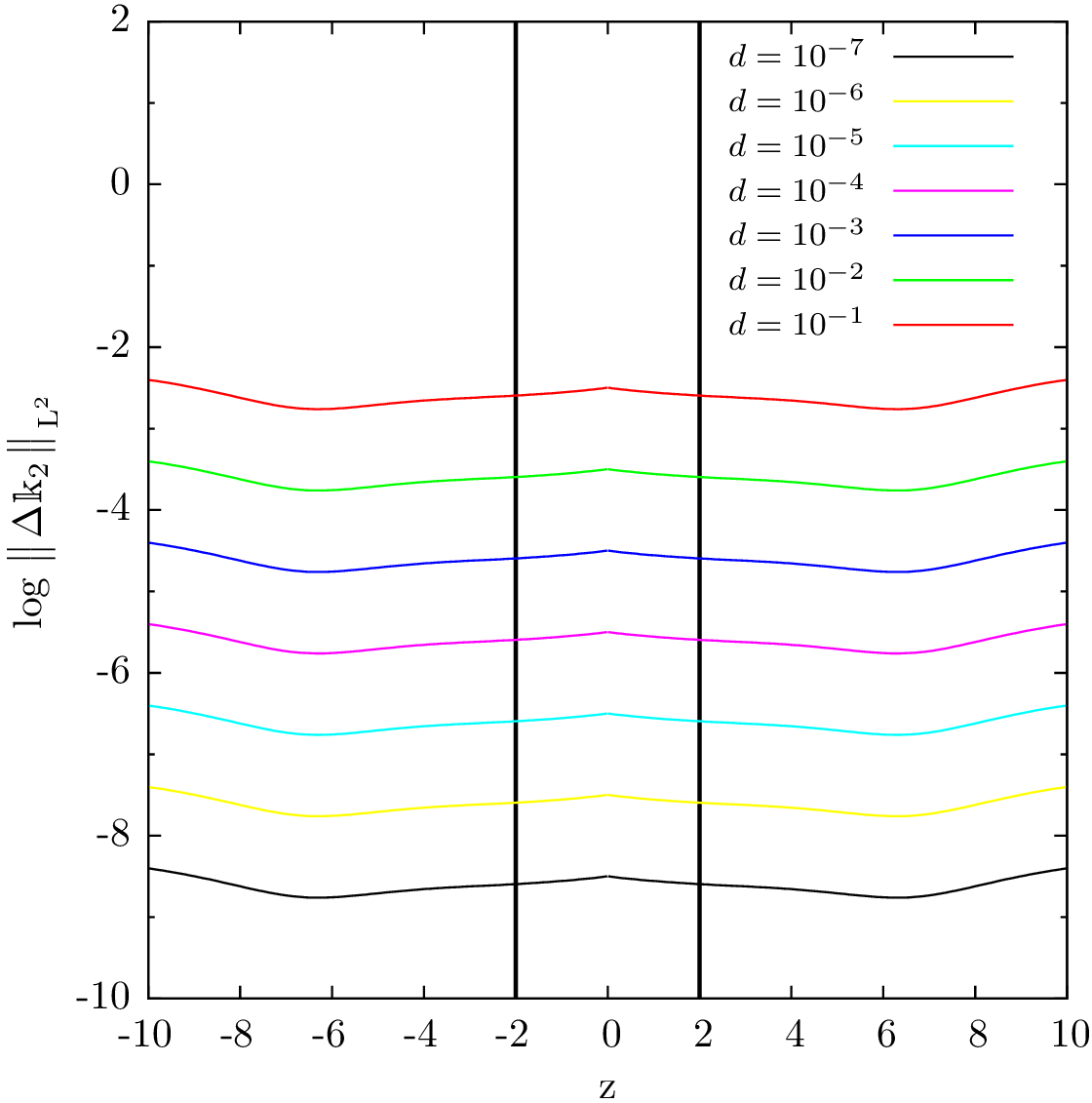}\\
\includegraphics[width=0.45\textwidth]{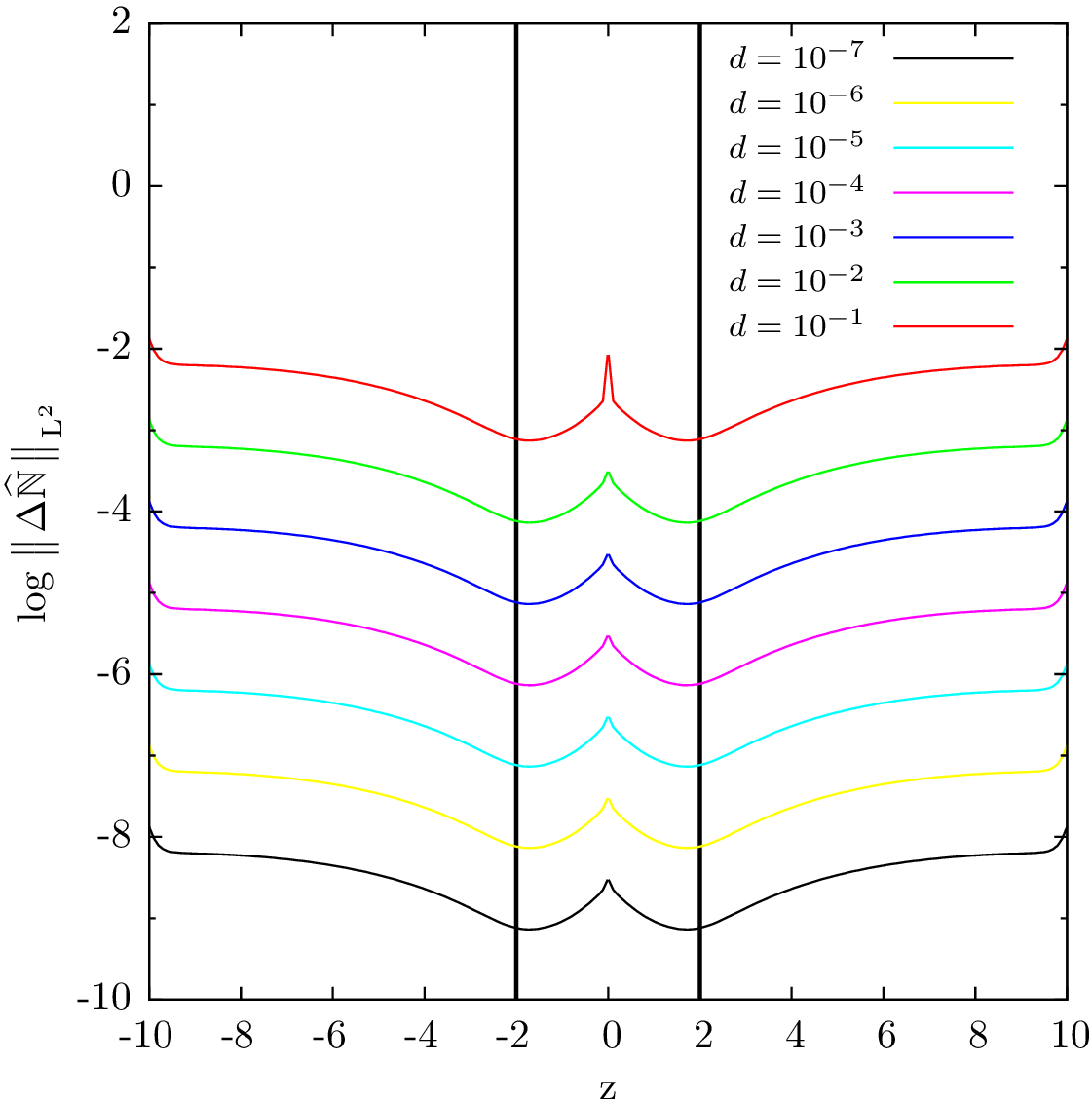}
\includegraphics[width=0.45\textwidth]{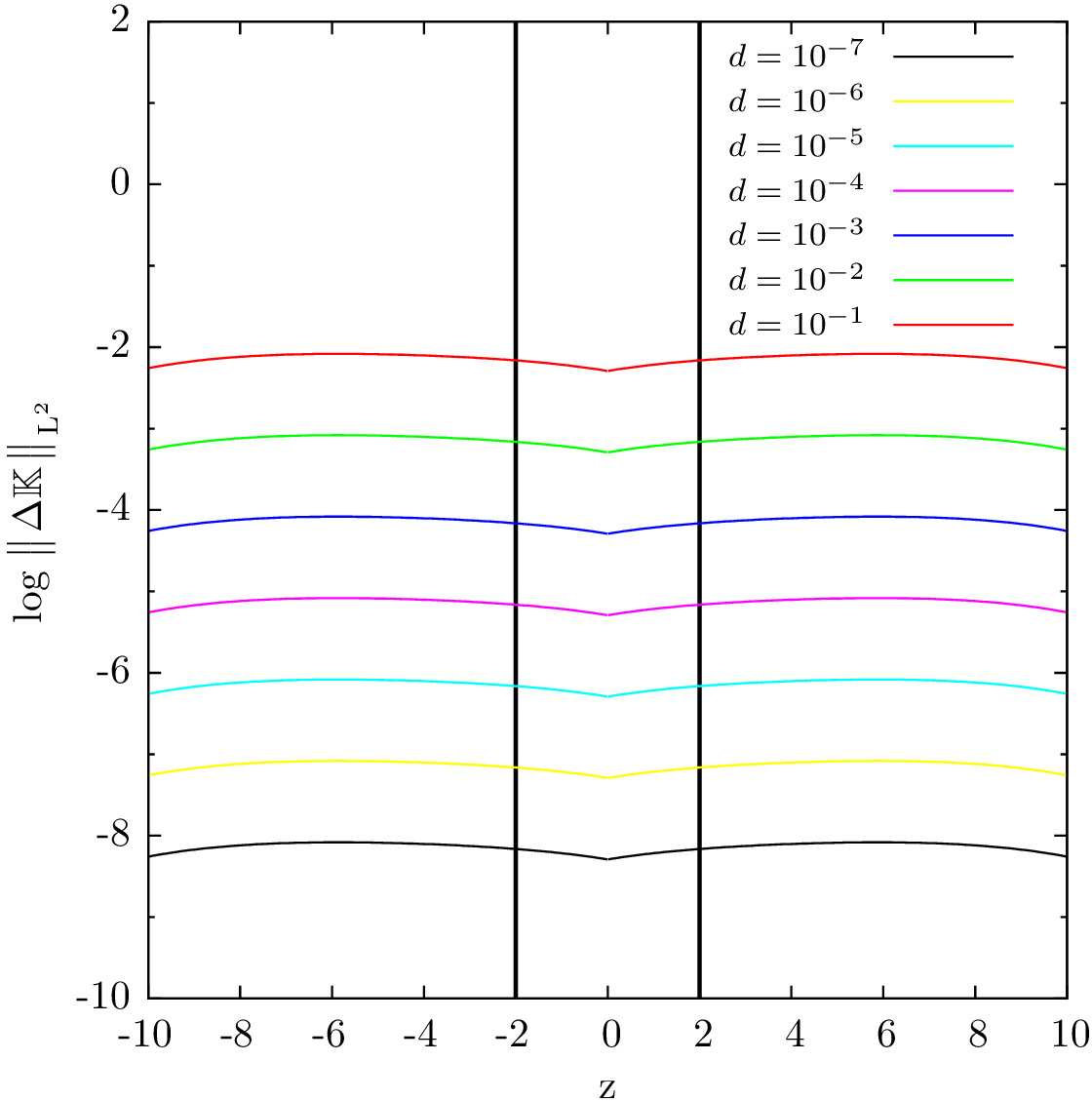}
\caption{The $z$-dependence of the logarithm of the L$^2$ norms of deviations, $\mathcal{F}^\Delta$, for a distorted Schwarzschild black hole. The background, $\preA\hspace{-0.07cm}\mathcal{F}$, corresponded to a black hole with $M=1$, while the initial-boundary data for the deviations corresponded to a displaced black hole with the same mass. The investigated displacements $d$ are listed on each plot. Thick black vertical lines are to indicate the location of the event horizon of the background black hole in the $z$-direction.}
\label{fig:devS-L2}
\end{figure}

\begin{figure}[H]
\centering
\includegraphics[width=0.45\textwidth]{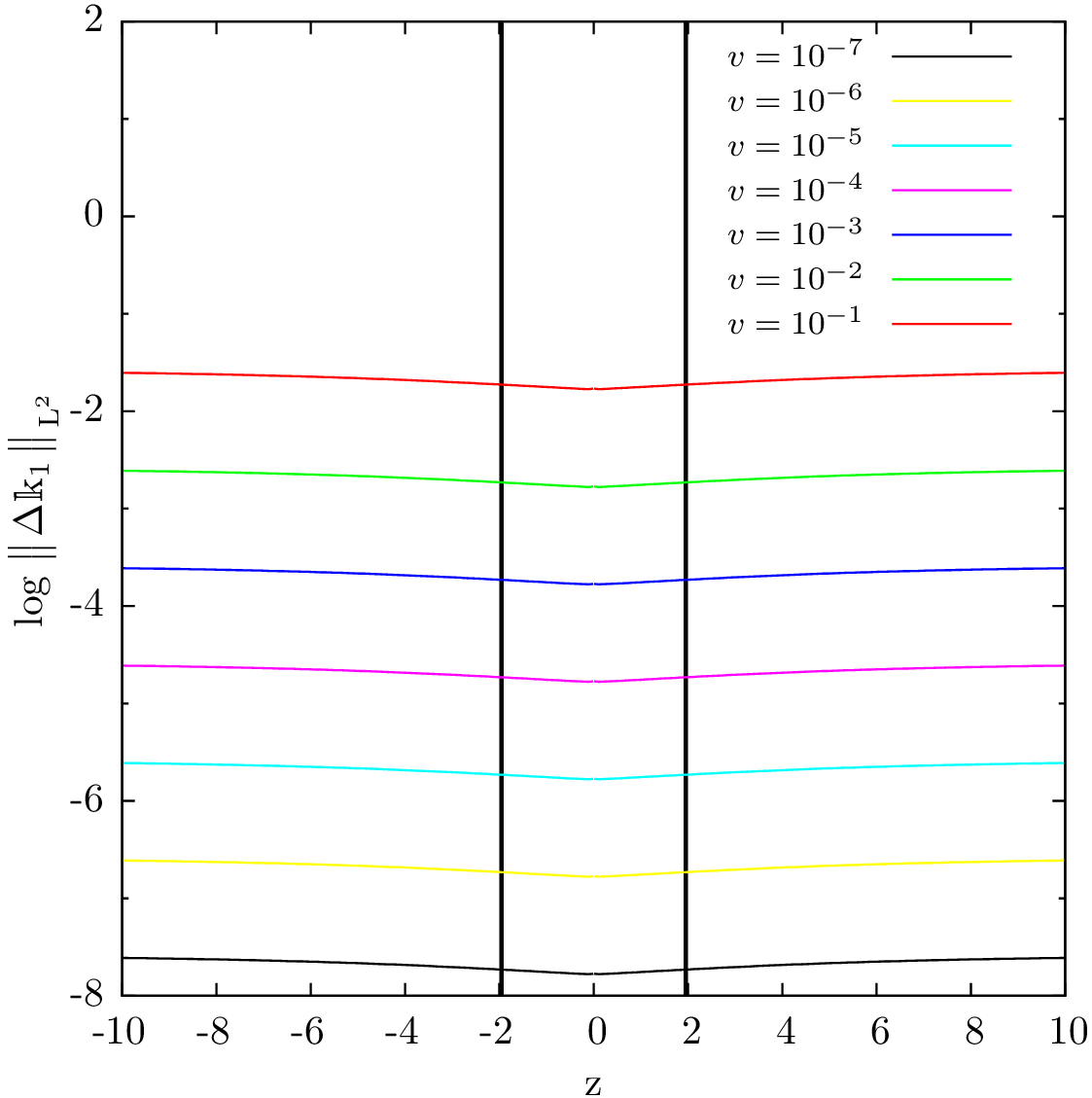}
\includegraphics[width=0.45\textwidth]{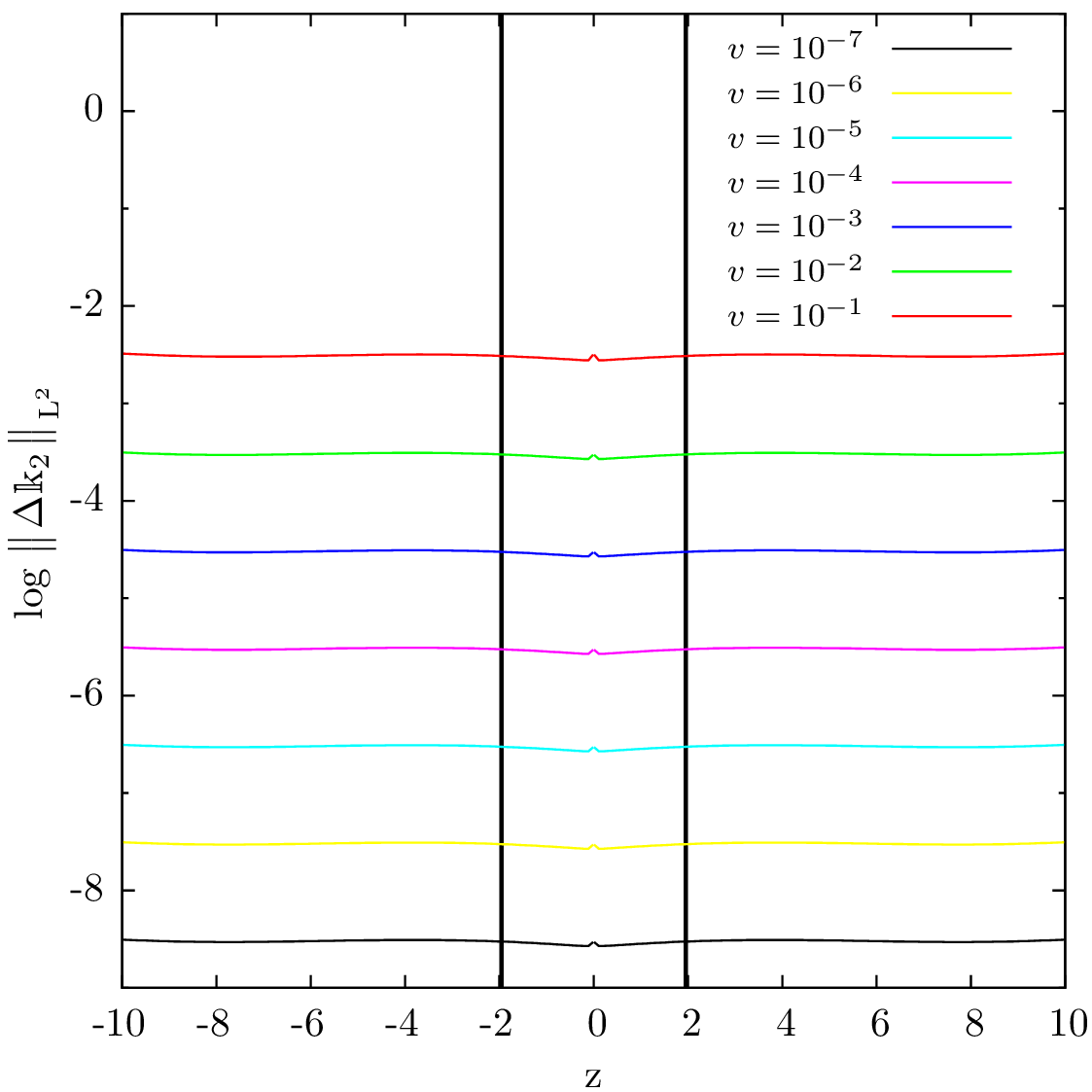}\\
\includegraphics[width=0.45\textwidth]{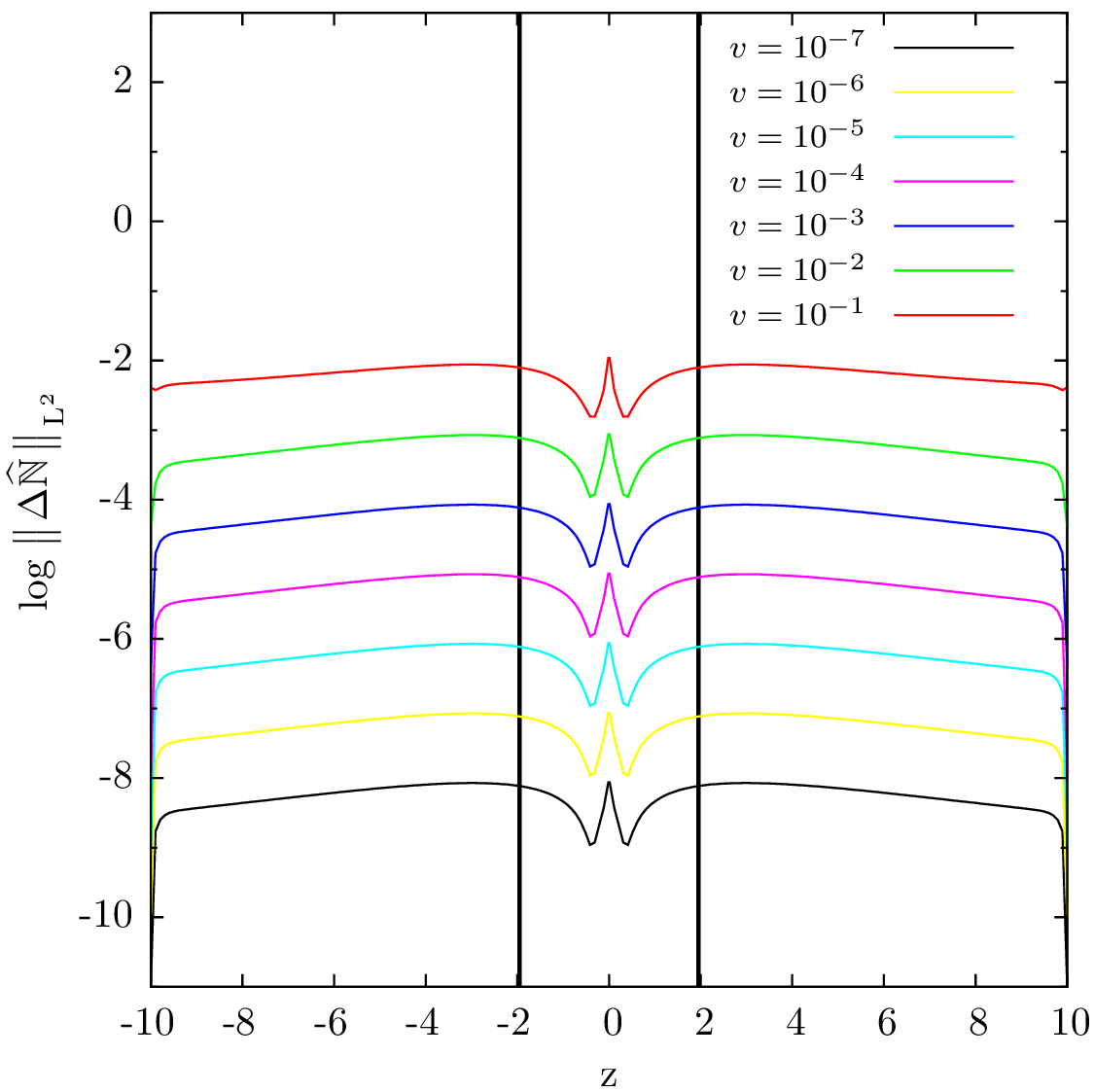}
\includegraphics[width=0.45\textwidth]{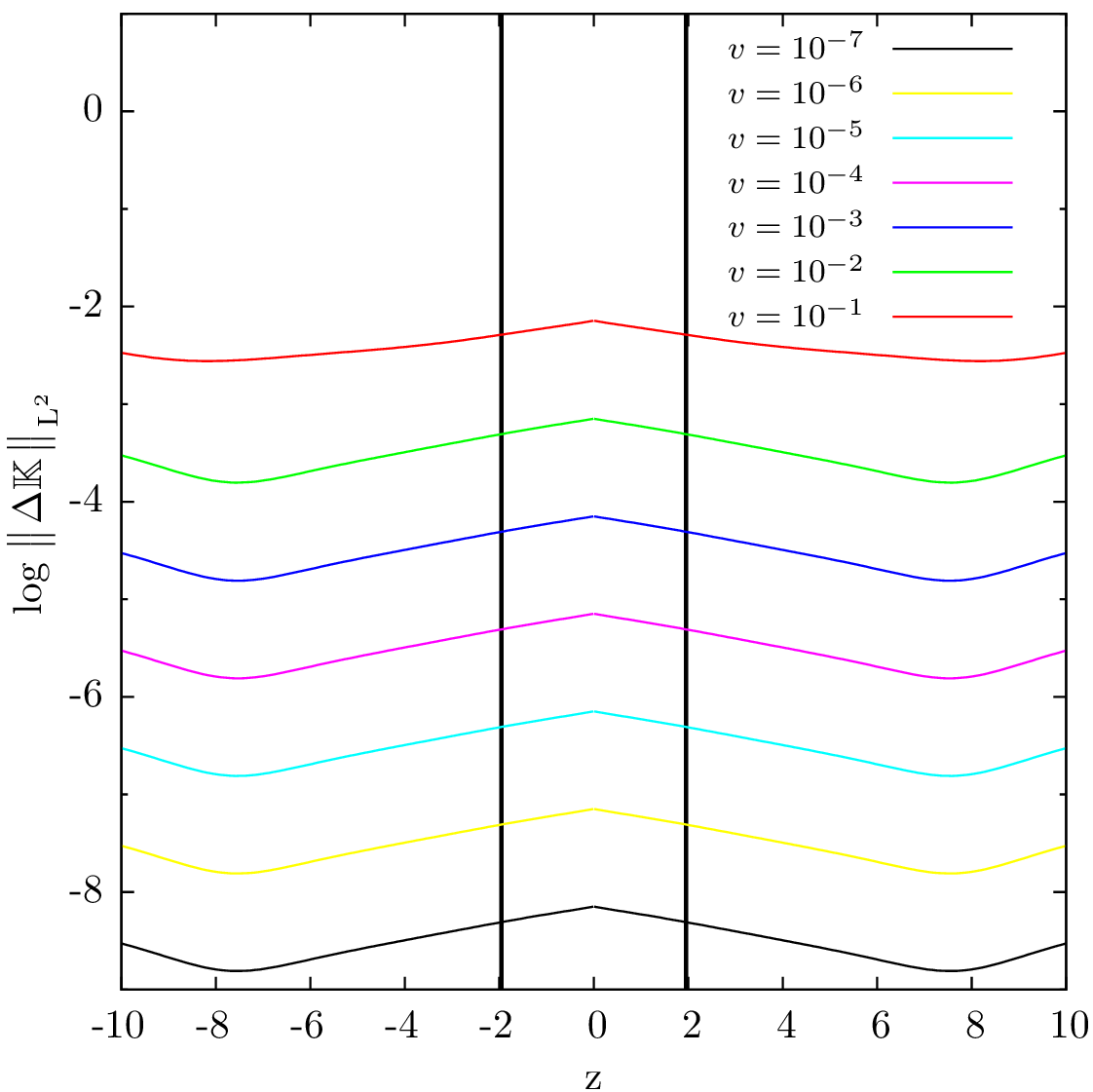}
\caption{The $z$-dependence of {the logarithm of the} L$^2$ norms of deviations, $\mathcal{F}^\Delta$, for a distorted Kerr black hole. The background, $\preA\hspace{-0.07cm}\mathcal{F}$, corresponded to a black hole with $M=1$ and $a=0.4$, while the initial-boundary data for the deviations corresponded to a boosted black hole with the same mass and spin. The investigated boosts $v$ are listed on each plot. Thick black vertical lines are to indicate the location of the event horizon of the background black hole in the $z$-direction.}
\label{fig:devK-L2}
\end{figure}

Figs.~\ref{fig:devS-L2} and~\ref{fig:devK-L2} present the L$^2$ norms of the variables as functions of the $z$-coordinate in the case of Schwarzschild and Kerr black holes, for which the parameters of discrepancies were chosen to be displacement and boost, respectively. Several values of these parameters of the discrepancies were examined, ranging between $10^{-7}$ and $10^{-1}$, what collectively served as a specific sort of a convergence test of the applied numerical method. In both cases the deviations remained small and of the same order of magnitude as the integration along $z$ proceeded. Moreover, their values were smaller as the parameter{s} of discrepancies were chosen to be smaller. This implies that the numerical solution tends to the analytic one as the distortion decreases. The above two observations prove the accuracy of the numerical code prepared for the deviation form of the constraints.

\medskip

\begin{figure}[H]
\centering
\includegraphics[width=0.45\textwidth]{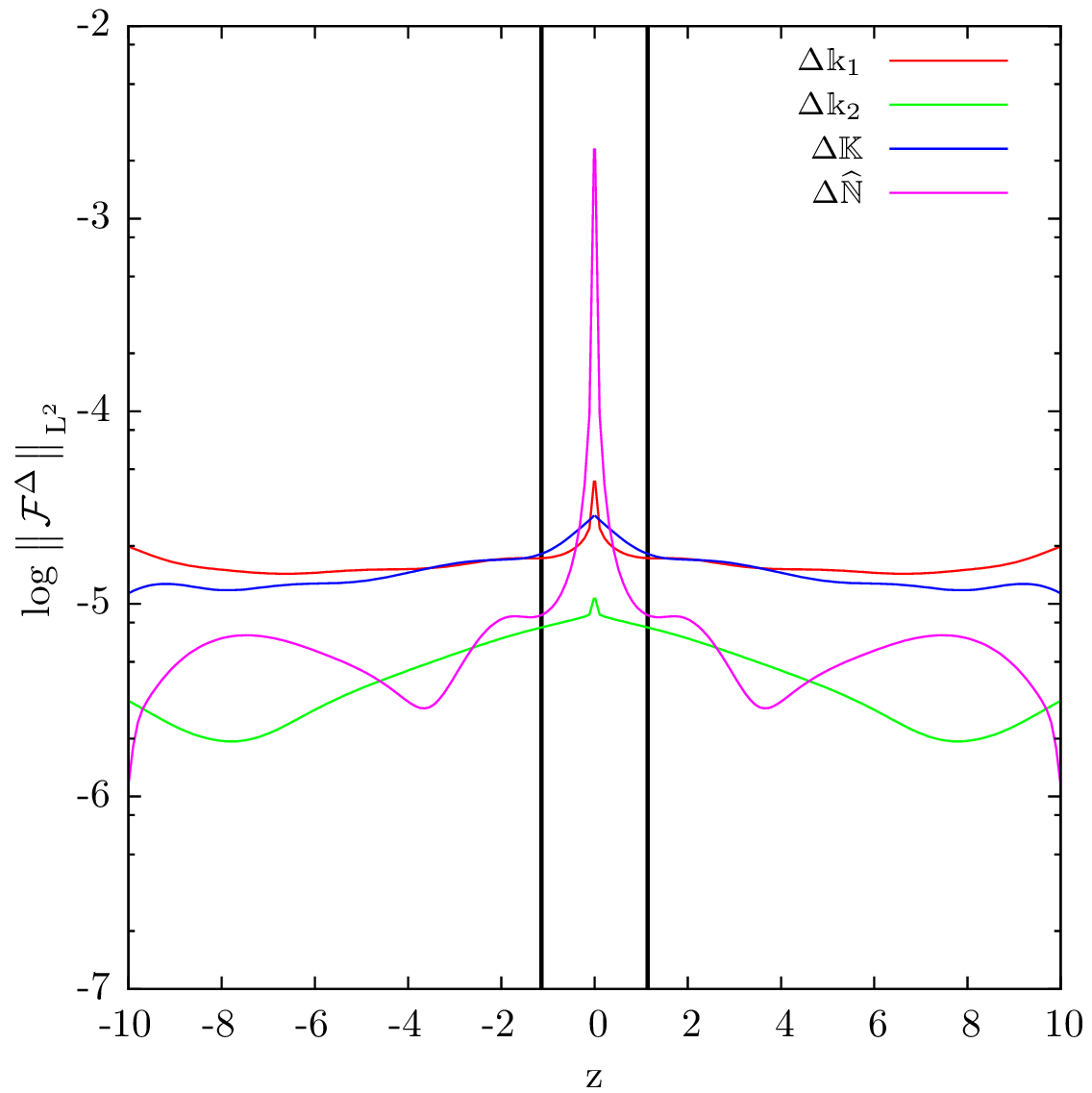}
\caption{The $z$-dependence of {the logarithm of the} L$^2$ norms of deviations, $\mathcal{F}^\Delta$, for a distorted highly boosted and highly spinning black hole. The background, $\preA\hspace{-0.07cm}\mathcal{F}$, corresponded to a black hole with $M=1$, $a=0.99$ and $v=0.99$, while the initial-boundary data for the deviations corresponded to a displaced black hole with the same mass, spin and boost and $d$ equal to $10^{-4}M$. Thick black vertical lines are to indicate the location of the event horizon of the background black hole in the $z$-direction.}
\label{fig:devK-L2bb}
\end{figure}

\begin{figure}[H]
\centering
\includegraphics[width=0.45\textwidth]{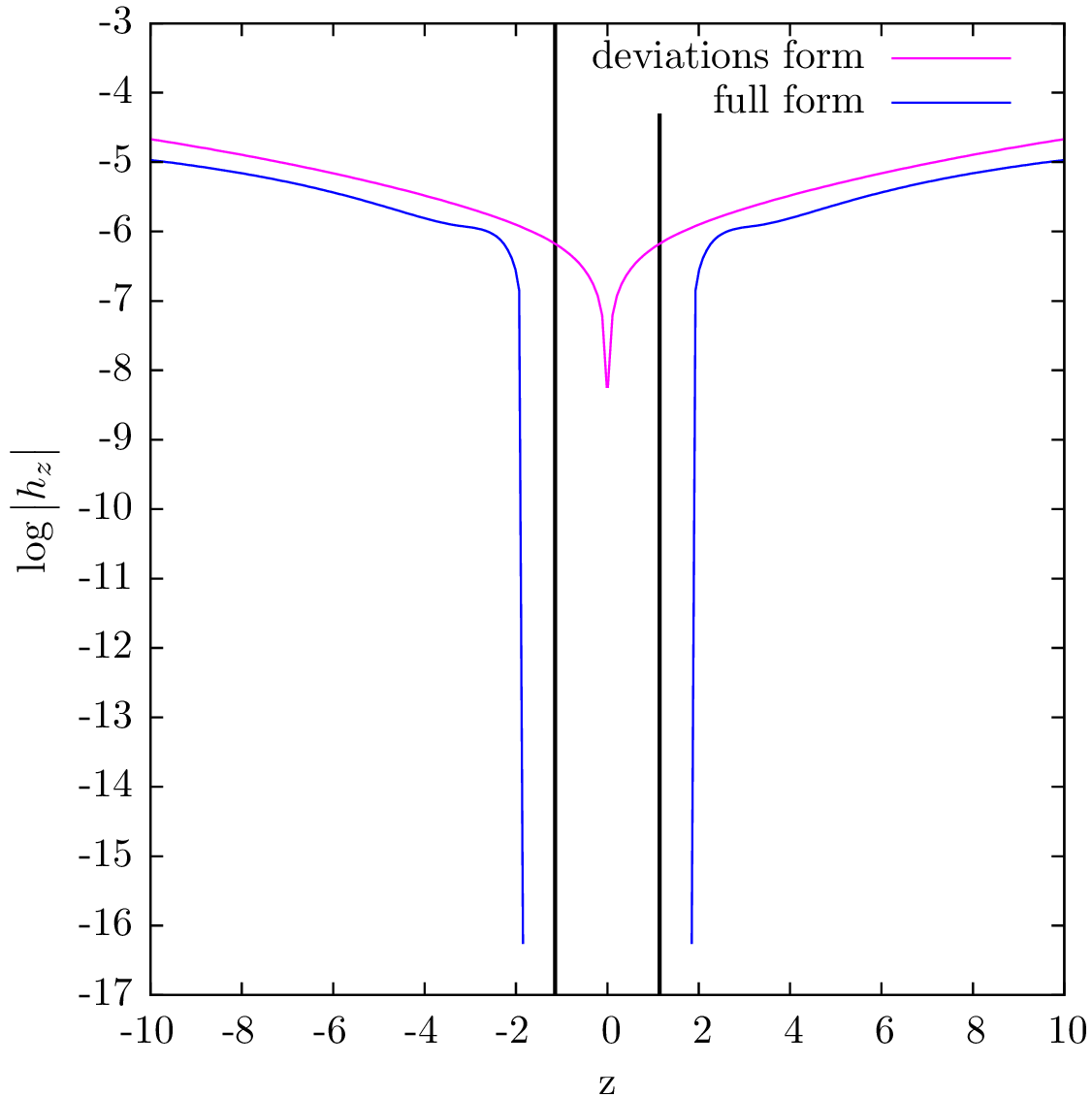}
\caption{The {logarithm of the} adaptive $z$-step size, $h_z$, versus $z$ is depicted for a highly boosted, highly spinning and displaced Kerr black hole with $M=1$, $a=0.99M$, $v=0.99$, in the deviations formulation of the constraints with $d=10^{-4}M$ and their full form with $d=3M$. Thick black vertical lines are to indicate the location of the event horizon of the background black hole in the $z$-direction.}
\label{fig:zsteps2-sing-Bav}
\end{figure}

The tests were also carried out for a highly boosted and highly spinning black holes. The plots of the L$^2$ norms analogous to the ones discussed above for a selected representative case of this type are presented in Fig.~\ref{fig:devK-L2bb}. These results verify that the method is accurate also for extreme values of physical parameters of the involved black holes, as the deviations also remain small within the entire domain. Moreover, in contrast to the full form case discussed in Section~\ref{ssec:fullform-relevance}, the computations involving big boosts do not require the use of a significantly dense{r} grid in the $x$ and $y$ directions. This is due to the fact that there are no big gradients in the constrained functions, what enforced using a denser grid in order to get a satisfactory accuracy of the derivatives when using the full form of the parabolic-hyperbolic equations. Consequently, the $z$-step {is} not required to decrease as extremely as in the full case while the $z$-integration is performed. Thus, the time of computations remains comparable to the cases involving smaller boosts. This observation justifies the superiority of the deviations formulation over the full form of the constraints in practical use. A comparison of the $z$-steps for the 'time'-integration within the full and deviations forms of the constraint equations is shown in Fig.~\ref{fig:zsteps2-sing-Bav}.

\begin{figure}[H]
\centering
(a)\hspace{7cm}(b)\\
\includegraphics[width=0.495\textwidth]{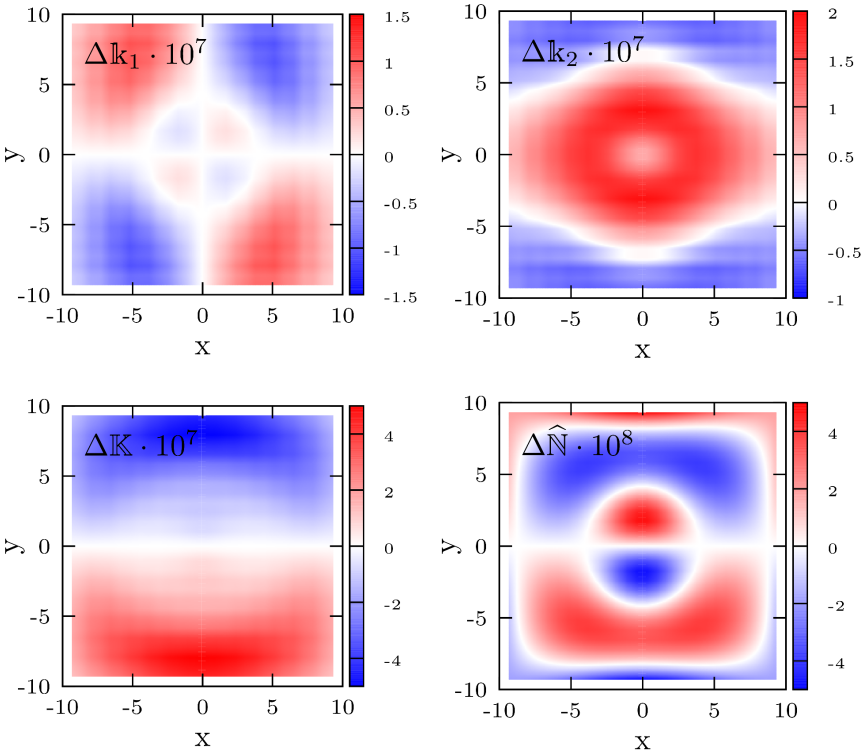}
\includegraphics[width=0.495\textwidth]{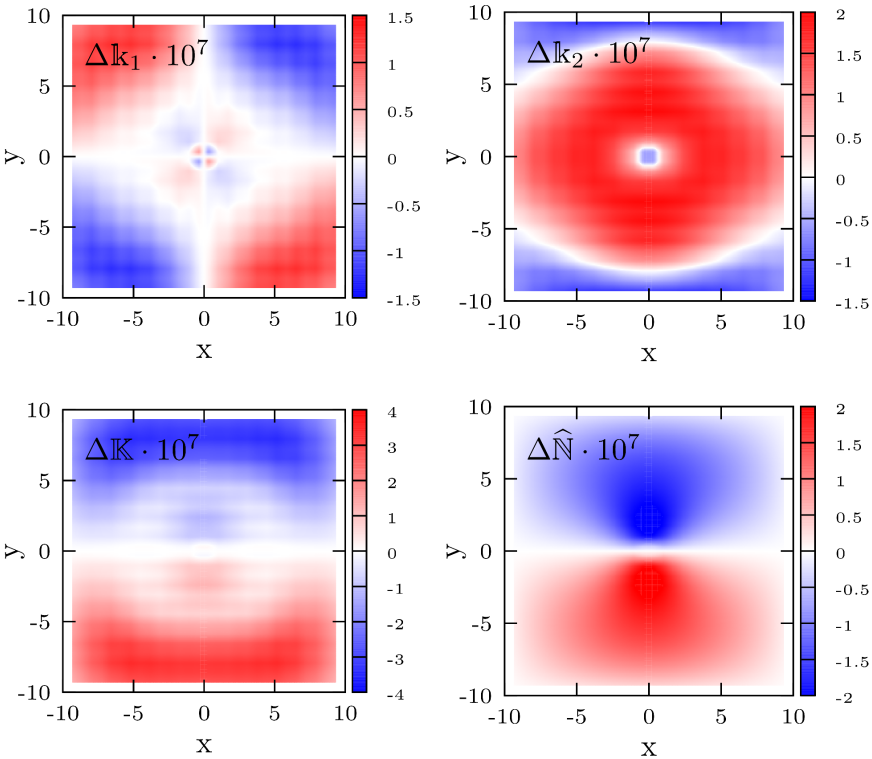}
\caption{The deviations, $\mathcal{F}^\Delta$, for a distorted Schwarzschild black hole within the setup described in the caption of Fig.~\ref{fig:devS-L2} with the displacement $d$ equal to $10^{-4}$ at (a)~$z=r_H$ and (b)~$z=\tfrac1{40} r_H$.}
\label{fig:devS-devz}
\end{figure}

\begin{figure}[H]
\centering
(a)\hspace{7cm}(b)\\
\includegraphics[width=0.495\textwidth]{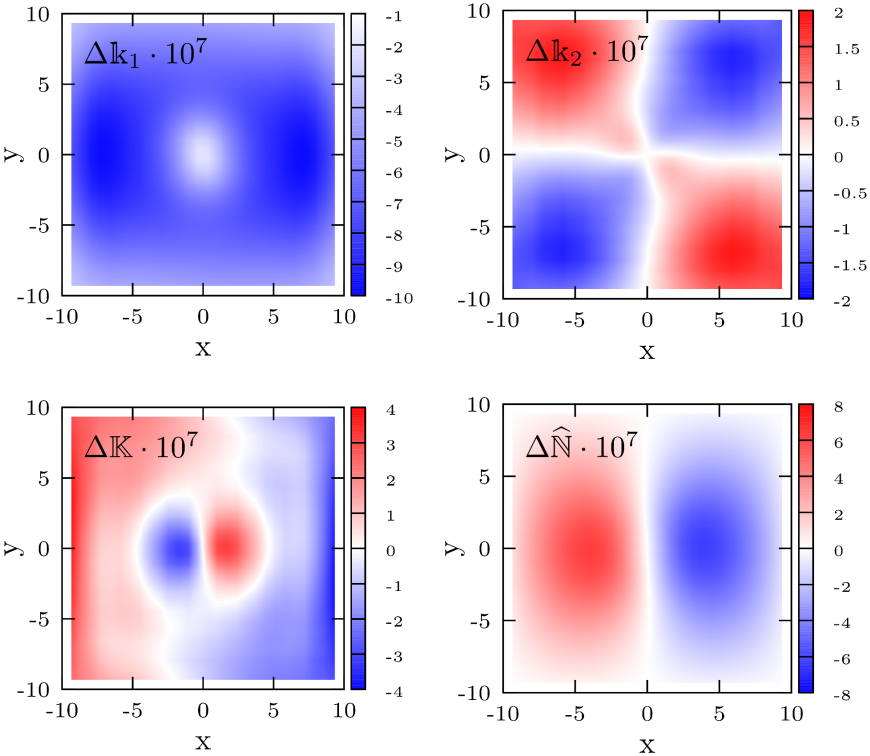}
\includegraphics[width=0.495\textwidth]{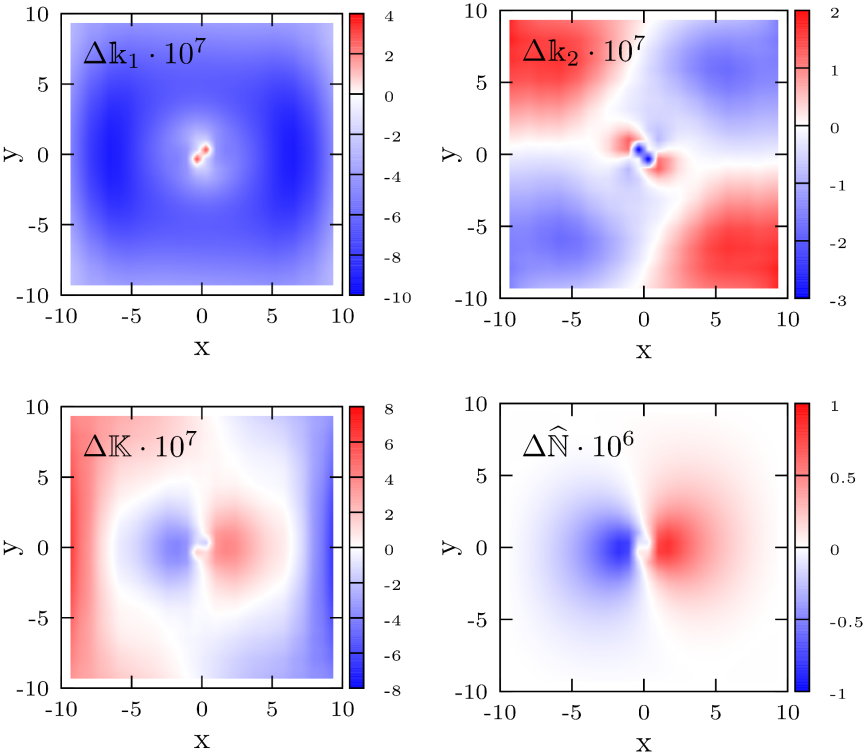}
\caption{The deviations, $\mathcal{F}^\Delta$, for a boosted Kerr black hole within the setup described in the caption of Fig.~\ref{fig:devK-L2} with the boost $v$ equal to $10^{-4}$ at (a)~$z=r_H$ and (b)~$z=\tfrac1{40} r_H$.}
\label{fig:devK-devz}
\end{figure}

\begin{figure}[H]
\centering
(a)\hspace{7cm}(b)\\
\includegraphics[width=0.495\textwidth]{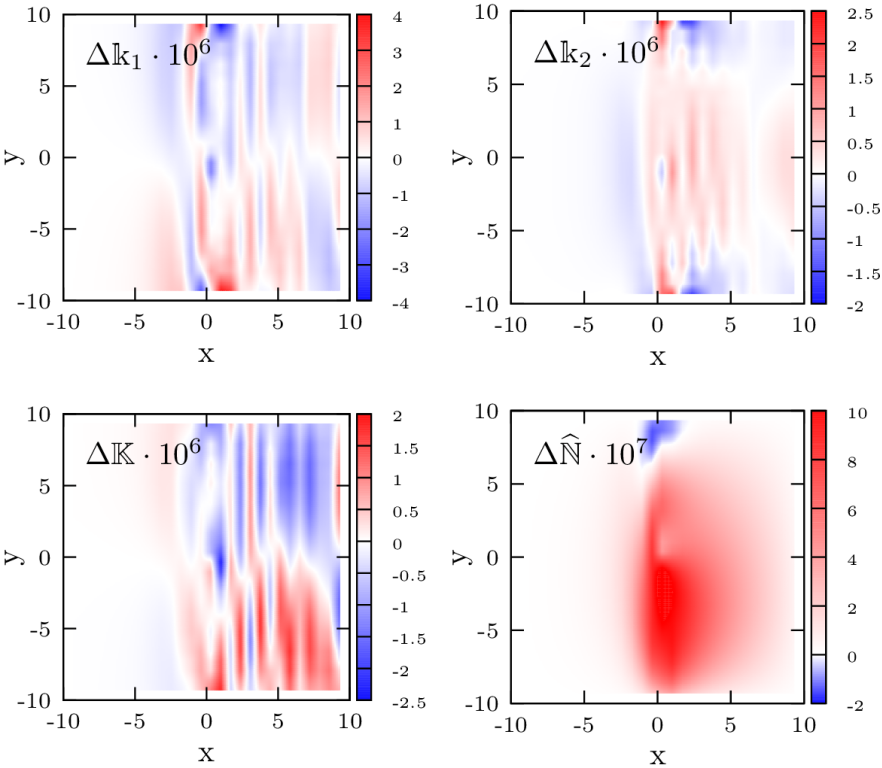}
\includegraphics[width=0.495\textwidth]{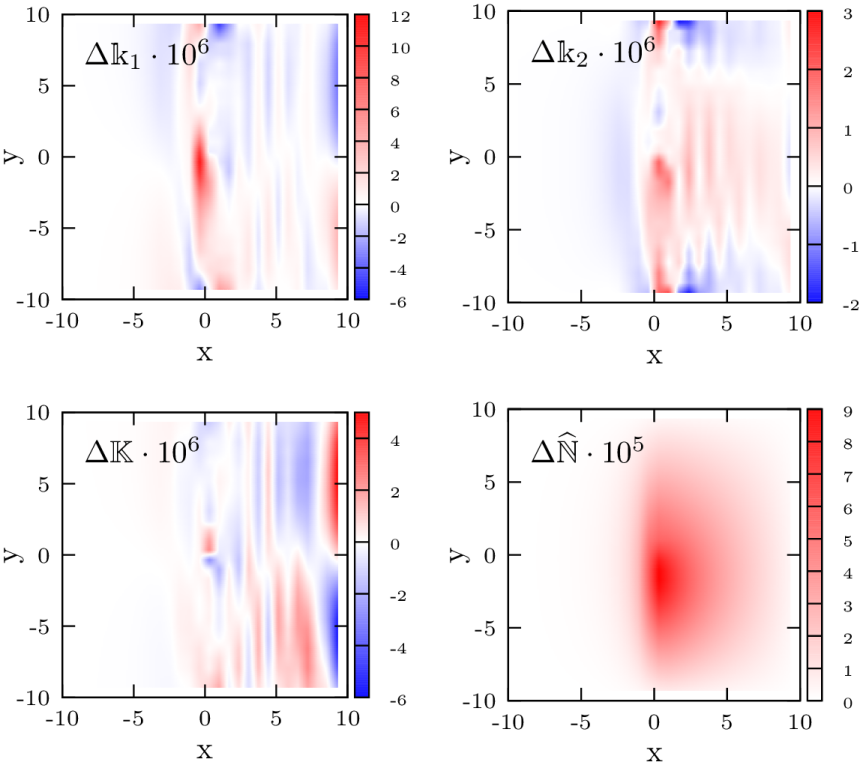}
\caption{The deviations, $\mathcal{F}^\Delta$, for a highly boosted and highly spinning black hole within the setup described in the caption of Fig.~\ref{fig:devK-L2bb} at (a)~$z=r_H$ and (b)~$z=\tfrac1{40} r_H$.}
\label{fig:devK-devzbb}
\end{figure}

Projections of the constrained functions deviations from the respective Kerr-Schild solution recorded on the $z=const$ slices corresponding to $r_H$ and $\tfrac1{40}r_H$ for the above-discussed distorted black holes with the values of the discrepancy parameters equal to $10^{-4}$ are shown in Figs.~\ref{fig:devS-devz} and~\ref{fig:devK-devz}. The corresponding results relevant for the case of a highly boosted and highly spinning black hole are presented in Fig.~\ref{fig:devK-devzbb}.

\subsection{The study of distorted black holes configurations}
\label{ssec:devs-results}

\begin{figure}[H]
\centering
(a)\hspace{7cm}(b)\\
\includegraphics[width=0.495\textwidth]{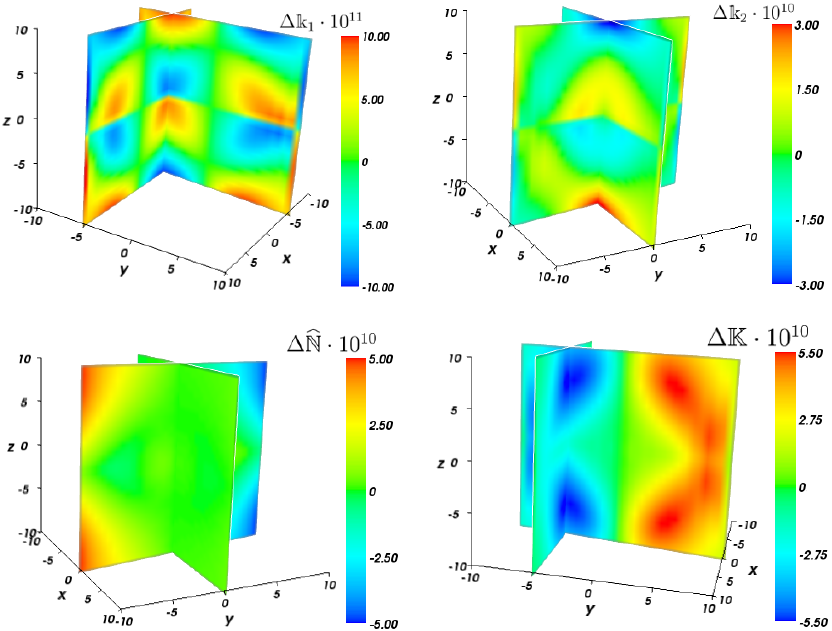}\hspace{0.5cm}
\includegraphics[width=0.495\textwidth]{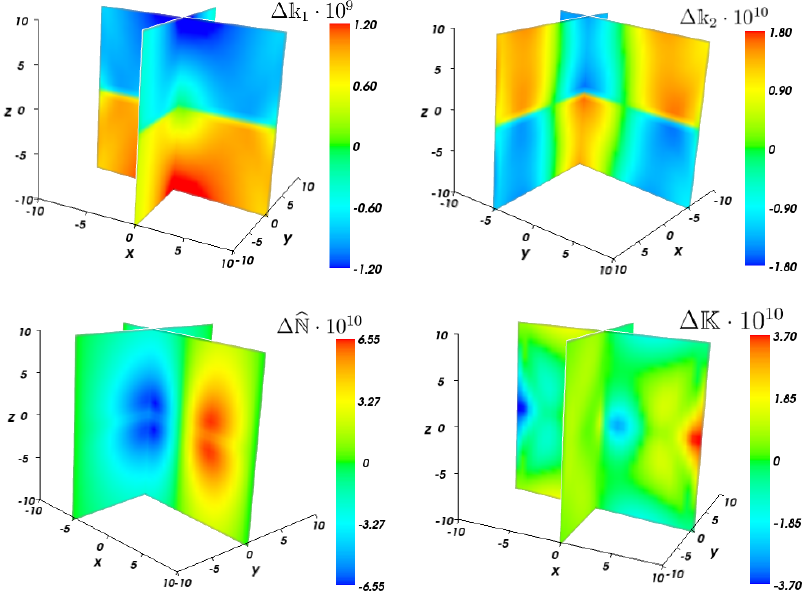}
\caption{The deviations, $\mathcal{F}^\Delta$, for distorted black holes with (a)~background functions $\preA\hspace{-0.07cm}\mathcal{F}$ corresponding to $M=1$ and the deviated initial-boundary data related with the same $M$ parameter and $d=10^{-7}M$ and (b)~background functions $\preA\hspace{-0.07cm}\mathcal{F}$ corresponding to $M=1$, $a=0.4M$ and the deviated initial-boundary data related with the same $M$ and $a$ parameters and $v=10^{-7}$.}
\label{fig:dev-devs}
\end{figure}

The constrained variables {of the deviation form of the constraints} are shown for two distorted black holes in Fig.~\ref{fig:dev-devs}. Different values of the deviating parameters lead to the same distribution of the functions, differing only in their order of magnitude.

\begin{figure}[H]
\centering
(a)\hspace{7cm}(b)\\
\includegraphics[width=0.3\textwidth]{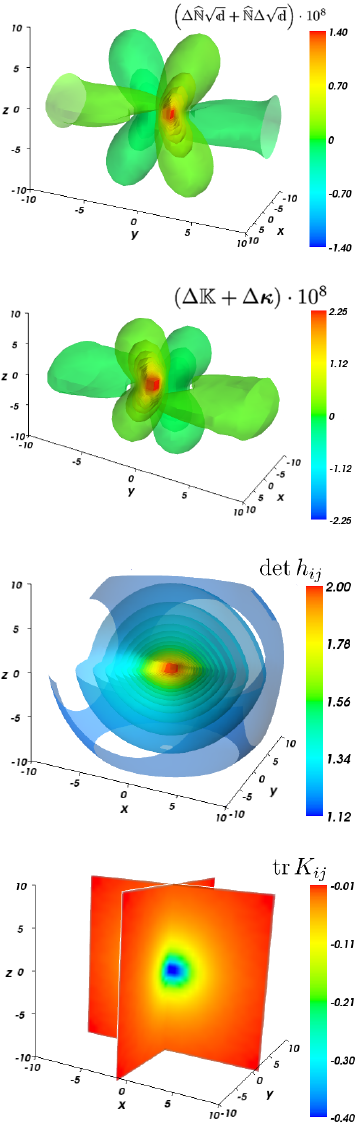}\hspace{0.5cm}
\includegraphics[width=0.3\textwidth]{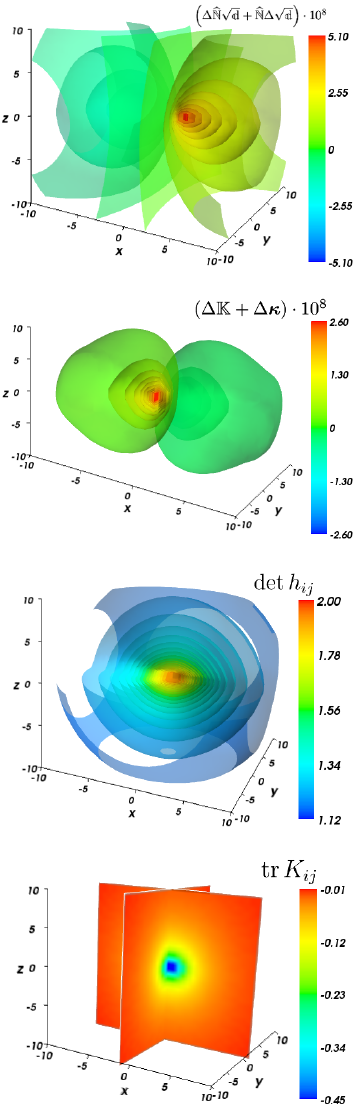}
\caption{Quantities {$\DNh\sqrt{\mathbbm{d}}+\Nh\Delta\sqrt{\mathbbm{d}}$}, $\DKK+\Delta{\boldsymbol{\kappa}}$, $\det h_{ij}$ and $\textrm{tr}K_{ij}$ for black holes with parameters listed in the caption of Fig.~\ref{fig:dev-devs}.}
\label{fig:dev-quant}
\end{figure}

Linearized versions of the gauge invariant quantities, i.e. $\DNh{\sqrt{\mathbbm{d}}}+\Nh\Delta{\sqrt{\mathbbm{d}}}$  and $\DKK+\Delta{\boldsymbol{\kappa}}$, together with $\det h_{ij}$ and $\textrm{tr}K_{ij}$, are depicted in Fig.~\ref{fig:dev-quant}. The quantities $\Delta\sqrt{\mathbbm{d}}$ and $\Delta\boldsymbol{\kappa}$ denote differences between {the particular freely specifiable functions of the} two analytic solutions of the {constraints that correspond to pure Kerr-Schild black holes} -- the {distorted initial-boundary and} background {ones}.

\section{Conclusions}
\label{sec:concl}
\setcounter{equation}{0}

This paper is to present results concerning the numerical integration of the parabolic-hyperbolic form of the constraints as applied to single vacuum black hole configurations. The parabolic-hyperbolic form of the constraints {was} used in two alternative approaches. In the full form and in their deviations based form. In either case the constraints were solved as an initial-boundary value problem using the analytically known Kerr-Schild black holes as reference solutions. 

\medskip

In the case of the full form of the parabolic-hyperbolic system the accuracy of our code was tested against analytically known initial data relevant for single non-distorted vacuum Kerr-Schild black holes. As accuracy indicators we used L$^2$ norms, absolute and relative errors. The tests revealed that the developed code is capable to be used in investigating black holes---with high precision, up to a few percent in the entire domain of integration apart from a small neighborhood of the singularity---even though they possess arbitrarily large spins and boosts. 
Our numerical code was found as expected to be $4^{th}$ order accurate. In particular, the accuracy of the code in applying it in the deviations based formalism was checked via tracing the order of magnitude of the deviations for distorted black holes, which is expected to remain small within the entire domain if the initial-boundary distortion is chosen to be small. It turned out that it is indeed the case, verifying thereby the appropriateness of the algorithm applied in the deviations based form.

\medskip

The numerical integration of the constraints turned out to be much more efficient in the form based on deviations. This confirmed our expectations--based on some analytic arguments--that there have to occur (fortunate) regularizations in some of the coefficients blowing up at the Cauchy horizon in the full setup. As a direct consequence of this some of the calculations were considerably less time-consuming in the deviations based case. In particular, the deviation method turned out to be significantly more time efficient--in comparison {to} the full setup--when highly boosted black holes were involved. This observation is of crucial importance concerning the construction of initial data for black hole binaries--which are in the forehead of our coming investigations--in applying the evolutionary form of the constraints.

\medskip

As already indicated the combination of the parabolic-hyperbolic form of the constraints and our numerical methods allowed us to investigate black holes with spins and boosts taking values from their entire physically allowed domains. In particular, the applied numerical method is capable {of} investigat{ing} black holes possessing spin and speed close to the maximal values of their physically allowed ranges. 

\medskip

The most important new feature provided by the applied initial-boundary value formulation of the parabolic-hyperbolic form of the constraints is that it does not require the use of boundary conditions in the strong field regime, i.e.~in the vicinity of the singularity of the Kerr-Schild black hole. This is in contrast to all the alternative methods, e.g.~those based on the elliptic (or conformal) method, as the `time'-integration of the underlying parabolic-hyperbolic system determines the values of the constrained variables in the strong field regime explicitly.

\medskip

As was mentioned in the Introduction, the long-term goal of the research is applying the parabolic-hyperbolic formulation along with the proposed numerical approach to obtain initial data sets in the case of binary black hole systems. On the analytical side, the generalization will require substituting the Kerr-Schild form of the black hole configurations (\ref{eq:ksm}) presented in section~\ref{sec:boo:spin:bh} by its binary version, i.e., the superposed Kerr-Schild data~\cite{Racz1605.01669,raczsuppl2017}
\begin{equation}\label{eqn:bin}
g_{\alpha\beta}=\eta_{\alpha\beta}+2 H^{[1]} \ell^{[1]}_{\alpha} \ell^{[1]}_{\beta} +2 H^{[2]} \ell^{[2]}_{\alpha} \ell^{[2]}_{\beta}.
\end{equation}
The subsequent analytical steps will remain the same, that is both the freely specifiable background functions within the whole computational domain and the constrained variables on the boundary should be imposed according to (\ref{eqn:bin}). Numerical implementation of the modification will require including separate parameters for both black holes of the binary. The most challenging part in solving the equations numerically will be maintaining adequate accuracy as the plane containing singularities will be approached. The variations of the constrained functions in the binary case are expected to be bigger in comparison to the single black hole cases and thus the grid in the spatial directions should be finer, what entails a decrease of the integration step in the `temporal' direction. For this reason, and having in mind relations presented in Fig.~\ref{fig:zsteps2-sing-Bav}, the deviations based formalism of the parabolic-hyperbolic formulation of the initial data seems more promising in the context of investigating binary black hole systems. Moreover, obtaining satisfactory error levels combined with reasonable computing time would possibly require using advanced numerical approaches such as parallel or distributed programming when constructing binary black hole initial data.

The generalization of the presented approach to binary black hole systems would also come with a need of investigating the production and propagation and ultimately also reducing the impact of the 'junk' radiation that is typical for the elliptic construction and may also appear in the outcoming 'parabolic-hyperbolic' initial data.

\section*{Appendix: The explicit form of the derivative operators
	}
	\setcounter{equation}{0}
\label{app:derivs}
\renewcommand{\theequation}{A.\arabic{equation}}
\renewcommand{\thelemma}{A.\arabic{lemma}}

As mentioned in Section~\ref{sec:numcomp}, our numerical implementation required the use of an appropriate coordinate system in solving the constraint equations in their parabolic-hyperbolic form. In doing so Cartesian coordinates $\left(x,y,z\right)$ were used, along with the following representation of partial derivatives operators:
\begin{align}\label{substpart}
\partial_{\rho} &\rightarrow \partial_{z}\,, \nonumber\\
\pmb{\partial} &\rightarrow \px + \mathbbm{i}\, \py\,, \nonumber\\
\pmb{\bar{\partial}} &\rightarrow \px - \mathbbm{i}\, \py\,, \\
\pmb{ \bar{\partial} \partial} &\rightarrow \px^{\,2} + \py^{\,2}\,, \nonumber\\
\pmb{\partial \partial} &\rightarrow \px^{\,2} - \py^{\,2} + 2\,\mathbbm{i}\, \px \py\,, \nonumber\\
\pmb{\bar{\partial} \bar{\partial}} &\rightarrow \px^{\,2} - \py^{\,2} - 2\,\mathbbm{i}\, \px\py\,. \nonumber
\end{align}
These operators were used in an algebraically equivalent rephrased form of the applied system of parabolic-hyperbolic equations such that they involved only real variables,  real coefficients and real source terms.

\medskip

\begin{figure}[H]
\centering
\includegraphics[width=0.7\textwidth]{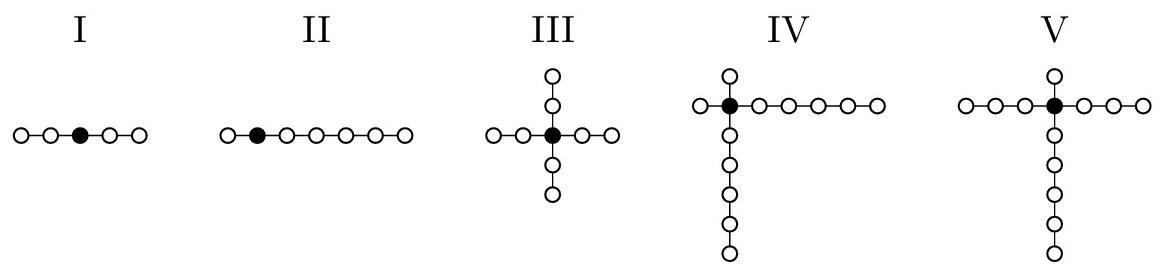}
\caption{Stencils used in calculations of first (I and II) and second (III--V) $x$- and $y$-derivatives of variables.}
\label{fig:stencils}
\end{figure}

The first and second $x$ and $y$ derivatives of the variables were calculated numerically using stencils sketched in Fig.~\ref{fig:stencils}. 
In the inner part of the $z=const$ slices, i.e., apart from the grid points next to the boundary, the fully centered 4$^{th}$ order accurate derivatives calculated on symmetric stencils I and III were used
\ben
\px\mathcal{F}\big|_{i,j} &=& \frac{1}{12h_x} 
\left( \mathcal{F}\big|_{i-2,j}-8\mathcal{F}\big|_{i-1,j}+8\mathcal{F}\big|_{i+1,j}-\mathcal{F}\big|_{i+2,j} \right), \\
\partial^{\: 2}_x\mathcal{F}\big|_{i,j} &=& \frac{1}{12h^{\,2}_x} 
\left( -\mathcal{F}\big|_{i-2,j}+16\mathcal{F}\big|_{i-1,j}-30\mathcal{F}\big|_{i,j}+16\mathcal{F}\big|_{i+1,j} -\mathcal{F}\big|_{i+2,j} \right),\hspace{0.25cm} \\
\partial^{\: 2}_{xy}\mathcal{F}\big|_{i,j} &=& \frac{1}{144h_x h_y} 
\bigg[ 8\left(\mathcal{F}\big|_{i+1,j-2}+\mathcal{F}\big|_{i+2,j-1}+\mathcal{F}\big|_{i-2,j+1}+\mathcal{F}\big|_{i-1,j+2} +\right. \nonumber\\
&&\left. -\mathcal{F}\big|_{i-1,j-2}-\mathcal{F}\big|_{i-2,j-1}-\mathcal{F}\big|_{i+1,j+2}-\mathcal{F}\big|_{i+2,j+1}\right) +\nonumber\\
&&+\mathcal{F}\big|_{i-2,j-2}+\mathcal{F}\big|_{i+2,j+2}-\mathcal{F}\big|_{i+2,j-2}-\mathcal{F}\big|_{i-2,j+2} +\nonumber\\
&&+64\left(\mathcal{F}\big|_{i-1,j-1}+\mathcal{F}\big|_{i+1,j+1}-\mathcal{F}\big|_{i+1,j-1}-\mathcal{F}\big|_{i-1,j+1} \right) \bigg],
\een
where $i$ and $j$ number the gridpoints in $x$ and $y$ directions, respectively.

The vicinity of the boundary enforced using asymmetric stencils II, IV and V, together with the 6$^{th}$ order accurate derivatives, which in the case of the forward, forward-forward and forward-centered versions, are
\ben
\px\mathcal{F}\big|_{i,j} &=& \frac{1}{60h_x} 
\left( -2\mathcal{F}\big|_{i-5,j}+15\mathcal{F}\big|_{i-4,j}-50\mathcal{F}\big|_{i-3,j}+\right.\nonumber\\
&&\left.+100\mathcal{F}\big|_{i-2,j}-150\mathcal{F}\big|_{i-1,j}+77\mathcal{F}\big|_{i,j}+10\mathcal{F}\big|_{i+1,j} \right), \\
\partial^{\: 2}_x\mathcal{F}\big|_{i,j} &=& \frac{-1}{180h^{\,2}_x} 
\left( 13\mathcal{F}\big|_{i-5,j}-93\mathcal{F}\big|_{i-4,j}+285\mathcal{F}\big|_{i-3,j}+\right.\nonumber\\
&&\left.-470\mathcal{F}\big|_{i-2,j}+255\mathcal{F}\big|_{i-1,j}+147\mathcal{F}\big|_{i,j}-137\mathcal{F}\big|_{i+1,j} \right), \\
\partial^{\: 2}_{xy}\mathcal{F}\big|_{i,j} &=& \frac{1}{3600h_x h_y} 
\left(\Xi_{-5}-7.5\Xi_{-4}+25\Xi_{-3}-50\Xi_{-2}+75\Xi_{-1}+\right. \nonumber\\
&&\left.-38.5\Xi_0-5\Xi_1 \right),
\een
where for forward-forward
\ben
\Xi_k &=& 4\mathcal{F}\big|_{i+k,j-5}-30\mathcal{F}\big|_{i+k,j-4}+100\mathcal{F}\big|_{i+k,j-3}+\nonumber\\
&&-200\mathcal{F}\big|_{i+k,j-2}+300\mathcal{F}\big|_{i+k,j-1}-154\mathcal{F}\big|_{i+k,j}-20\mathcal{F}\big|_{i+k,j+1}
\een
and for forward-centered
\ben
\Xi_k &=& 2\mathcal{F}\big|_{i+k,j-3}-18\mathcal{F}\big|_{i+k,j-2}+90\mathcal{F}\big|_{i+k,j-1}+\nonumber\\
&&-90\mathcal{F}\big|_{i+k,j+1}+18\mathcal{F}\big|_{i+k,j+2}-2\mathcal{F}\big|_{i+k,j+3}.
\een
The backward, backward-backward and backward-centered versions of the above relations require a reversing of signs.

\section*{Acknowledgments}

The authors gratefully acknowledge the support of the Albert Einstein Institute (AEI) in Golm, Germany, by providing us access to the IT resources to carry out our numerical investigations.
AN was supported by the Polish National Science Centre under a postdoctoral scholarship DEC-2016/20/S/ST2/00368. 
{\L}N was supported by the Polish National Science Centre under a postdoctoral scholarship DEC-2014/12/S/ST2/00332. IR was supported by the POLONEZ programme of the National Science Centre of Poland (under the project No. 2016/23/P/ST1/04195) which has received funding from the European Union's Horizon 2020 research and innovation programme under the Marie Sk{\l}odowska-Curie grant agreement No.~665778. IR is also grateful for partial support by NKFI grant K-115434 and the kind hospitality of AEI where parts of the theoretical backgrounds were carried out.

\includegraphics[scale=0.42, height = 42pt]{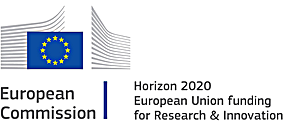} 

\bibliographystyle{nicestyle}
\bibliography{par-fosh-Nak.bib}

\end{document}